# Control of Flight

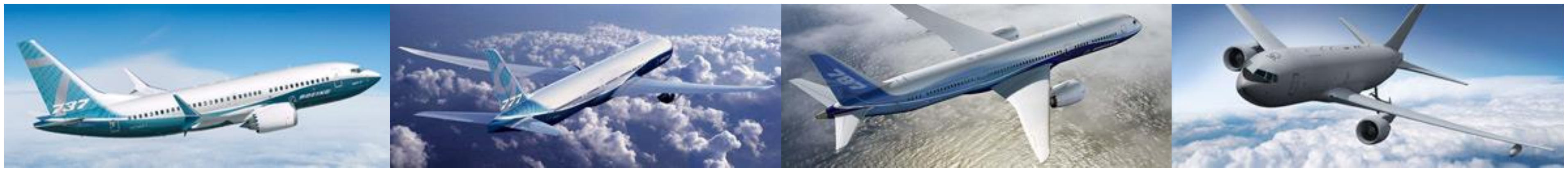

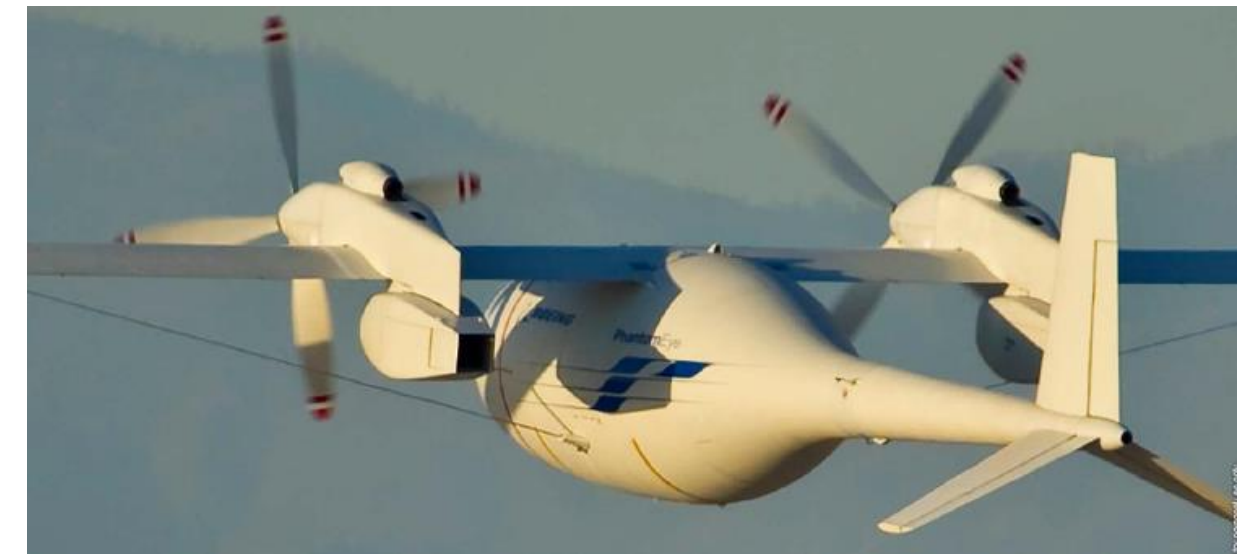

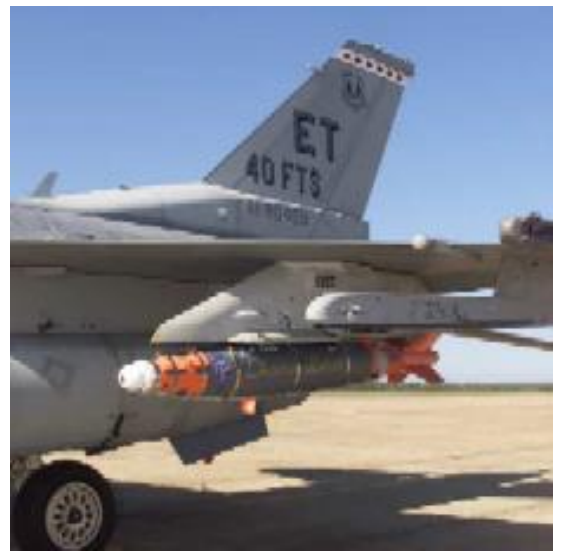

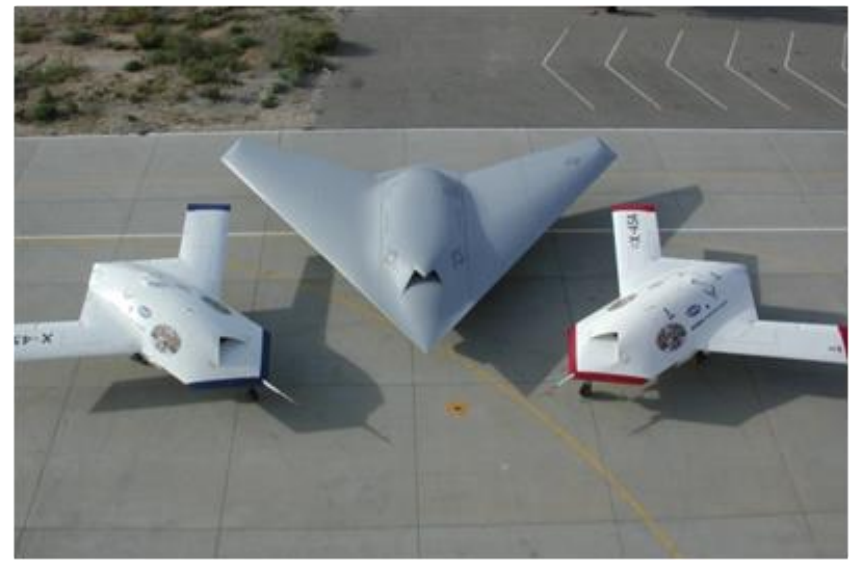

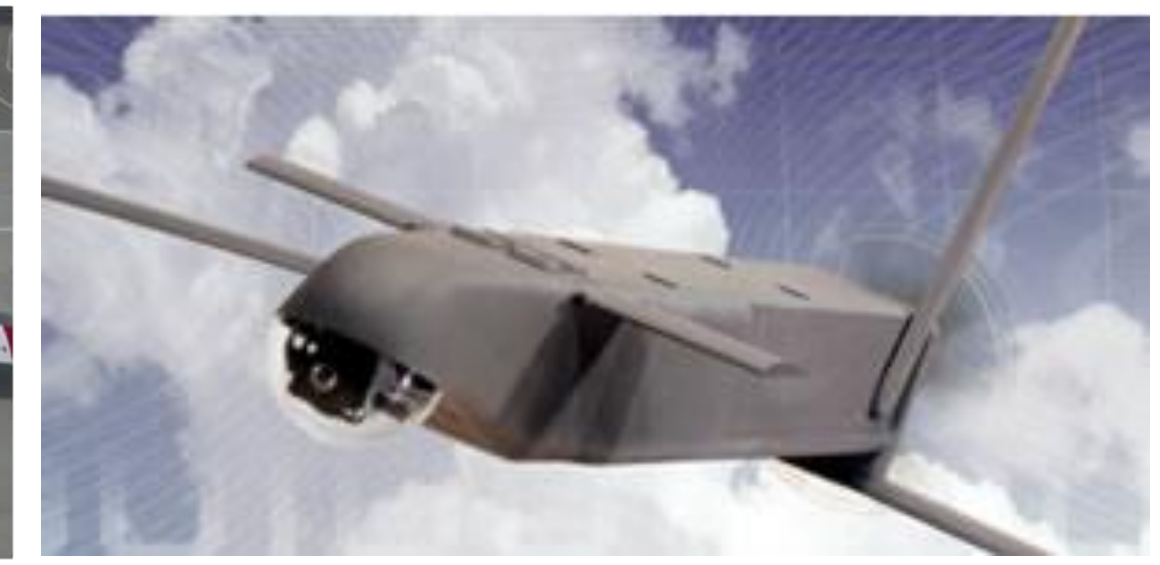

**Eugene Lavretsky, Ph.D.**
**Principal Senior Technical Fellow**

**The Boeing Company**
**Guidance, Control & Autonomy**

**06-17-2026**

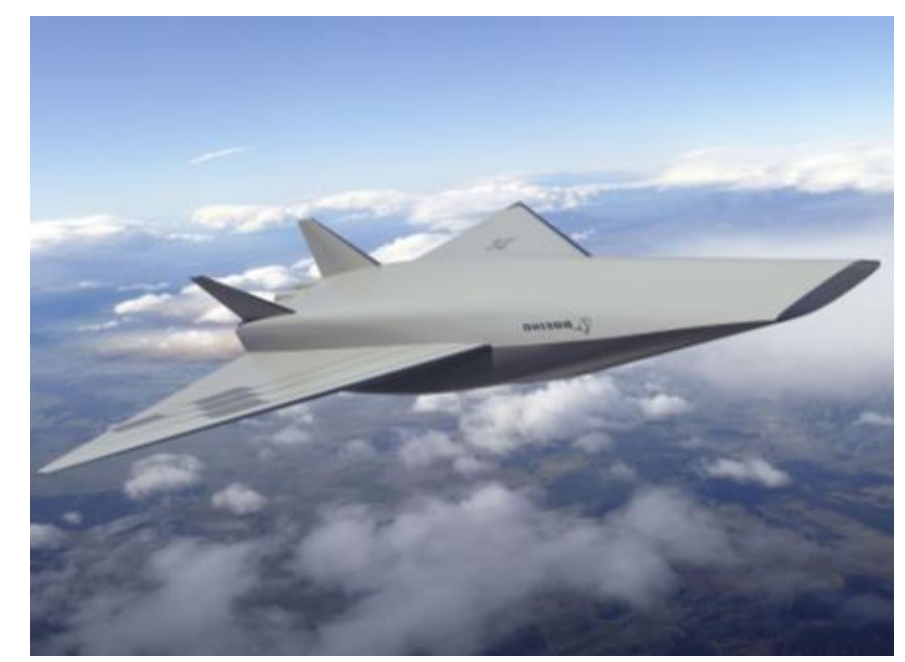

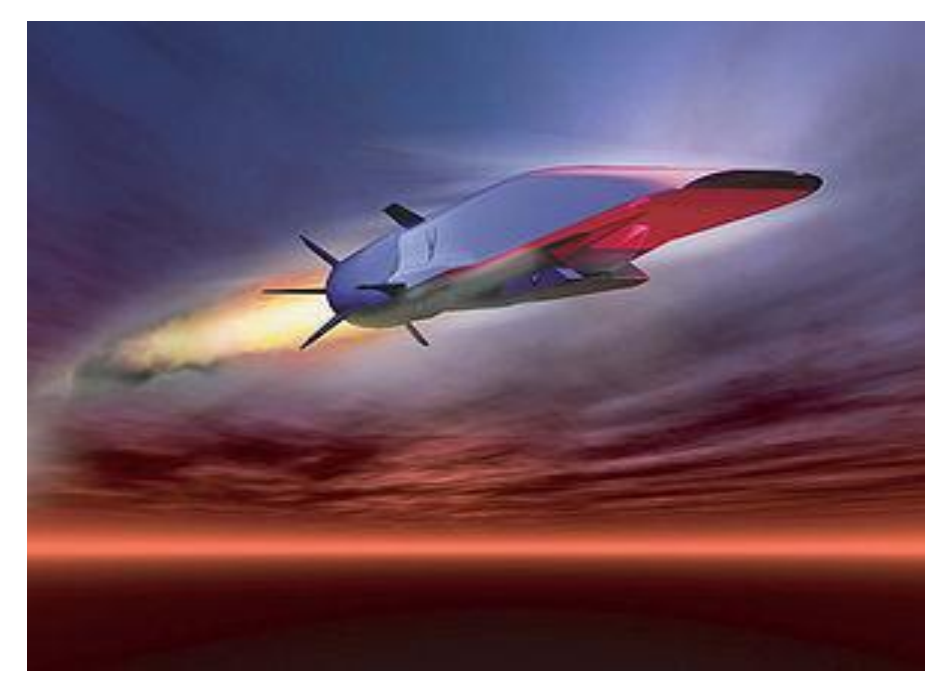

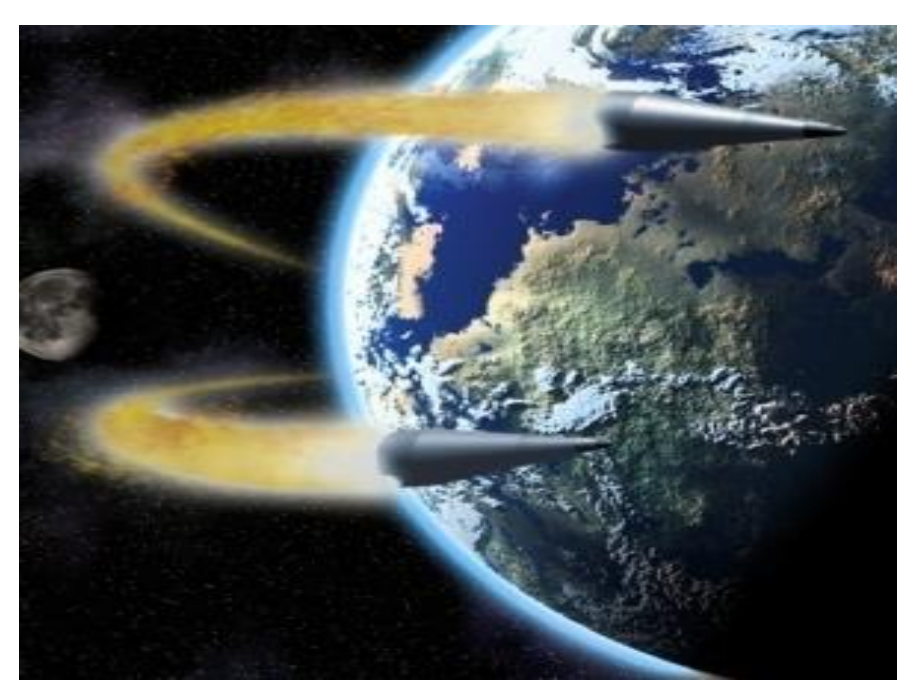

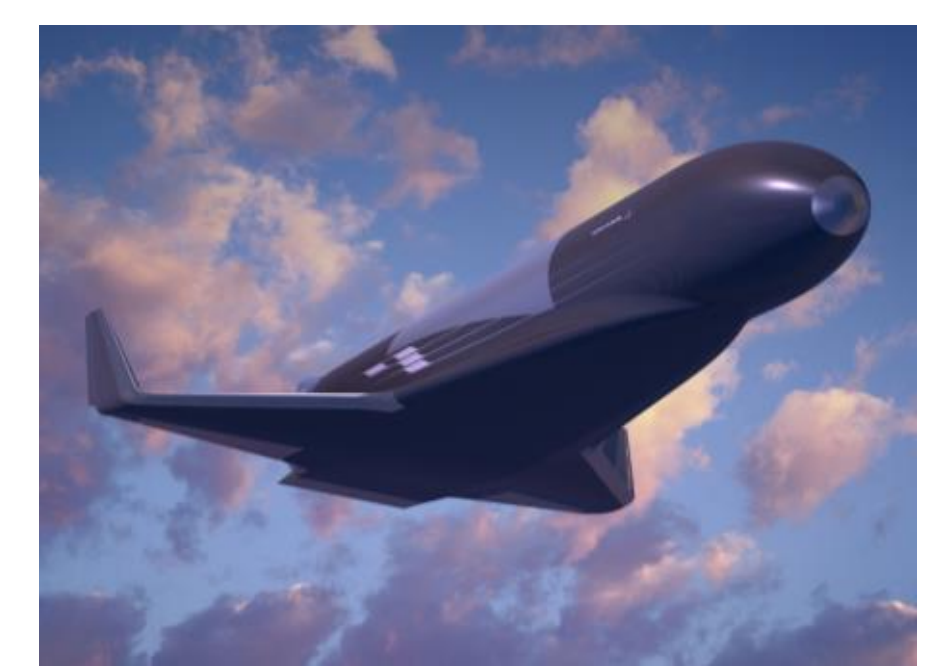

# Presentation Overview

- **Control of Flight**
  - Introduction, Definitions, State-of-the-Art Methods, and Technical Challenges

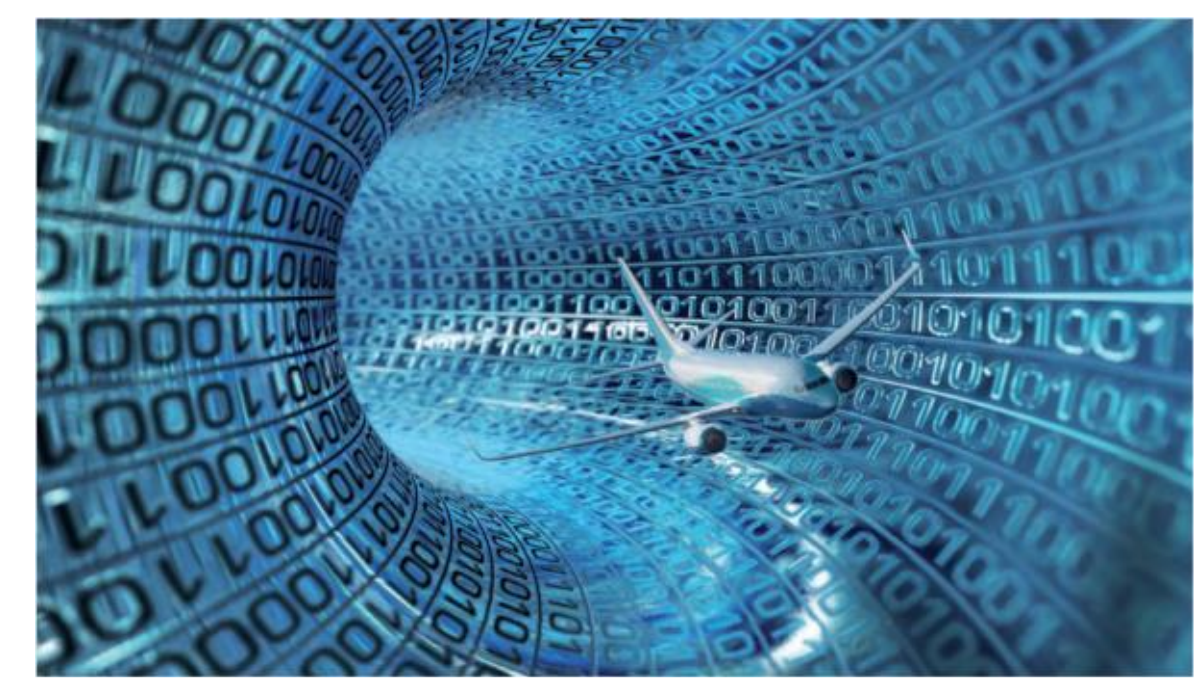

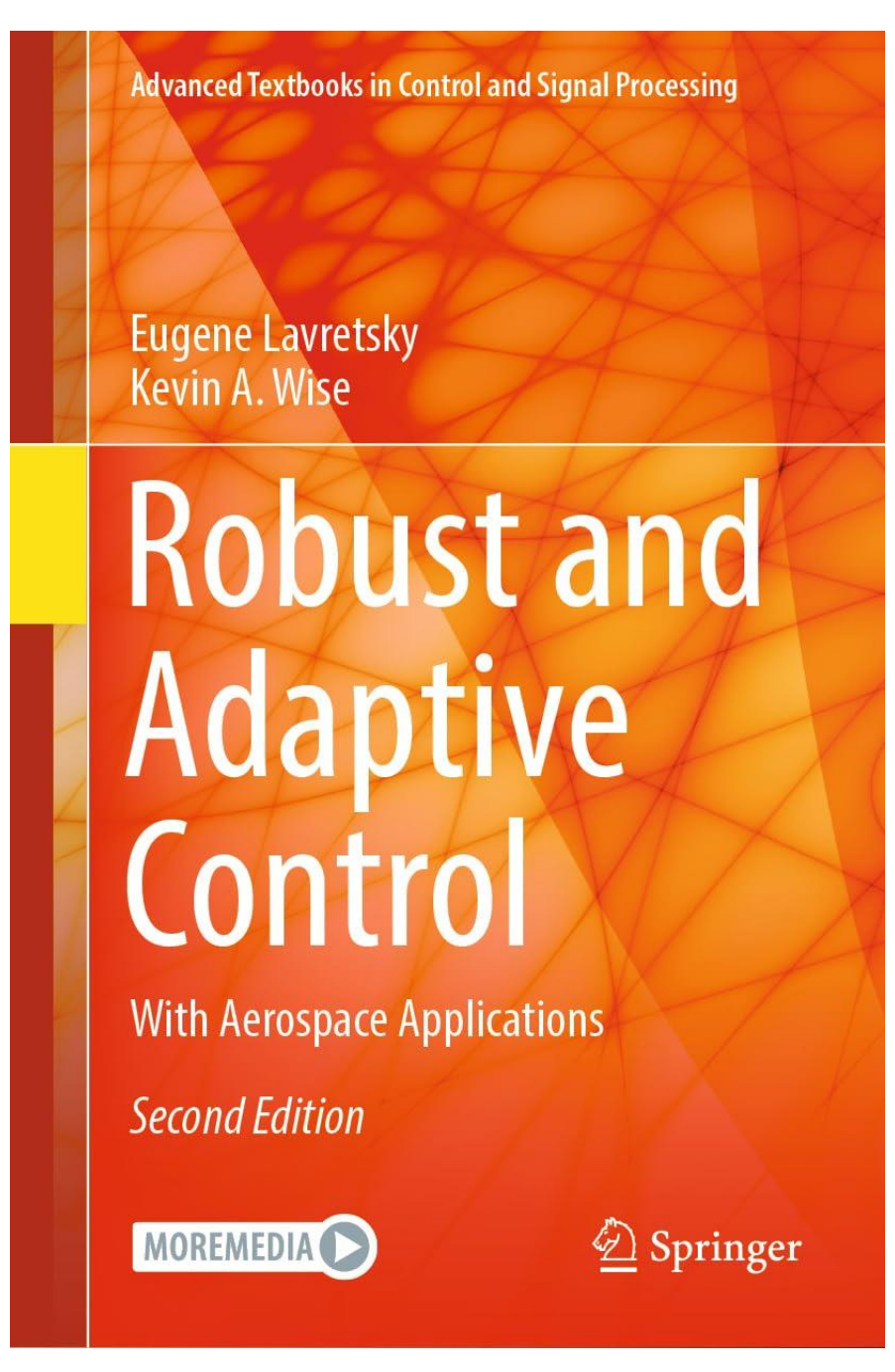

- **My Recent Contributions**
  - Observer Based Loop Transfer Recovery (OBLTR) Output Feedback Controllers with Direct Adaptive Augmentation

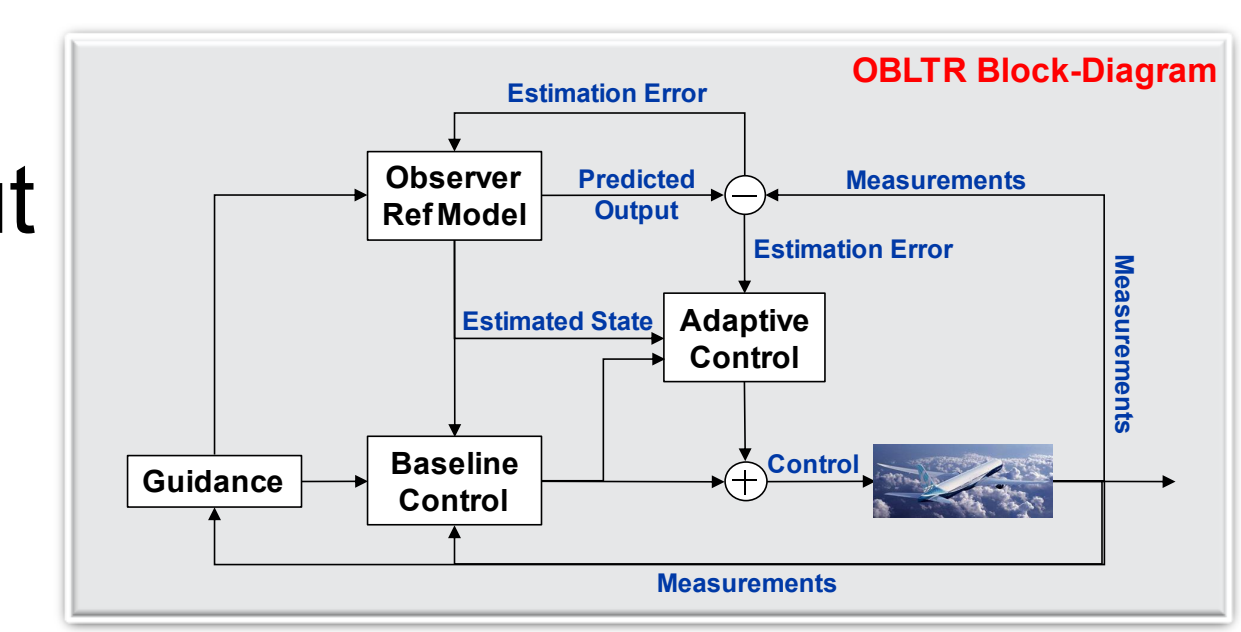

  - Control Augmentation Design in the Presence of Operational Constraints

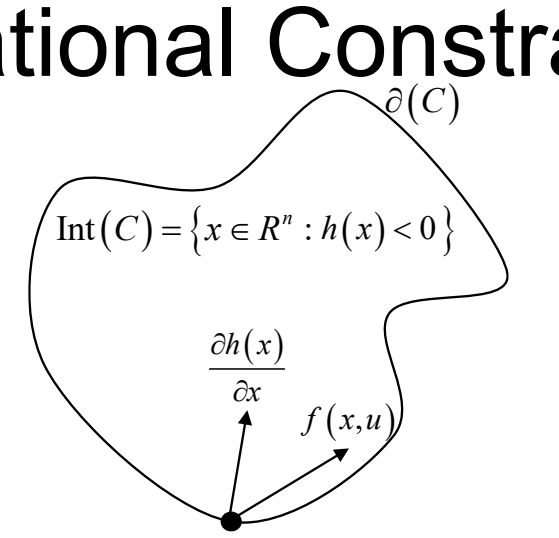

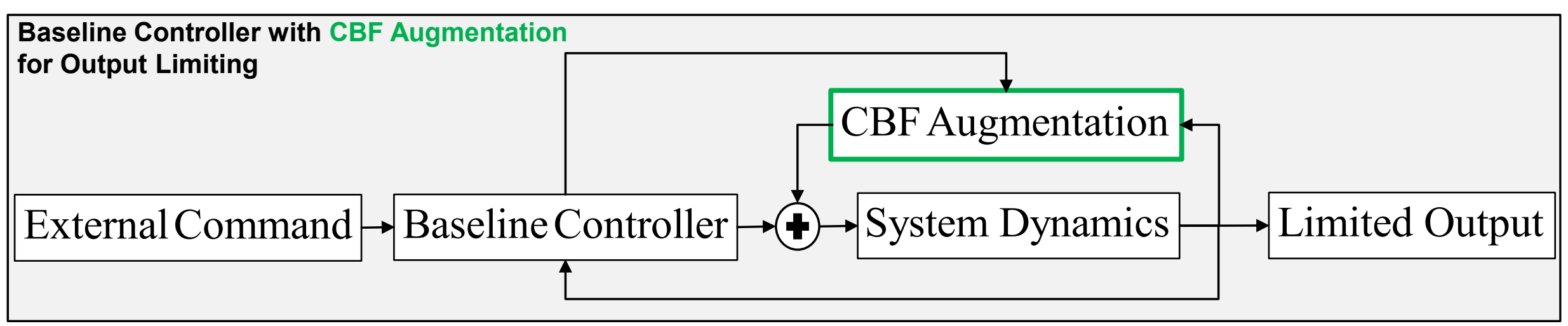

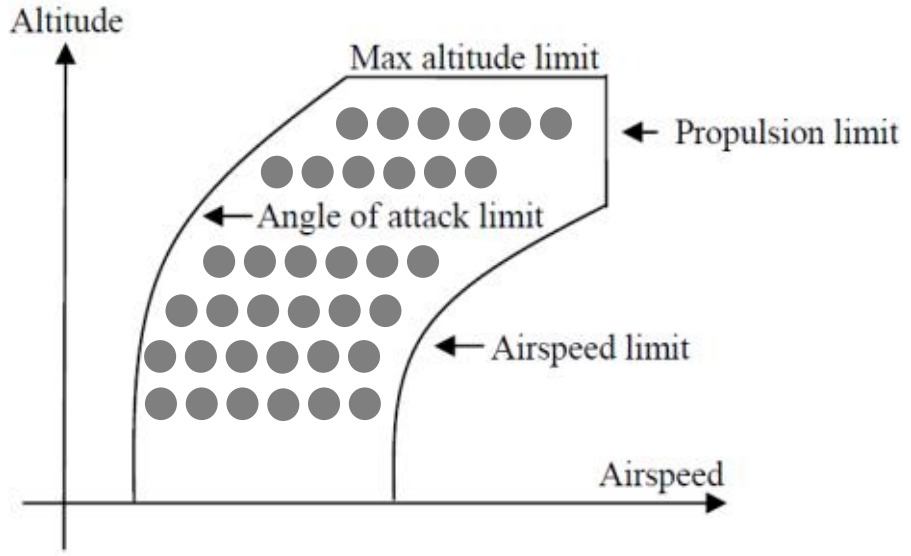

  - Unsteady Aerodynamics Modeling for Control Applications

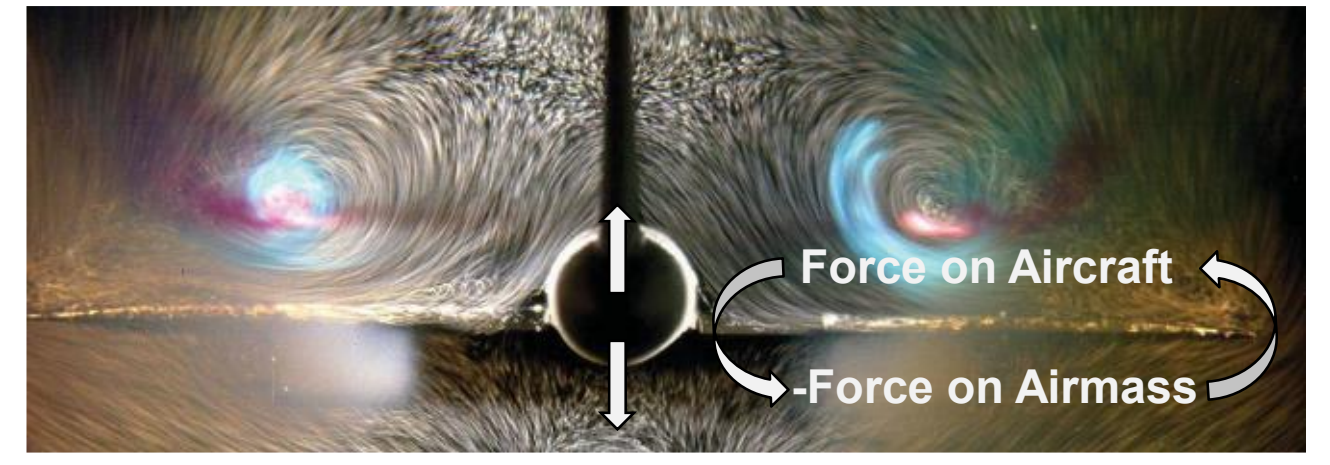

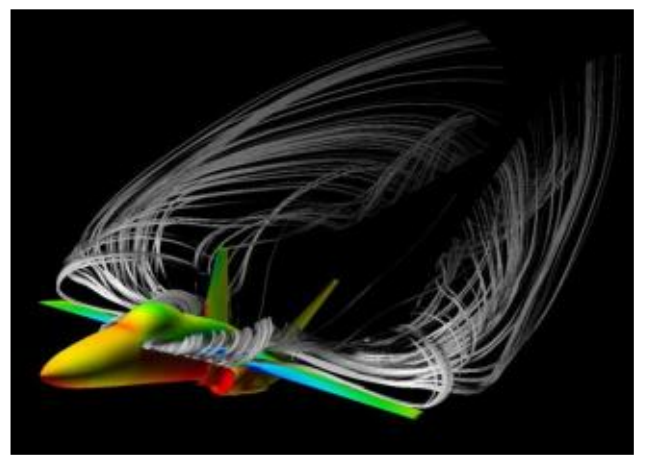

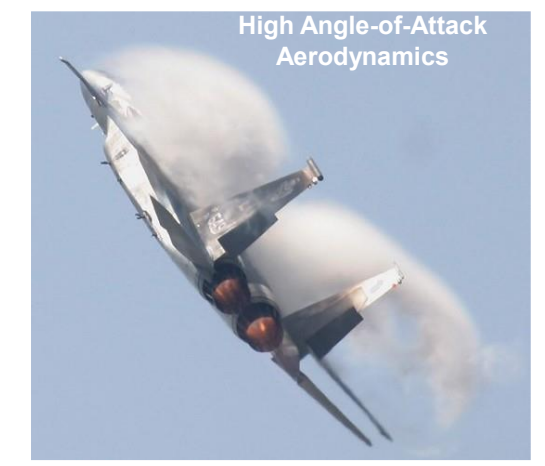

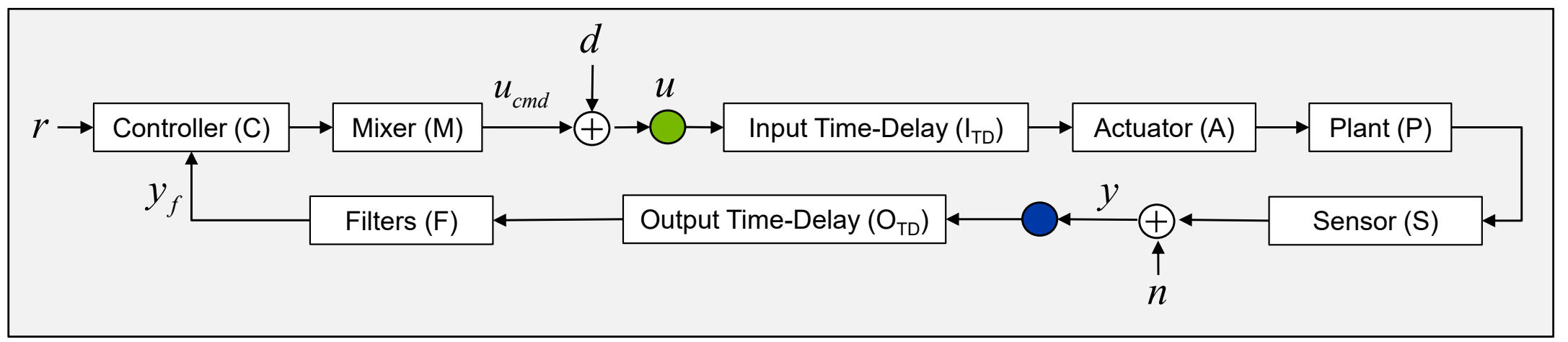

# Control of Flight

**Stable ⟶ Unstable**

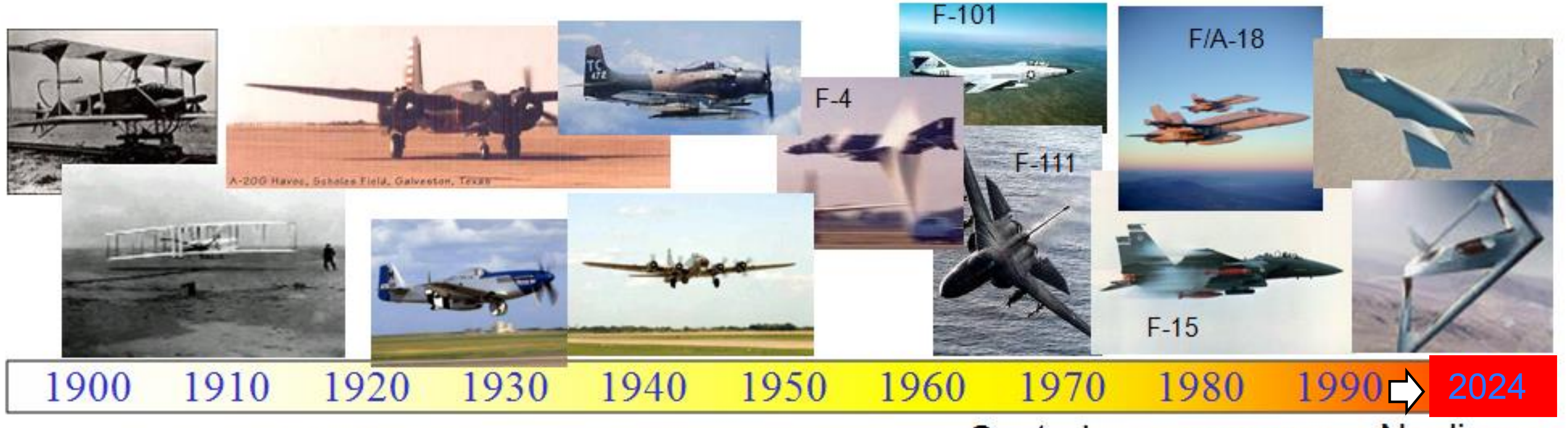

Mechanical Systems Operated By Pilot

Autopilot Reduces Pilot Workload

Control Augmentation Systems

Fly-By-Wire Controls

Nonlinear Adaptive Control

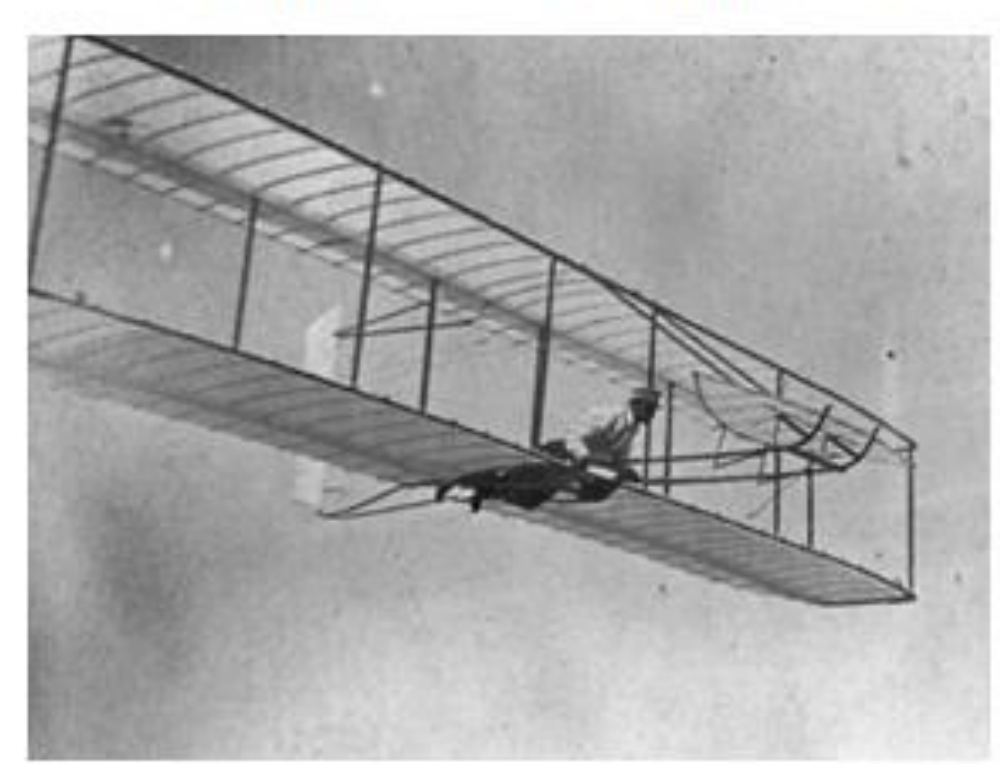

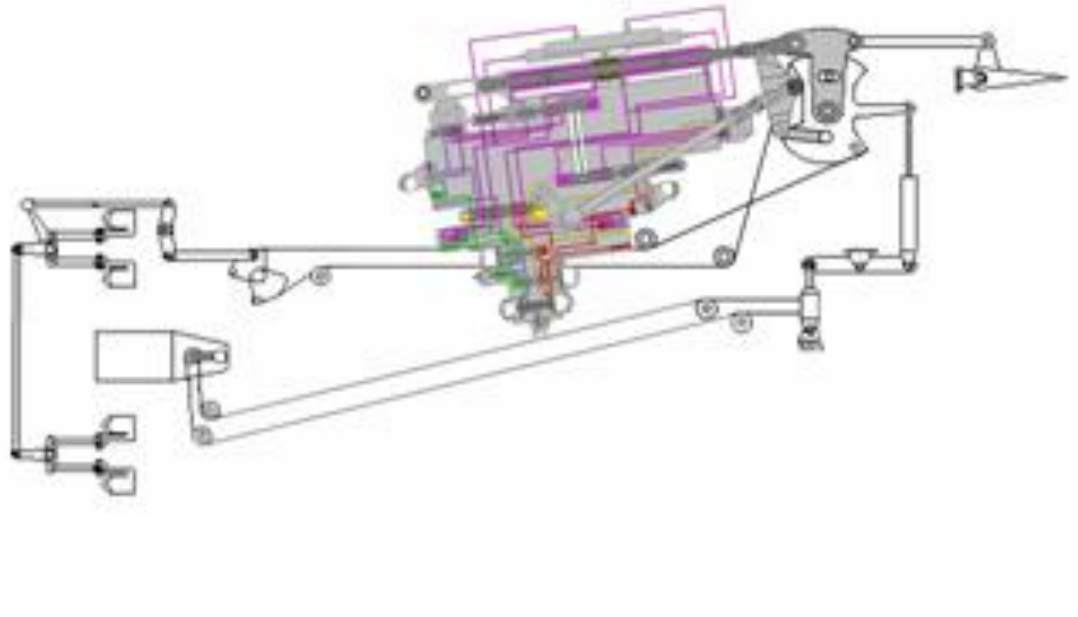

Mechanical To Digital

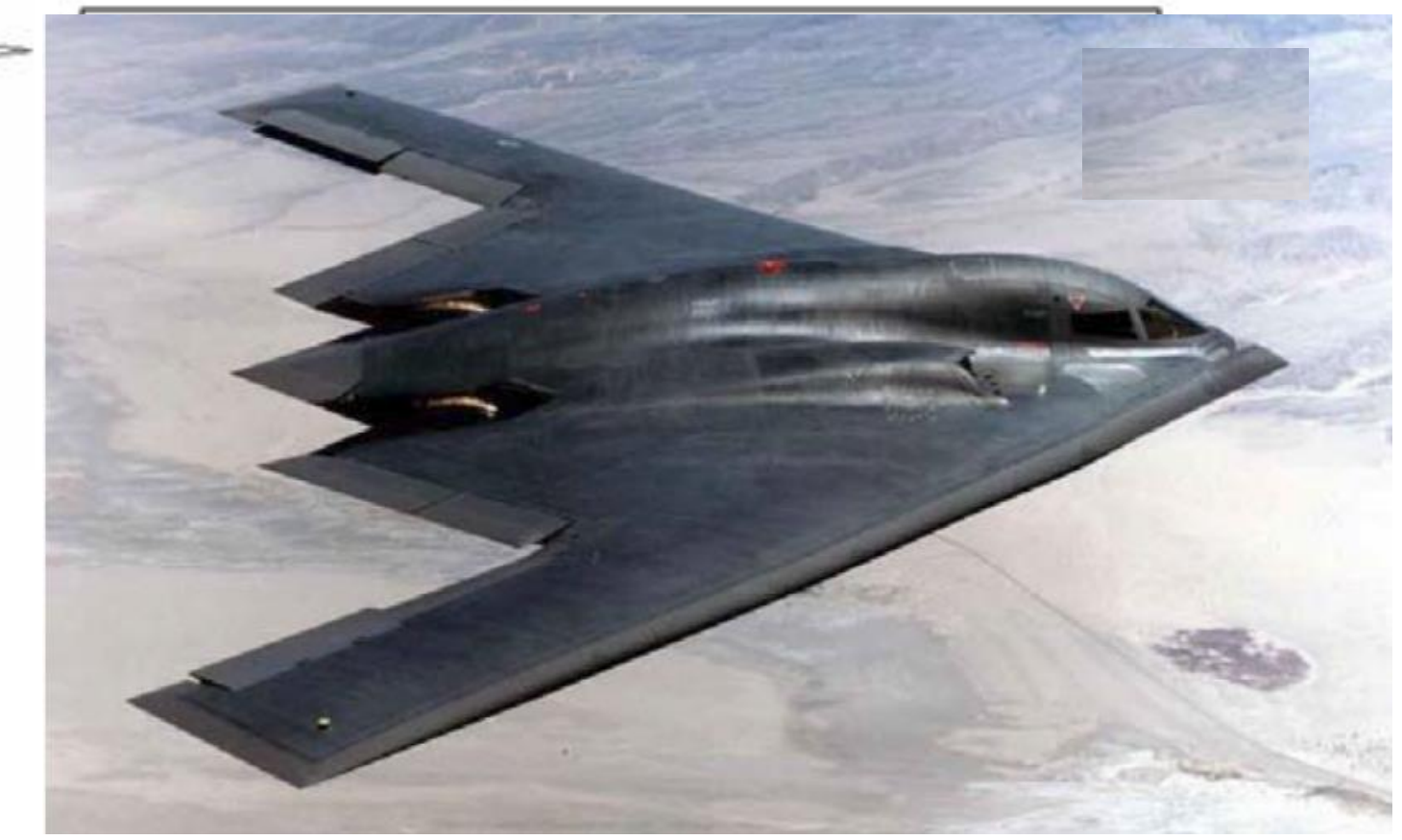

# What is Flight Control

- **Dynamic Output (Sensor-based) Feedback with Command-Feedforward Connections**
  - Change what you do (control actuation) based on what you see (sensed output) and what you are commanded to do (external commands from pilot or guidance system)

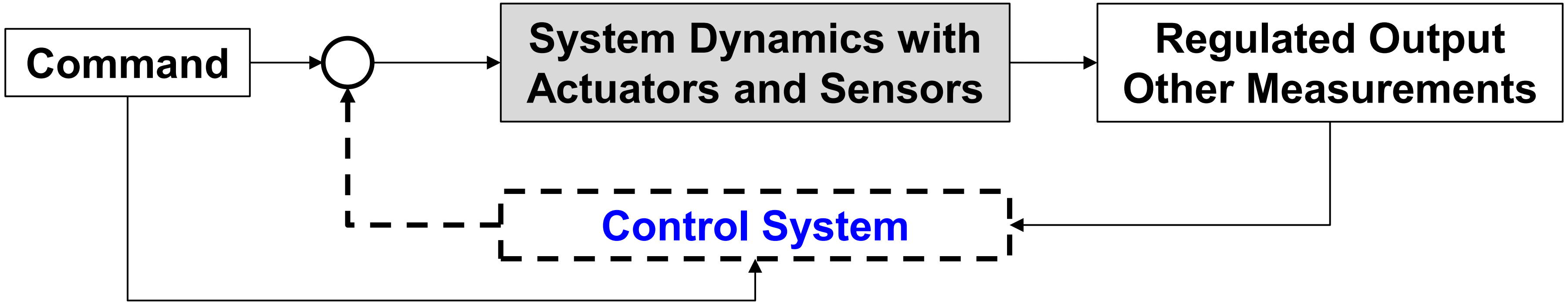

- **Must be able to operate in real time and within "unknown unknowns"**
  - Incorrect assumptions about system dynamics
  - Corrupted sensors
  - Degraded / failed controls
  - Environmental disturbances

- **How Hard Can it Be?**

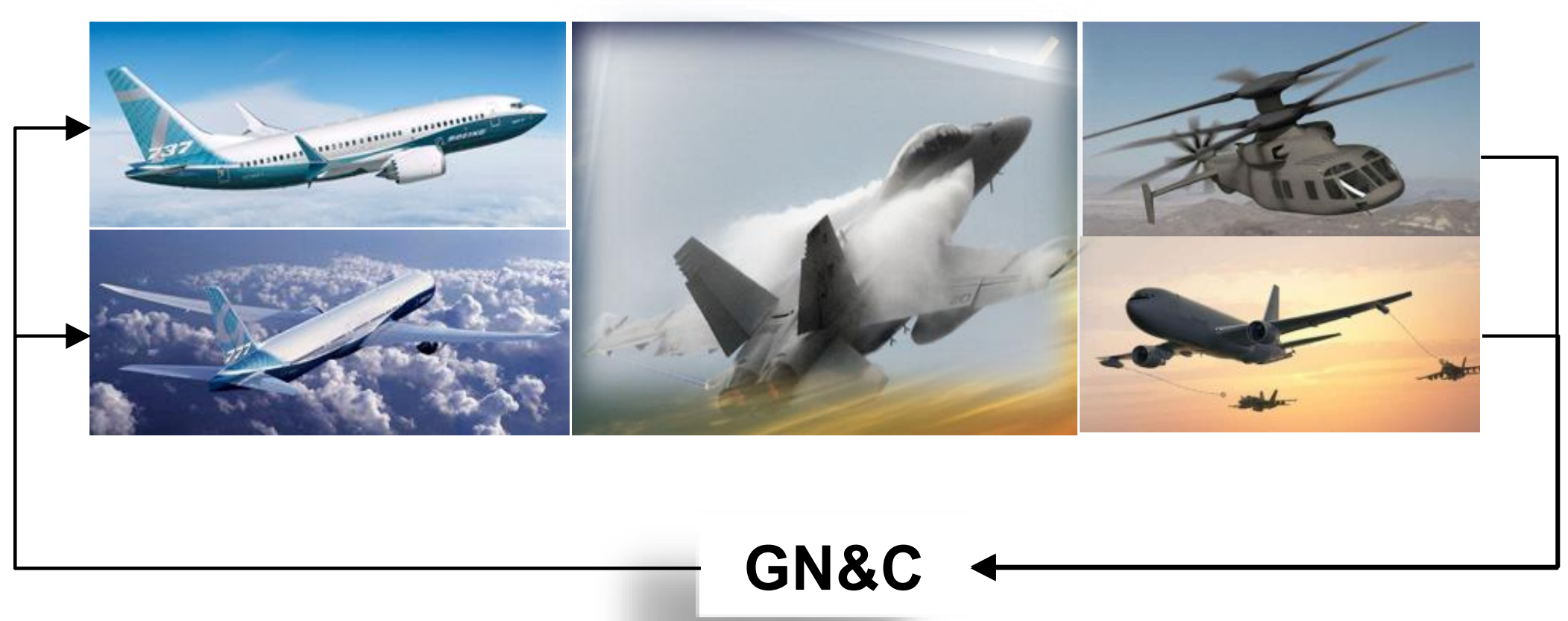

# What is Guidance, Navigation & Control (GN&C)?

Autonomous Flight

Battle Mgmt | Multi-Ship Mission Mgmt | Mission Mgmt | Guidance/ Steering | Flight Control

$K_5$ → $K_4$ → $K_3$ → $K_2$ → $\Delta$ Uncertainties, $K_1$

Human Operator/ Pilot Interaction

Contingencies and Uncertainties in All Loops

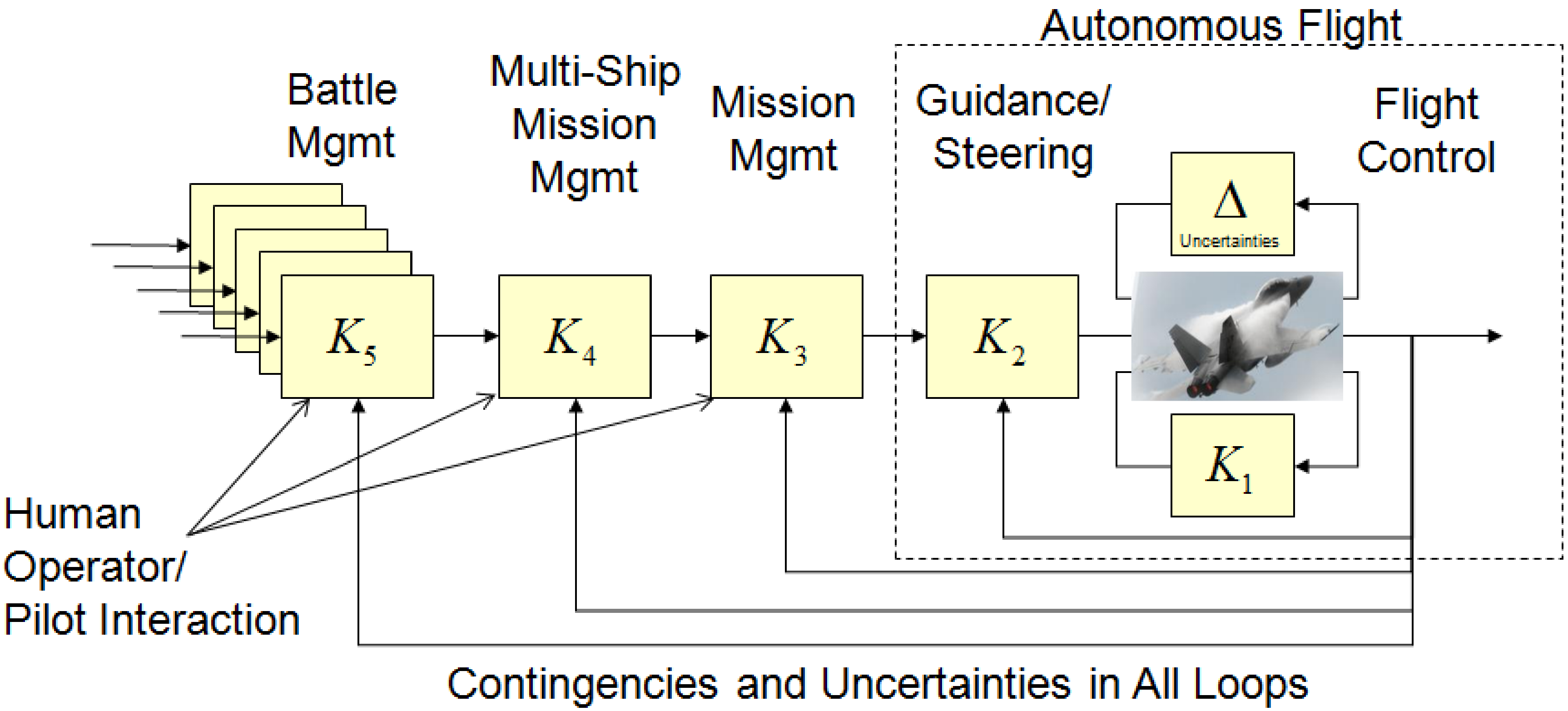

- **Model-based design, must work for operational systems**
- **Three "Must Haves": <u>Stability, Performance, Robustness</u>**
- **Software tools**
    - Automated Design, Analysis, and Autocode
- **Design process supports certification**
- **Documented best engineering practices**
- **Trained workforce to execute programs**

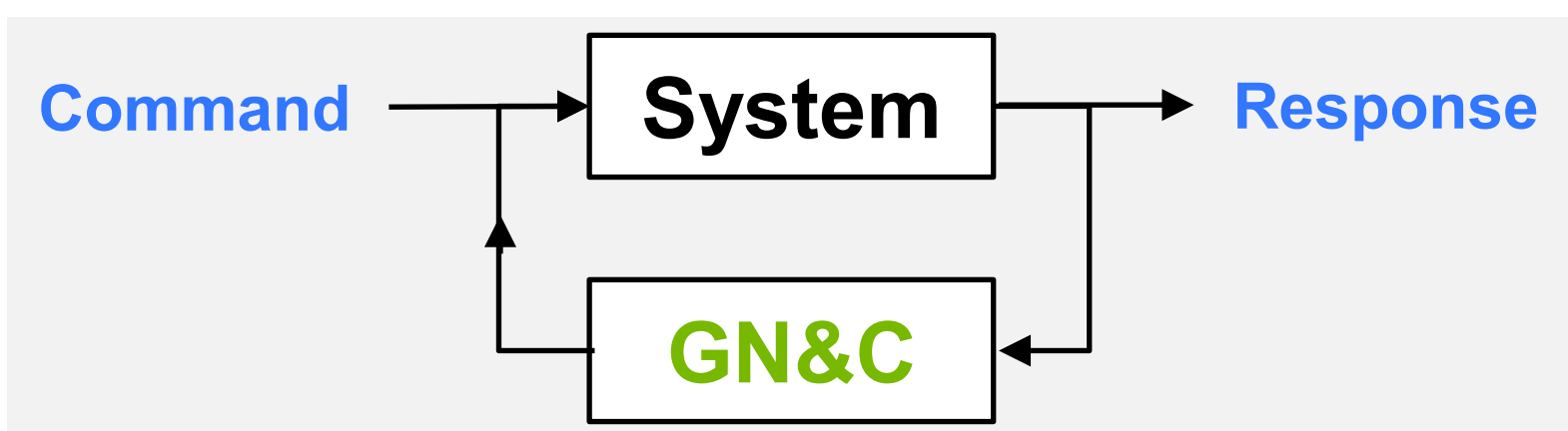

# What Makes “Good” Control Design in General ?

- **Simplified Models for Control Design**
  - Capture essential dynamics
  - Keep it “simple”: Complex design models lead to sensitive control solutions
  - The art of control design
    - Controllers designed using simplified models are guaranteed to work for real systems
- **Closed-Loop Stability**
  - Formal proof for simplified models
  - Relative stability analysis with subsystem models
  - Hardware-in-the-Loop (HIL) testing
- **Command Tracking Performance**
  - High Fidelity Flight Simulation
- **Robustness to Uncertainties**
  - Guaranteed by design for simplified models
  - Validated and verified in high fidelity simulation environment

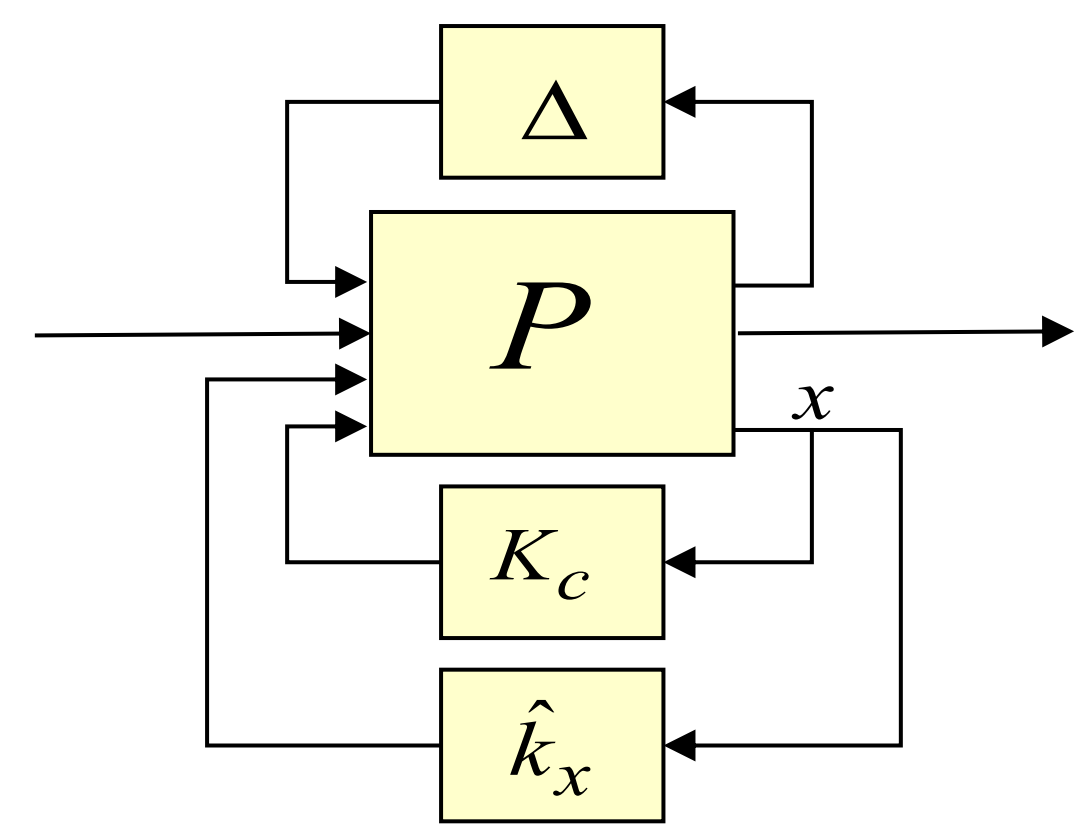

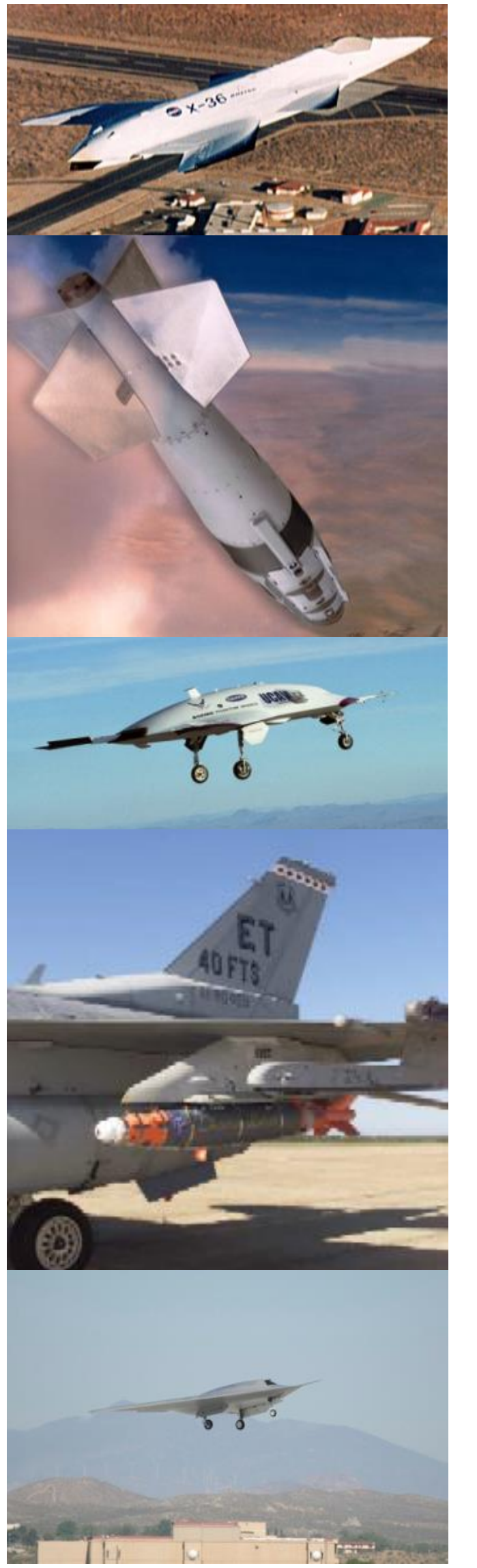

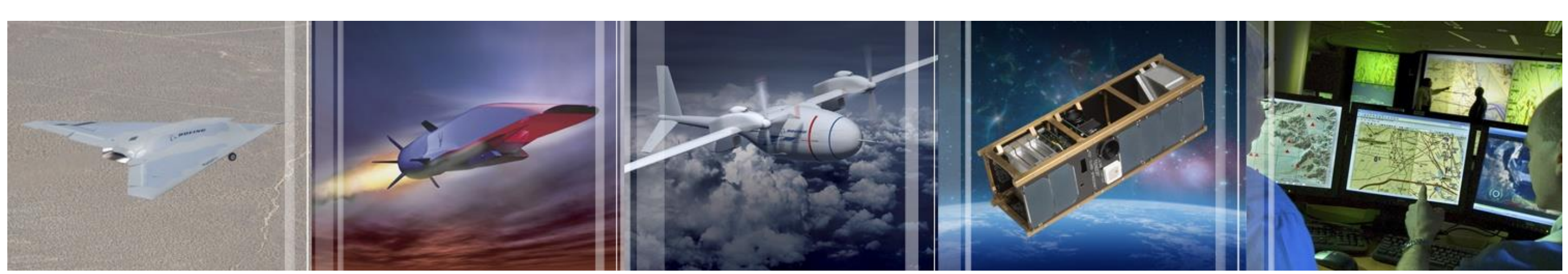

# Control Design, Analysis, Implementation and Process Workflow

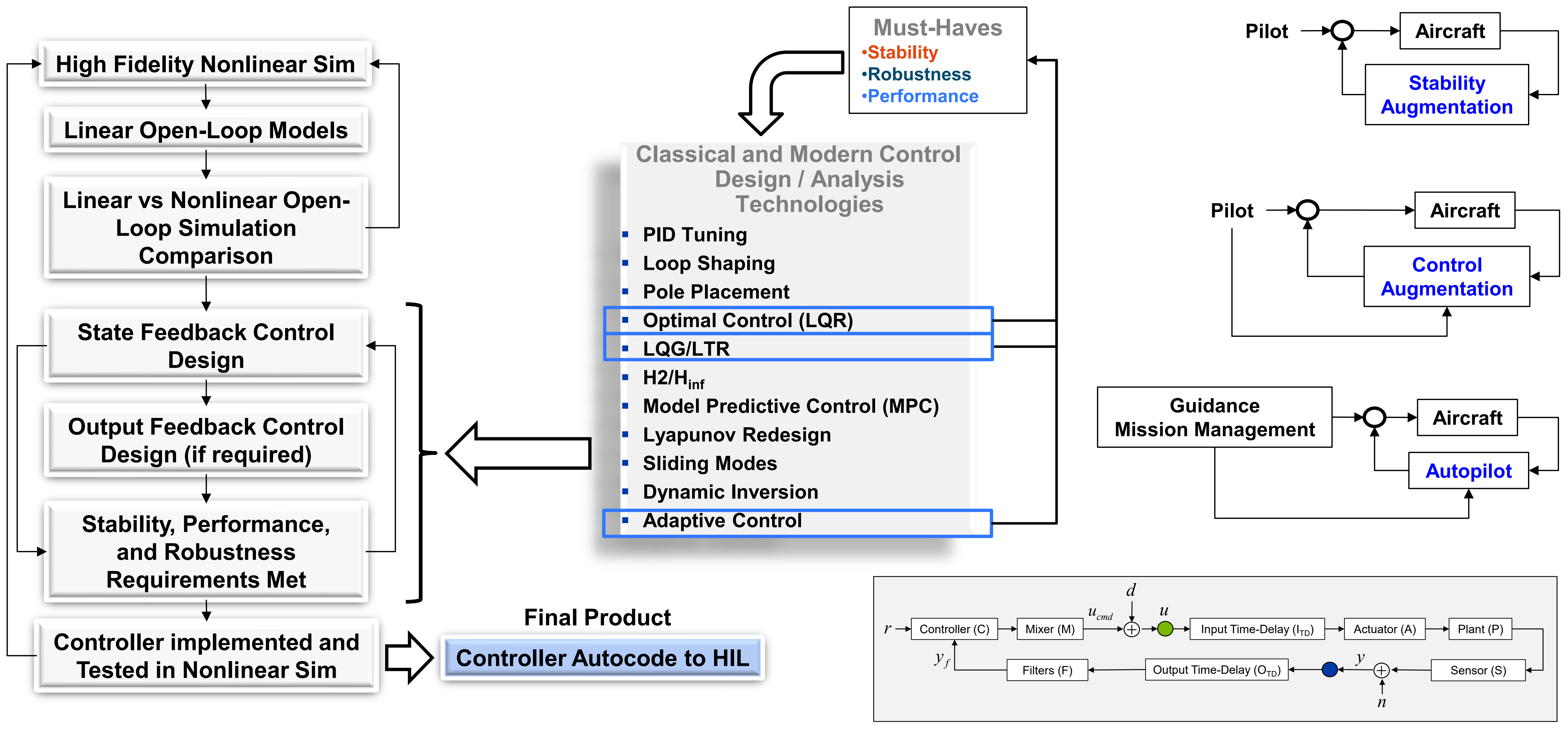

# Model-Based Control For Uncertain Systems

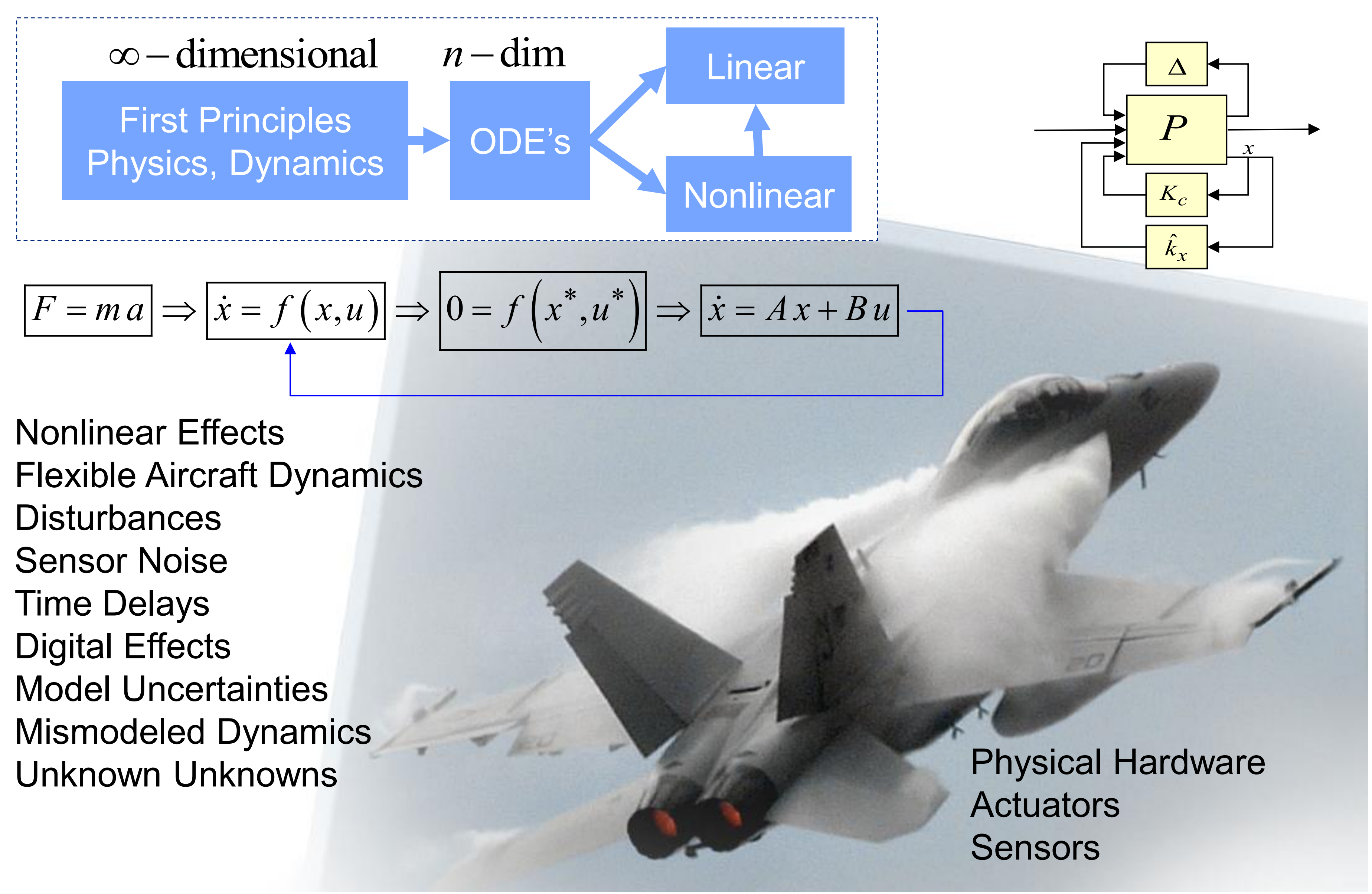

# Open-Loop 6-DoF Flight Simulation Block-Diagram

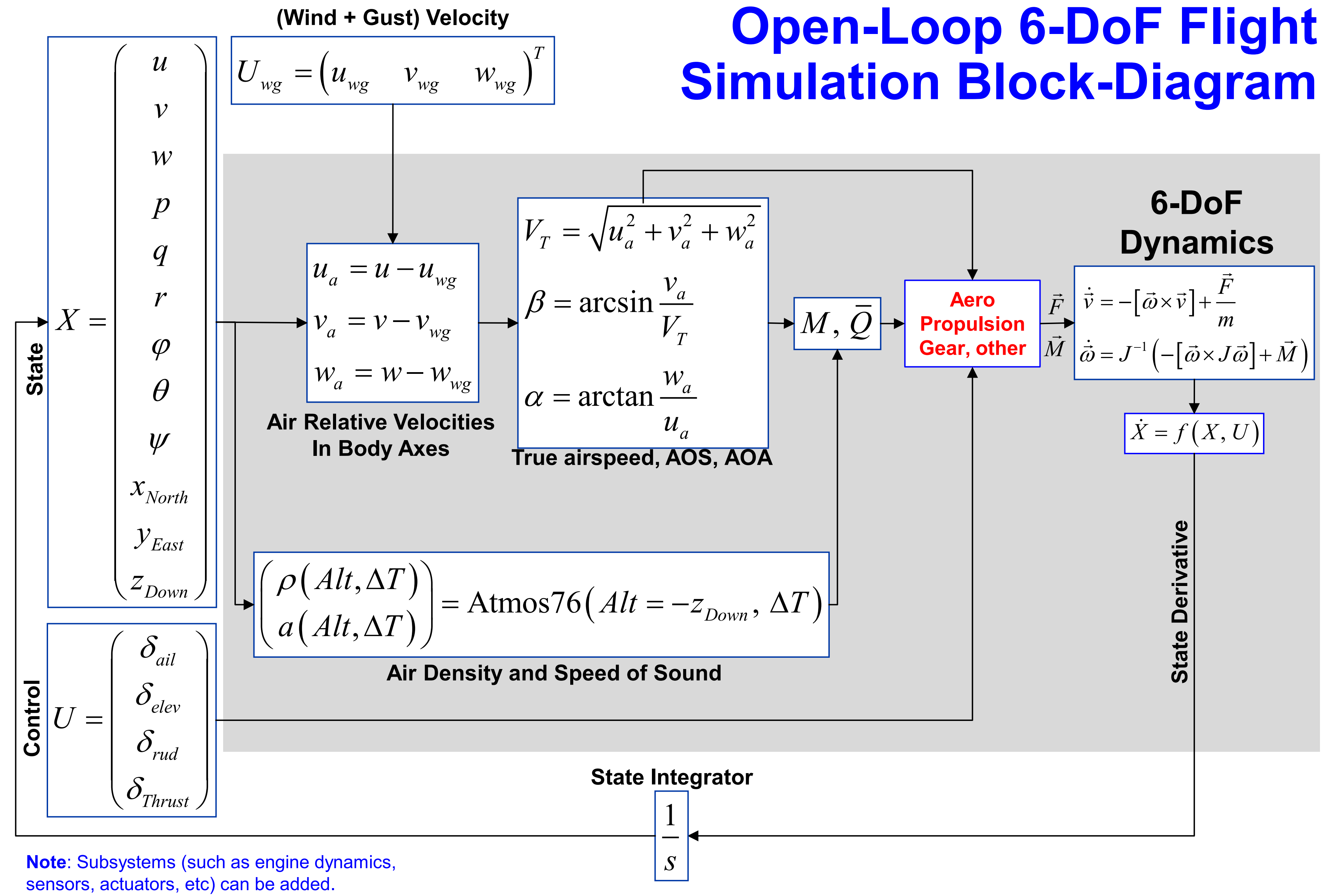

**Note**: Subsystems (such as engine dynamics, sensors, actuators, etc) can be added.

# Flight Dynamics Linearization and Gain-Scheduling

- **6-DoF EOMs**
  - Longitudinal
    - Short period
    - Phugoid
  - Lateral-Directional
    - Roll
    - Dutch roll
    - Spiral

$$\text{Dynamics}: \dot{\vec{x}} = f(\vec{x}, \vec{u})$$
$$\text{Output}: \vec{y} = h(\vec{x}, \vec{u})$$

$$\text{States}: \vec{x} = (u,\ v,\ w,\ p,\ q,\ r,\ \varphi,\ \theta,\ \psi) \in R^9$$
$$\text{Controls}: \vec{u} = (\delta_a, \delta_e, \delta_r . \delta_{th}) \in R^4$$
$$\text{Outputs}: \vec{y} = (A_x,\ A_y,\ A_z,\ V_T,\ \beta,\ \alpha,\ p,\ q,\ r,\ \varphi,\ \theta,\ \psi) \in R^{12}$$

$$\vec{X} = \Big(\underbrace{u,\ v,\ w,\ p,\ q,\ r,\ \varphi,\ \theta,\ \psi}_{\vec{x}},\ x,\ y,\ h\Big)^T \in R^{12}$$

$$\vec{Y} = \Big(\underbrace{A_x,\ A_y,\ A_z,\ V_T,\ \beta,\ \alpha,\ p,\ q,\ r,\ \varphi,\ \theta,\ \psi}_{\vec{y}},\ \dot{x},\ \dot{y},\ \dot{h},\ x,\ y,\ h\Big) \in R^{18}$$

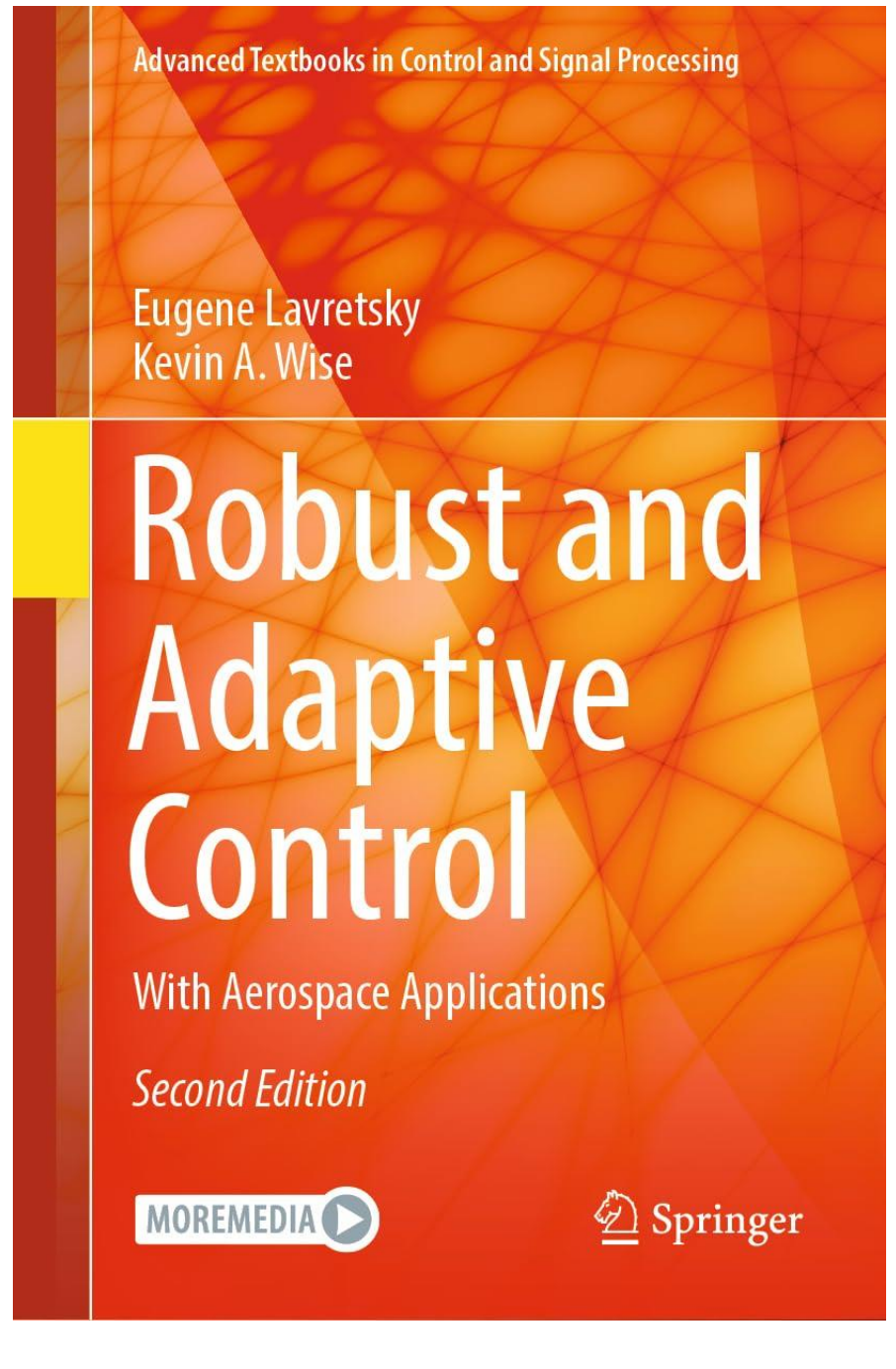

- **Extra States and Outputs can be added**

$$\left.\begin{array}{l}\text{Trimmed Dynamics}: \dot{\vec{x}}^* = f(\vec{x}^*, \vec{u}^*) \\ \text{Trimmed Output}: \vec{y}^* = h(\vec{x}^*, \vec{u}^*)\end{array}\right\} \Rightarrow \text{Linear Models}: \left\{\begin{array}{l}\dot{x} = A\ x + B\ u \\ y = C\ x + D\ u\end{array}\right.$$

- **Gain Scheduling Design Process**
  - Cover flight envelope with a dense set of trim points.
  - Write simplified linear models around each of the trim point.
  - Use these dynamics to design fixed-point flight controllers per point.
  - Interpolate linear controllers (gain-schedule based on flight conditions).

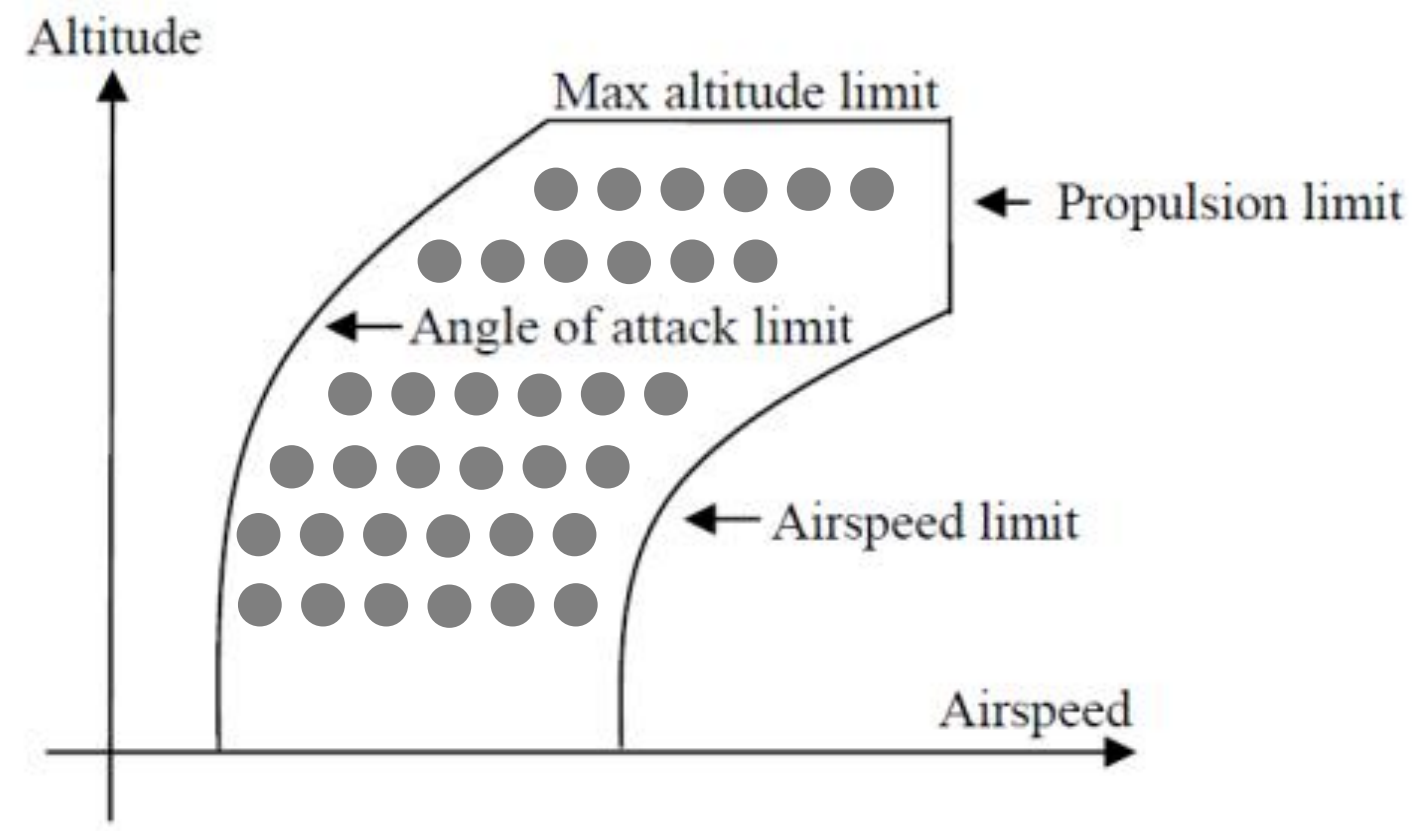

# Observer-Based Control with Loop-Transfer Recovery and Adaptive Augmentation (OBLTR)

OBLTR Block-Diagram

Estimation Error
Observer Ref Model
Predicted Output
Measurements
Estimation Error
Estimated State
Adaptive Control
Measurements
Guidance
Baseline Control
Control
Measurements

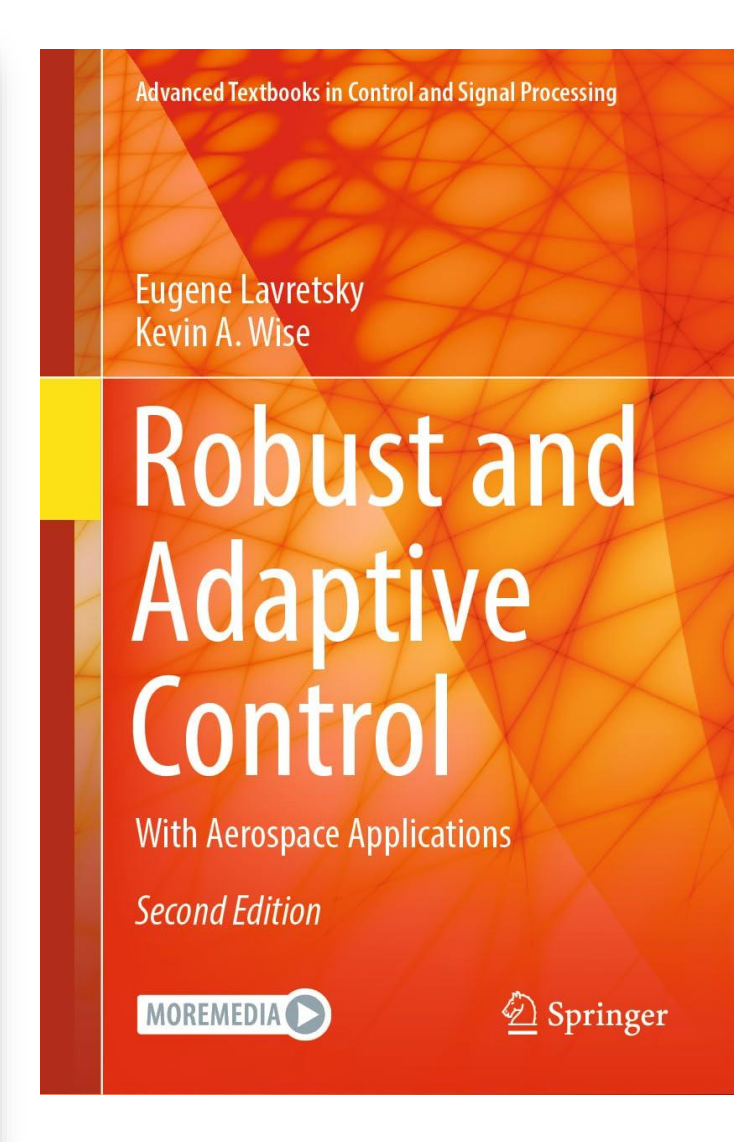

**Advancing State-of-the-Art via SW-Automated Flight Control Design and Analysis Technologies for Aerial Platforms and Systems**

# Robust & Adaptive Control for Aerial Platforms with "Unknown Unknowns"

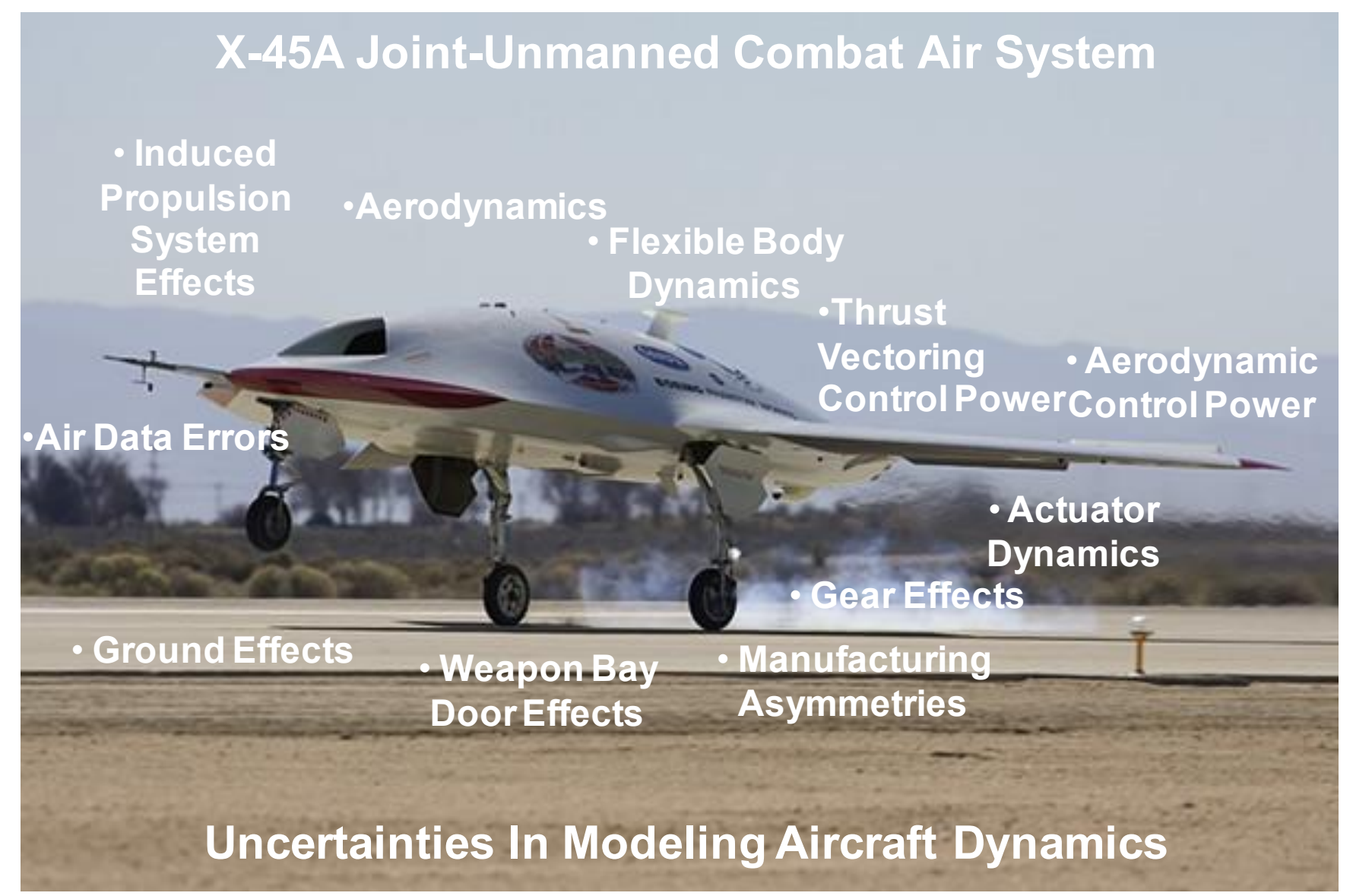

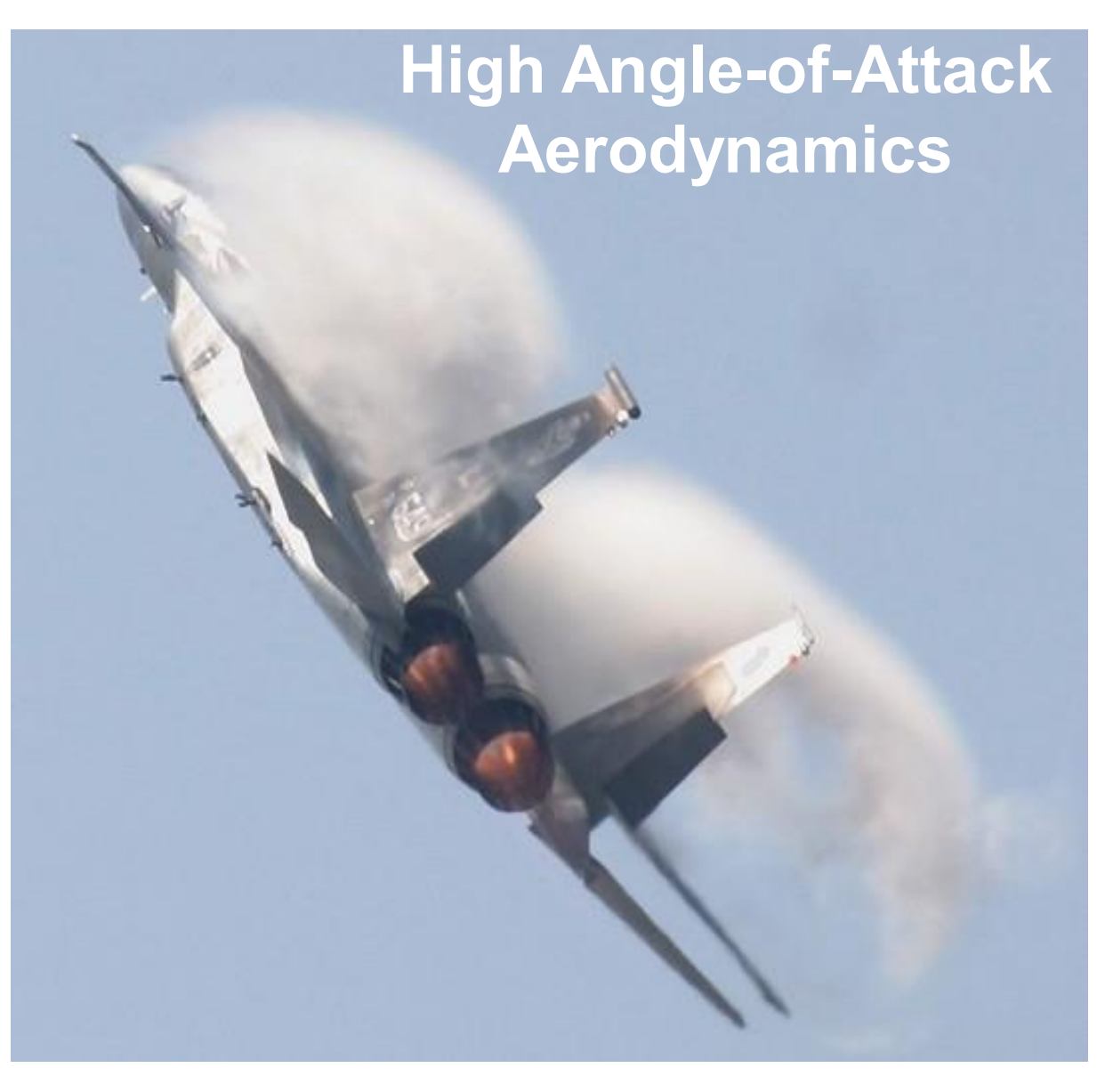

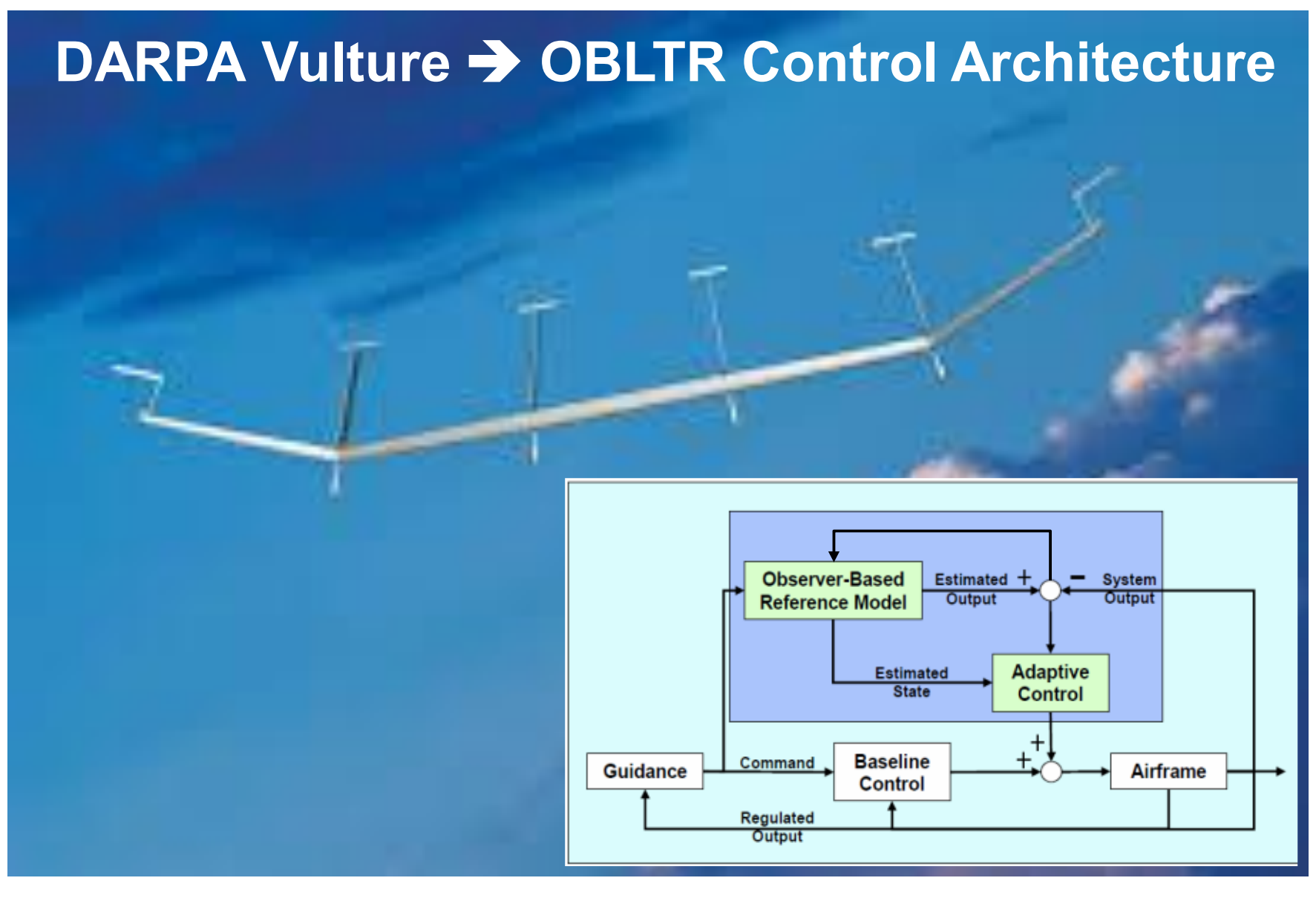

- **Must Address Uncertain Aerodynamics, Flex Modes, Propulsive, and Environmental Effects**
  - **DARPA VULTURE program led to the development of OBLTR design**
    - **Robust and adaptive output feedback control of very flexible aircraft**
- **Reduce Dependency On Accurate Models In Control Architectures**
- **Use Control Theory To Extend Robustness Properties To Achieve Mission Assurance**
- **More Testing / Wind Tunnel Is Not the Answer**

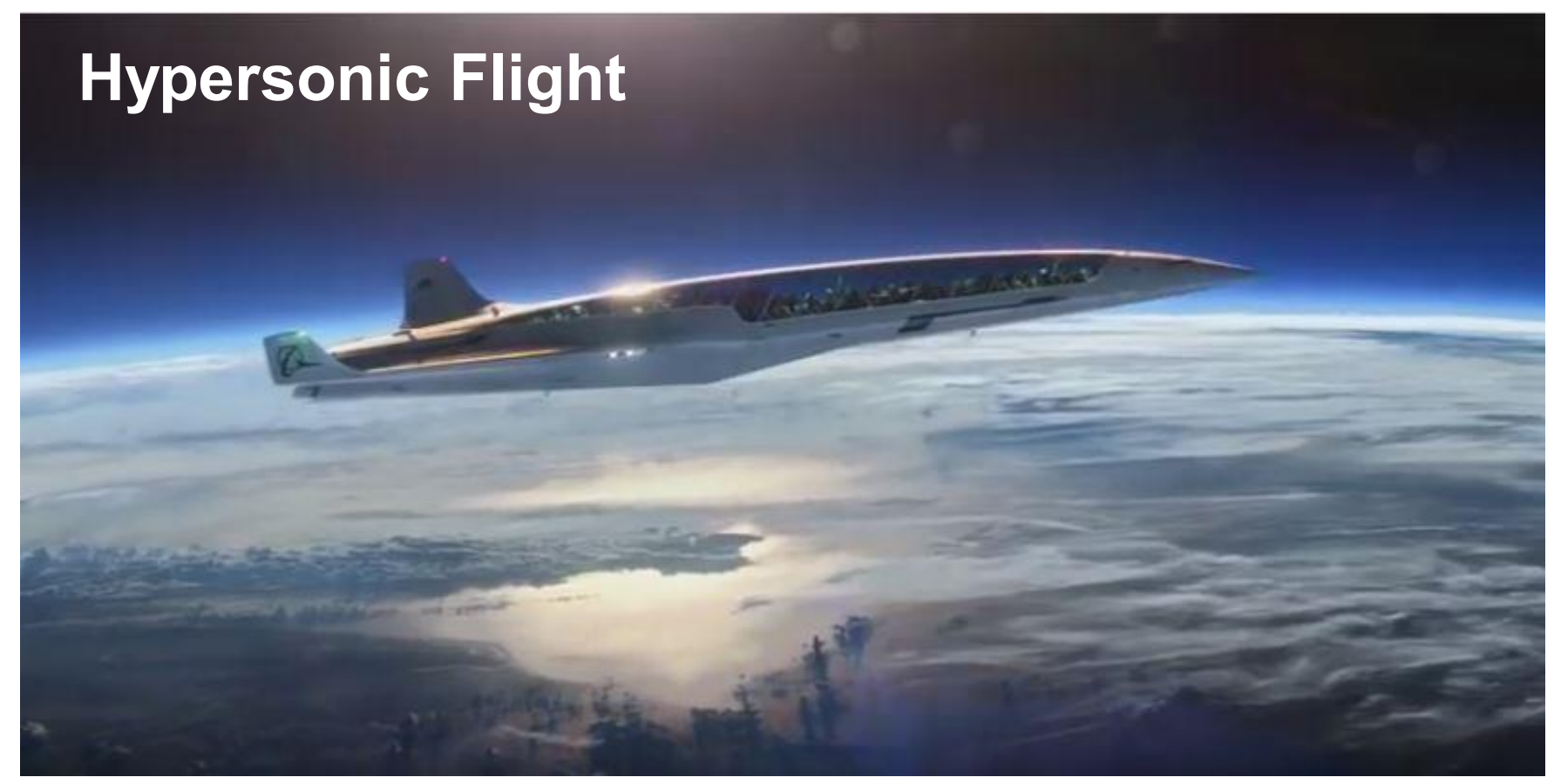

# Observer-Based Control with Loop Transfer Recovery & Adaptive Augmentation (OBLTR)

- **<u>Boeing Flight Proven Technology</u>**

- **Robust and Adaptive Flight Control System with Sensor Output Feedback**
  - **Recovers optimal stability margins at input**
  - **Adaptive augmentation maintains basaeline closed-loop dynamics in the presence of uncertainties and control failures**
  - **Closed-loop reference model concept to minimize transients**

- **Nonaffine Control Allocation Mixer Technology**
  - **Distributes (real-time) virtual into actual controls**

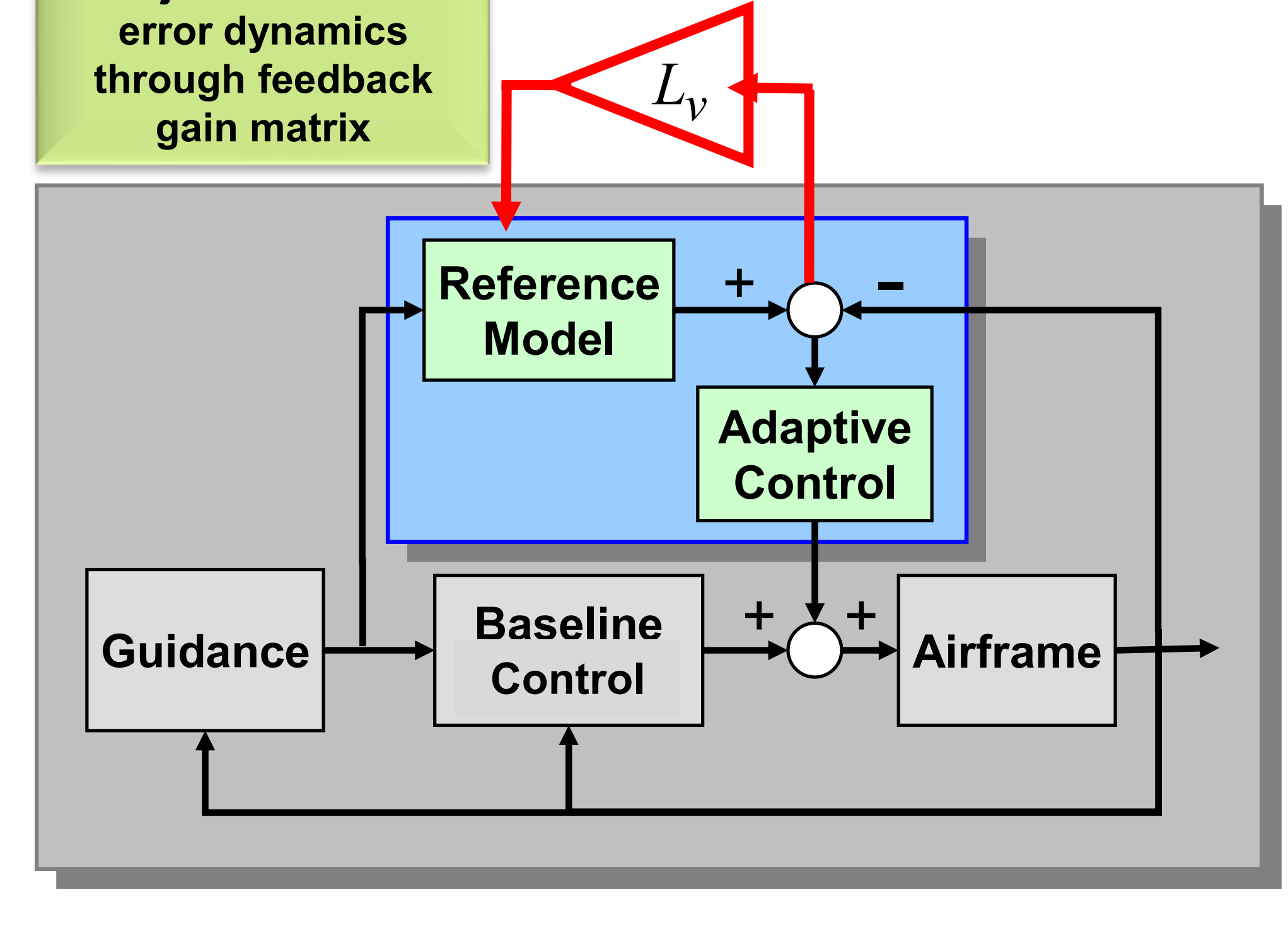

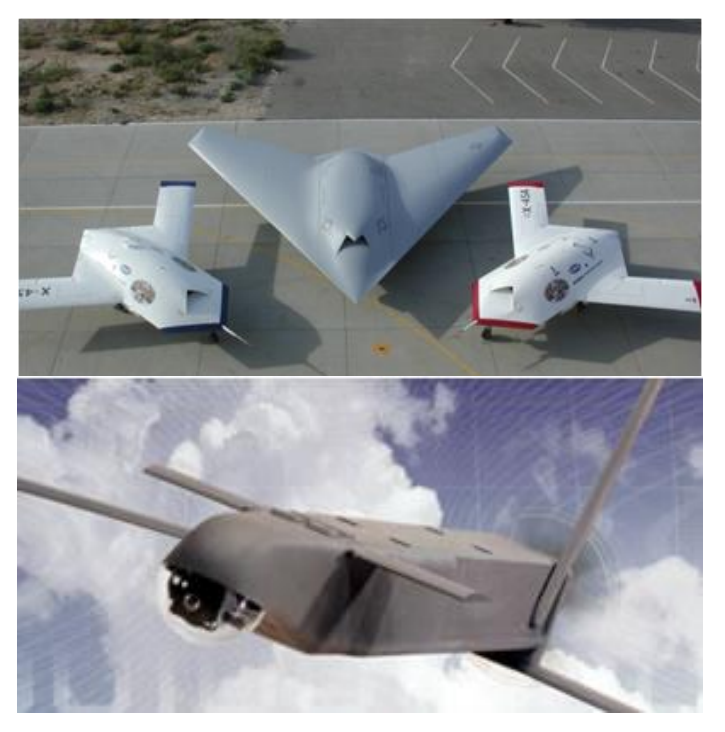

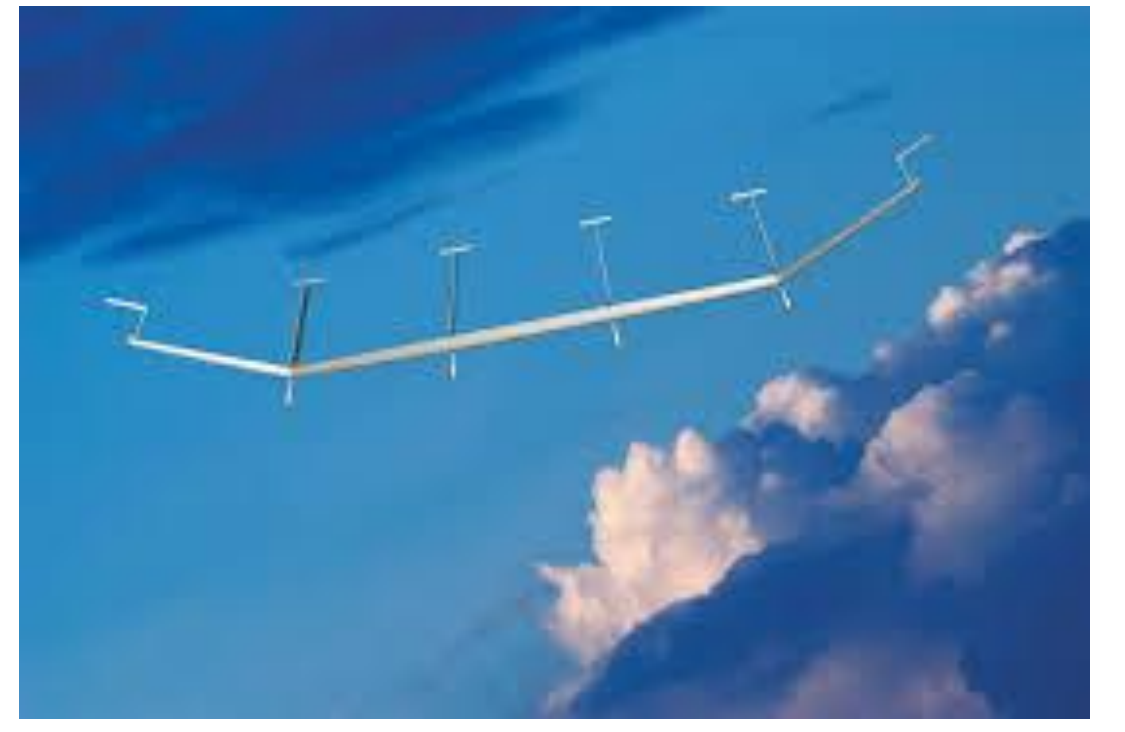

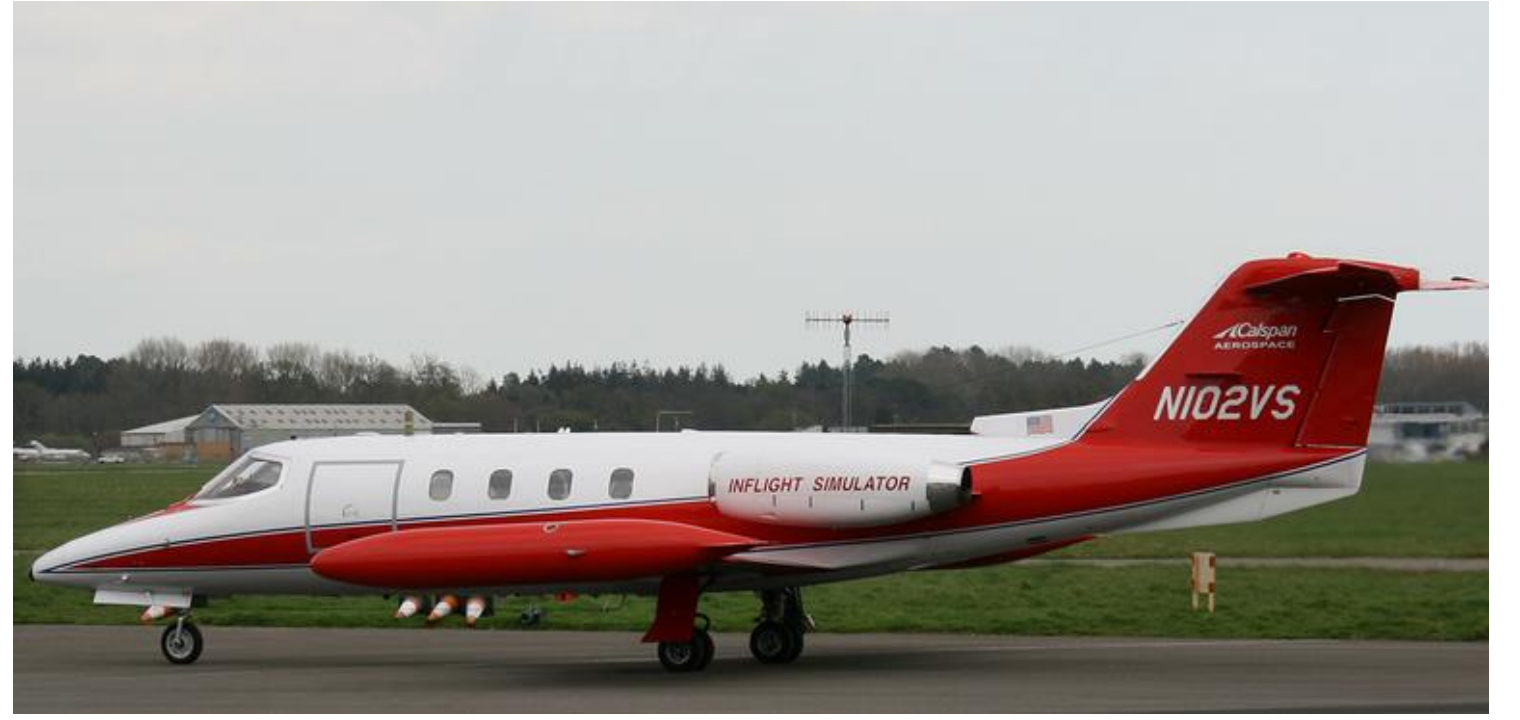

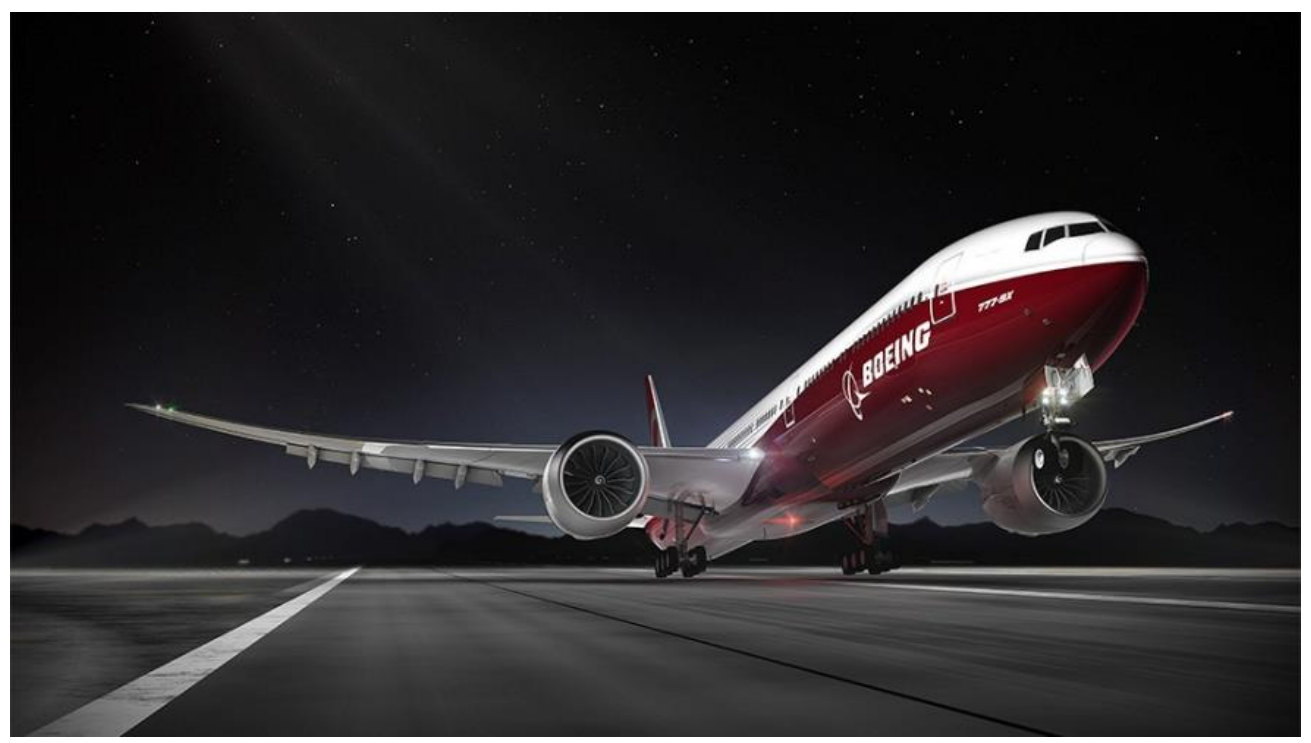

# OBLTR Basis: Asymptotic Properties of Algebraic Riccati Equation

- **For a parameter-dependent ARE in the "filter" form**

$$P_v A^T + A P_v - P_v C^T R_v^{-1} C P_v + Q_v = 0, \quad Q_v = Q_0 + \left(\frac{v+1}{v}\right) B B^T, \quad R_v = \frac{v}{v+1} R_0$$

$$P_v A^T + A P_v - P_v C^T R_0^{-1} C P_v + Q_0 + B B^T + \frac{1}{v}\left[B B^T - P_v C^T R_0^{-1} C P_v\right] = 0$$

- **The following relations hold true, as v→ 0:**

$$\text{Asymptotic Expansion}: P_v = P_0 + \varphi_1(v) P_1 + o(\varphi_1(v)), \quad \varphi_1(v) = O\left(\frac{v}{v+1}\right) = O(v)$$

$$p = m, \quad \det(C B) \neq 0, \quad \det\begin{pmatrix} sI - A & B \\ C & 0 \end{pmatrix} = 0 \Rightarrow s \in \mathbb{C}^- \quad \Rightarrow \quad \begin{aligned} & P_v = P_0 + P_1 v + O(v^2), \quad P_i^T = P_i > 0, \quad (i = 0,1) \\ & P_v C^T = B W^T \sqrt{R_0} + O(v) \\ & \lim_{v \to 0} x^T P_v x \geq \lambda_{\min}(P_0) > 0 \end{aligned}$$

  - where

$$\text{SVD}: B^T C^T R_0^{-\frac{1}{2}} = U \Lambda V \Rightarrow W = (U V)^T$$

- **The "Bridge" between Robust and Adaptive Control : As v → 0, a square system becomes almost Strictly Positive Real (SPR)**
  - Loop Transfer Recovery (LTR) for observer-based baseline control
  - Lyapunov-based asymptotic stability for output feedback model reference adaptive control augmentation

$$\tilde{P}_v = P_v^{-1} \Rightarrow \tilde{P}_v B = C^T R_0^{-\frac{1}{2}} W + O(v)$$

# Overcoming Minimum Phase Restrictions: The Squaring-up Modification

Misra, P., "A Computational Algorithm for Squaring-Up Linear Systems, Part I: Zero Input-Output Matrix", Proceedings of the 1992 IEEE CDC, Tucson, AZ, Dec. 1992, pp.149-150

- **Required sufficient conditions – RESTRICTIVE. To overcome – use Squaring-Up design.**

$$\underbrace{p = m}_{\text{Square Dynamics}}, \quad \underbrace{\det(CB) \neq 0}_{\text{Relative Degree One}}, \quad \underbrace{\det\begin{pmatrix} sI - A & B \\ C & 0 \end{pmatrix} = 0 \Rightarrow \boxed{s \in \mathbb{C}^-}}_{\text{Trnasmission Zeros are Stable = System is Minimum-Phase}}$$

- **"Tall" controllable observable minimum-phase systems with full rank C*B can be squared-up (fictitious inputs added for observer design only) to satisfy the above conditions**

$$m - \text{Inputs} \; \Downarrow \quad p - \text{Outputs} \Rightarrow \begin{pmatrix} A & B \\ C & 0_{p\times m} \end{pmatrix} \in R^{(n+p)\times(n+m)}$$

$$p - \text{Inputs} \; \Downarrow \quad p - \text{Outputs} \Rightarrow \begin{pmatrix} A & (B, B_2) \\ C & 0_{p\times p} \end{pmatrix} \in R^{(n+p)\times(n+p)}$$

**Tall System ⟶ Square System**

$$\bar{B} = (B,\, B_2) \Rightarrow \boxed{\det(C\bar{B}) \neq 0 \wedge \left(\det\begin{pmatrix} sI - A & \bar{B} \\ C & 0 \end{pmatrix} = 0 \Rightarrow \boxed{s \in \mathbb{C}^-}\right)} \Rightarrow \boxed{Q_v = Q_0 + \left(\frac{v+1}{v}\right)\bar{B}\,\bar{B}^T, \quad R_v = \frac{v}{v+1}R_0}$$

**OBLTR Weights**

- **How restrictive is the "minimum-phase" assumption for a Tall system ?**
  - **Not restrictive**: Most Tall systems do not have finite transmission zeros. Method extended to square non-minimum-phase systems, (2nd edition, Ch 15)
  - Individual transfer functions are allowed to be non-minimum phase
- **Relevance of the derived asymptotics to Robust & Adaptive Control**
  - Dynamic State Observers with Loop Transfer Recovery ➔ Adaptive Control with Closed-Loop Reference Models

# Observer-Based Loop Transfer (OBLTR) Design: Summary

- **Open-loop extended dynamics *without* feedforward control terms in measurements**

$$\dot{x} = A\,x + B\,u + B_{cmd}\,y_{cmd}, \quad y_{reg} = C_{reg}\,x + D_{reg}\,u$$

$$y_{meas} = \begin{pmatrix} e_{yI} \\ y_{p\,meas} \end{pmatrix} = \underbrace{\begin{pmatrix} I_{m\times m} & 0 \\ 0 & C_{p\,meas} \end{pmatrix}}_{C_{meas}} \underbrace{\begin{pmatrix} e_{yI} \\ x_p \end{pmatrix}}_{x} = C_{meas}\,x$$

- **LQR State Feedback**

$$(Q,\,R) \Rightarrow \boxed{P\,A + A^T\,P + Q - P\,B\,R^{-1}\,B^T\,P = 0} \Rightarrow \boxed{K_{lqr} = R^{-1}\,B^T\,P}$$

- Using integrated accelerations as measurements
  - Robustness to Noise sensitivity
  - Significant improvements in reducing Structural Mode Interactions (SMI)

- **Squared-up dynamics → Stable transmission zeros**

$$\bar{B} = (B,\,B_2) \Rightarrow \det\left(C_{meas}\,\bar{B}\right) \neq 0 \wedge \left(\det\begin{pmatrix} s\,I - A & \bar{B} \\ C_{meas} & 0 \end{pmatrix} = 0 \Rightarrow \boxed{s \in \mathbb{C}^-}\right)$$

- **Parameterized Weights → Filter Algebraic Riccati Equation → Observer Gains**

$$Q_v = Q_0 + \left(\frac{v+1}{v}\right)\bar{B}\,\bar{B}^T, \quad R_v = \frac{v}{v+1}\,R_0 \Rightarrow \boxed{P_v\,A^T + A\,P_v + Q_v - P_v\,C_{meas}^T\,R_v^{-1}\,C_{meas}\,P_v = 0} \Rightarrow \boxed{L_v = P_v\,C_{meas}^T\,R_v^{-1}}$$

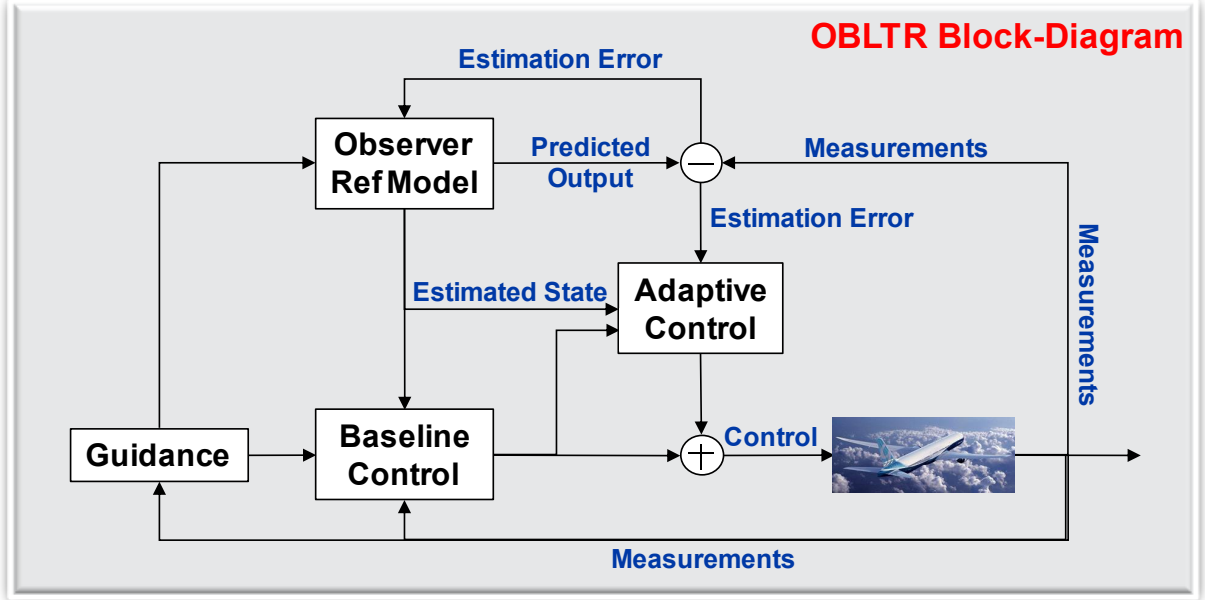

- **State Observer and Control feedback: Choose v sufficiently small to recover LQR margins @ Input**

$$\dot{\hat{x}} = A\,\hat{x} + B\,u + B_{cmd}\,y_{cmd} + L_v\left(y_{meas} - \hat{y}_{meas}\right)$$

$$\hat{y}_{meas} = C_{meas}\,\hat{x}, \quad \boxed{u = -K_{lqr}\,\hat{x}}$$

$$\dot{\hat{x}} = \underbrace{\left(A - B\,K_{lqr} - L_v\,C_{meas}\right)}_{A_{obs}}\hat{x} + B_{cmd}\,y_{cmd} + L_v\,y_{meas}$$

- State Observer = Closed-Loop Reference Model for Adaptive Control Augmentation
  - Transient minimization
- Output Feedback Adaptive Control
  - Adaptive law remains the same as in conventional MRAC

- **Output Feedback Dynamic Compensator Recovers LQR Gain & Phase Margins @ Input**

$$u = -K_{lqr}\,\hat{x} = -K_{lqr}\left(s\,I_{n\times n} - A + B\,K_{lqr} + L_v\,C_{meas}\right)^{-1}\left(B_{cmd}\,y_{cmd} + L_v\,y_{meas}\right)$$

# OBLTR Technology Summary

$$y_{meas} = \begin{pmatrix} e_{yI} \\ y_{p\,meas} \end{pmatrix} = \underbrace{\begin{pmatrix} I_{m\times m} & 0 \\ 0 & C_{p\,meas} \end{pmatrix}}_{C_{meas}} \underbrace{\begin{pmatrix} e_{yI} \\ x_p \end{pmatrix}}_{x} = C_{meas}\, x$$

$$A = \begin{pmatrix} 0_{m\times m} & C_{p\,reg} \\ 0_{n_p\times m} & A_p \end{pmatrix}, \quad B = \begin{pmatrix} D_{p\,reg} \\ B_p \end{pmatrix}, \quad B_{cmd} = \begin{pmatrix} -I_{m\times m} \\ 0_{n_p\times m} \end{pmatrix}$$

$$u = -K_{lqr}\,\hat{x} = -K_{lqr}\left(s\,I_{n\times n} - A + B\,K_{lqr} + L_v\,C_{meas}\right)^{-1}\left(B_{cmd}\,y_{cmd} + L_v\,y_{meas}\right)$$

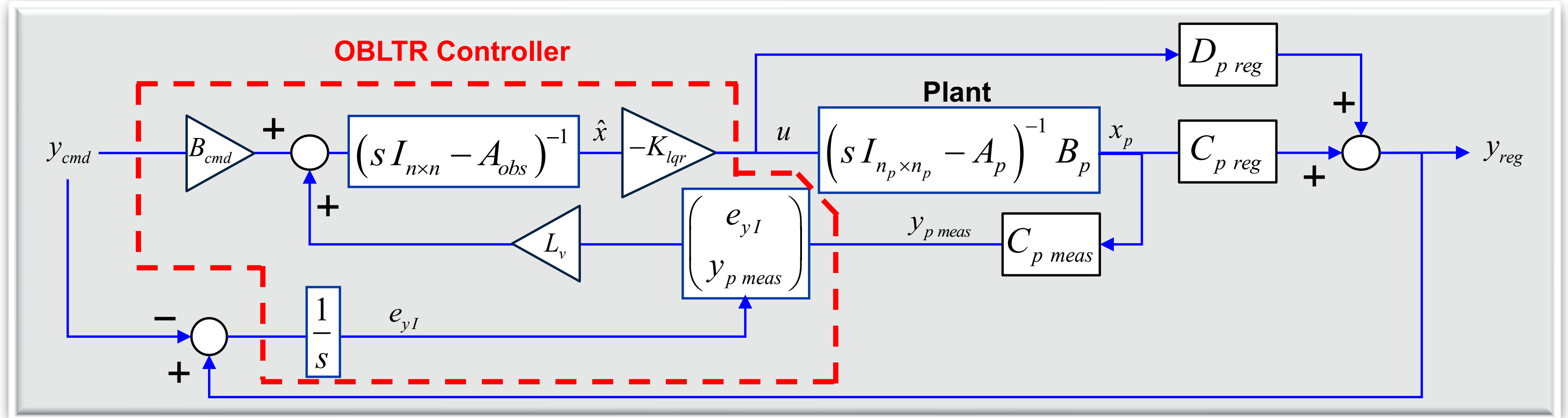

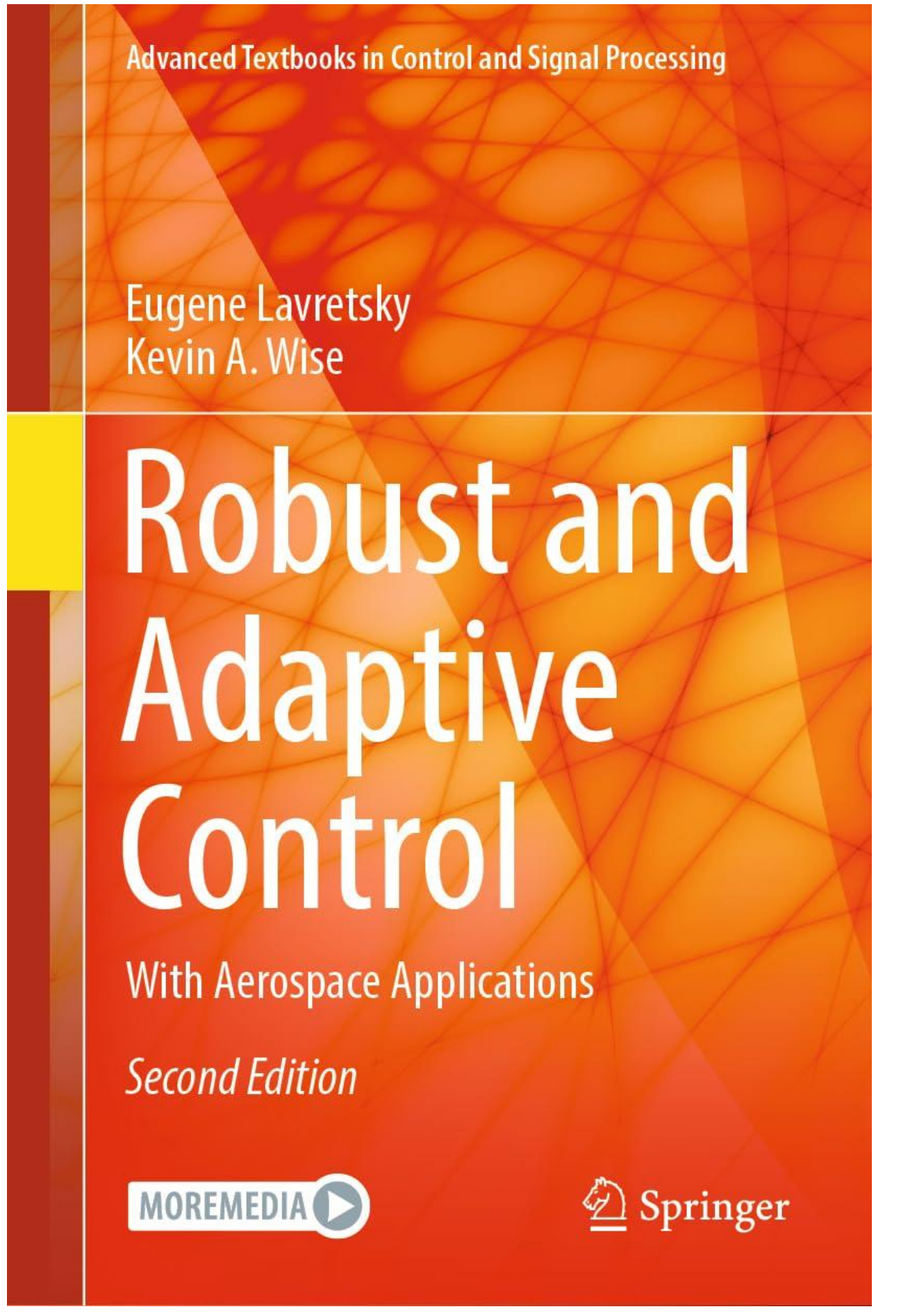

- Dynamic output feedback for command tracking
- Guaranteed closed-loop stability.
- Recovers LQR state feedback margins at plant input, as $v \rightarrow 0$
- Increased robustness at high frequencies ("better" than state feedback)
- Allows custom selection of proportional and integral feedback signals
- Directly supports adaptive output feedback control augmentation

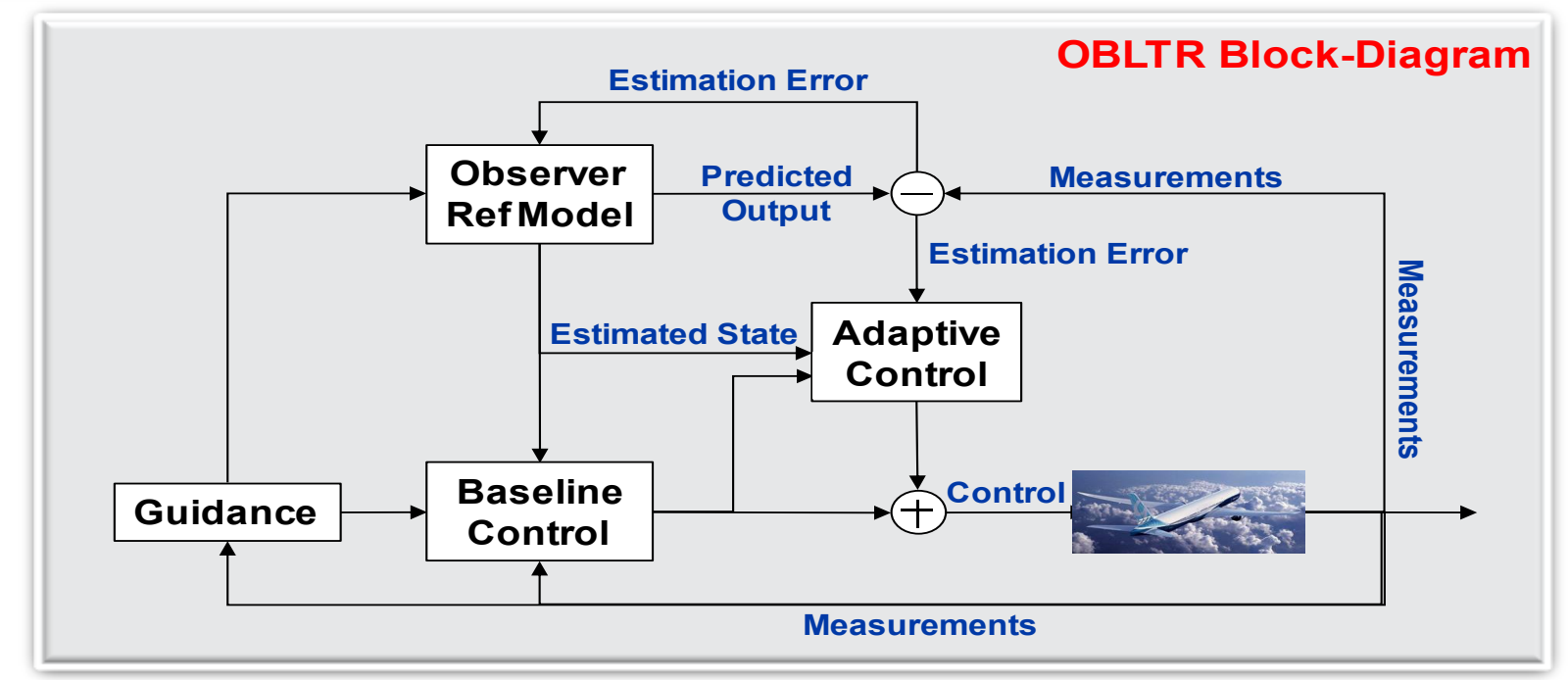

# Boeing Observer-Based Control with Loop Transfer Recovery (OBLTR) and Robust Adaptive Augmentation

- Improved Performance, Mission Reliability
- Closed-Loop Reference Model Adaptive Control
- Verified, Validated and Flight Proven

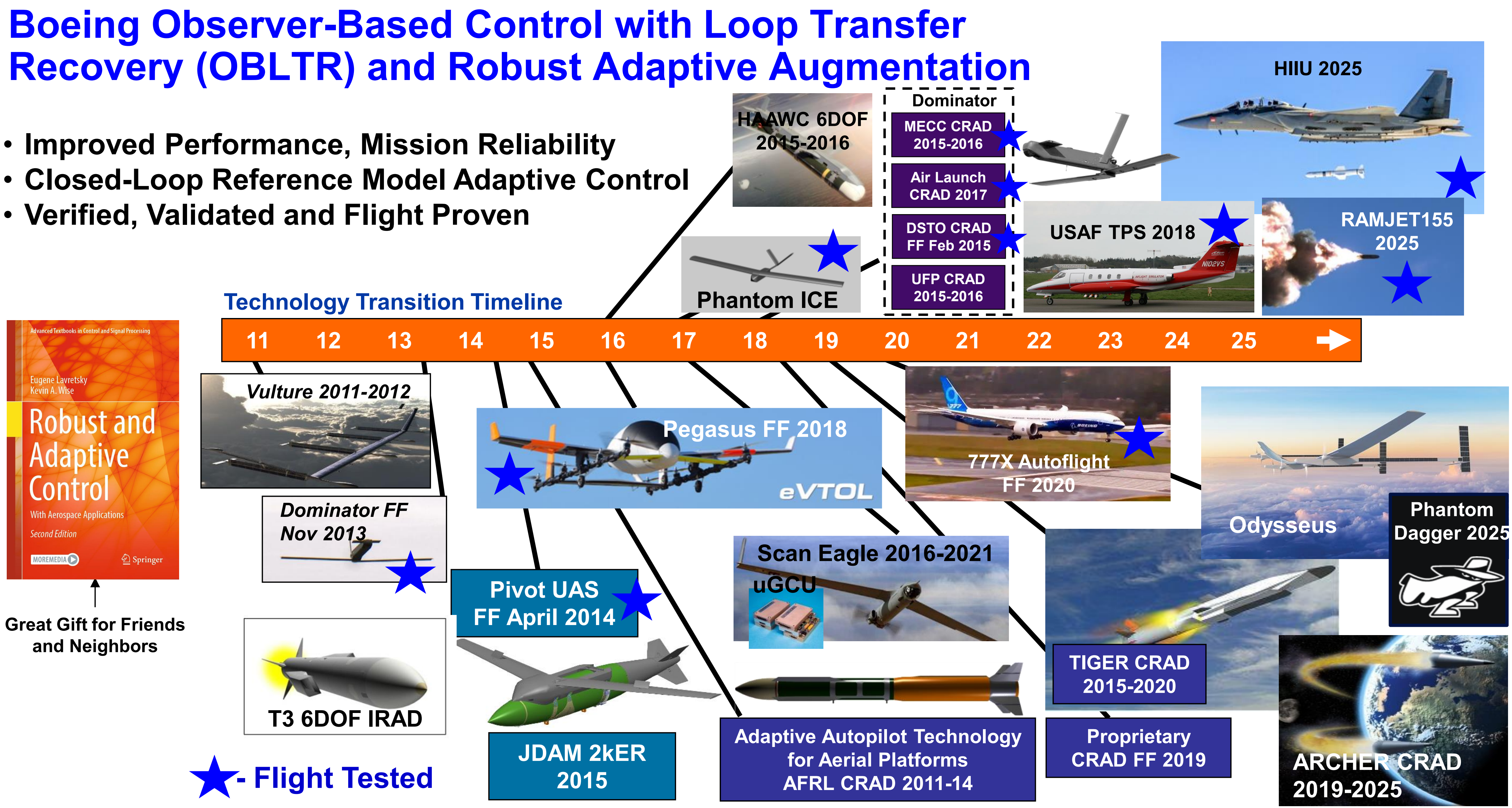

# Flight Control with Operational Constraints
# A Control Barrier Functions (CBF) Based Design

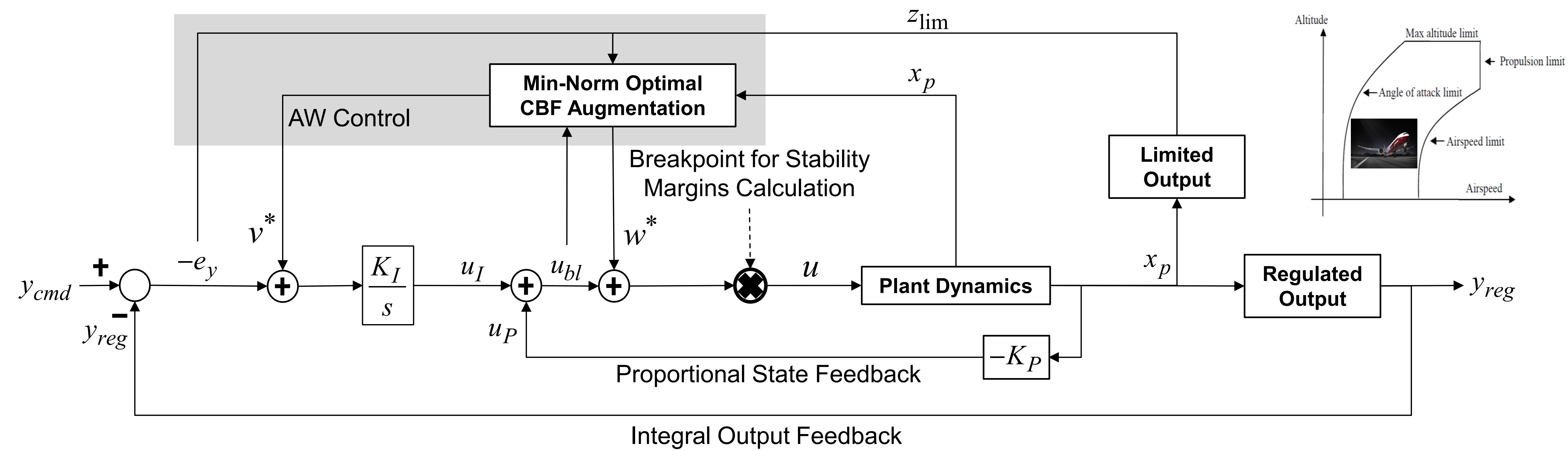

- **"Why": Formal design and analysis methods for enforcing prescribed limits on control inputs and outputs for flight critical systems with quantifiable measures of:**
  - Stability, Robustness, Command tracking performance

# Theoretical Basis

- **Nagumo's Theorem (Forward Invariance)**
  - M. Nagumo, "Über die Lage der Integralkurven gewöhnlicher differentialgleichungen," in Proc. Physico-Math. Soc. Jpn., in 3rd Series, vol. 24, Jan. 1942, pp. 551–559.
  - M. Menner, E. Lavretsky, "Translation of Nagumo's Foundational Work on Barrier Functions: On the Location of Integral Curves of Ordinary Differential Equations," June 2024, arXiv:2406.18614

- **Min-norm optimal control design via Quadratic Programming (QP)**
  - R.A. Freeman, P.V. Kokotovic, "Inverse optimality in robust stabilization," SIAM J. Control Optim., vol. 34, no. 4, pp. 1365–1391, 1996.
  - S. Boyd, L. Vandenberghe, Convex Optimization. Cambridge, U.K., Cambridge Univ. Press, 2004.

- **Control Barrier Functions (CBF-s)**
  - A.D. Ames, X. Xu, J.W. Grizzle, P. Tabuada, "Control barrier function based quadratic programs for safety critical systems," IEEE Trans. Autom. Control, vol. 62, no. 8, pp. 3861–3876, Aug. 2017.

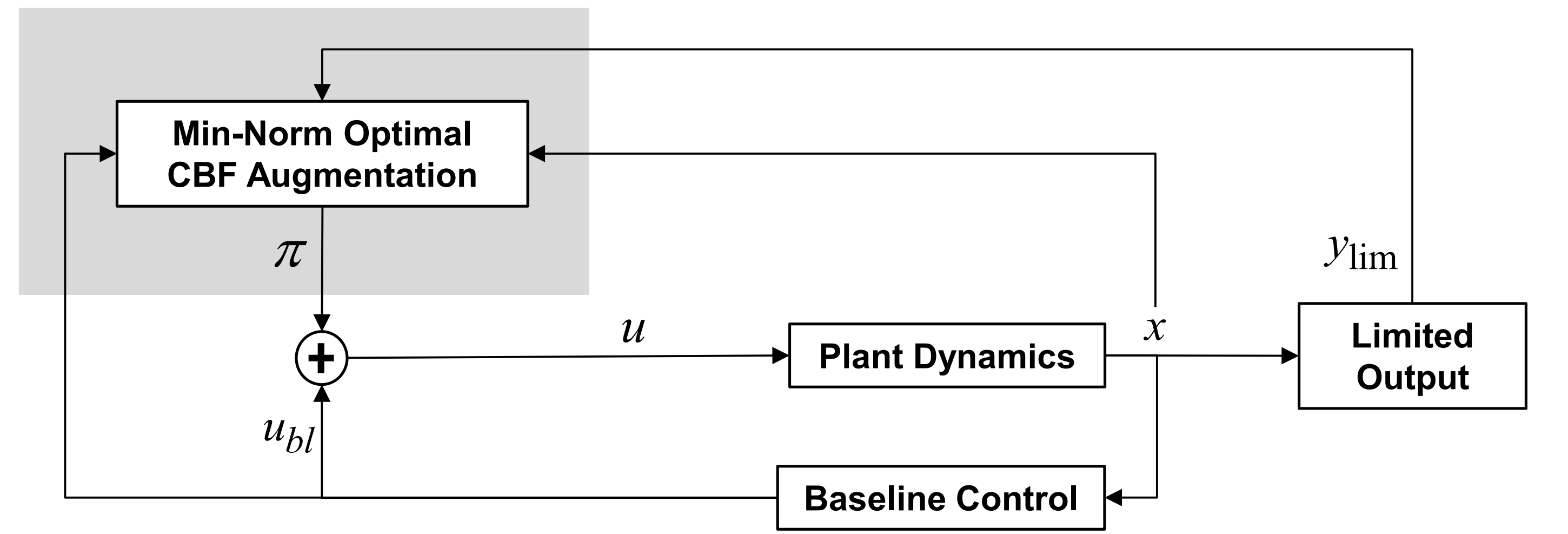

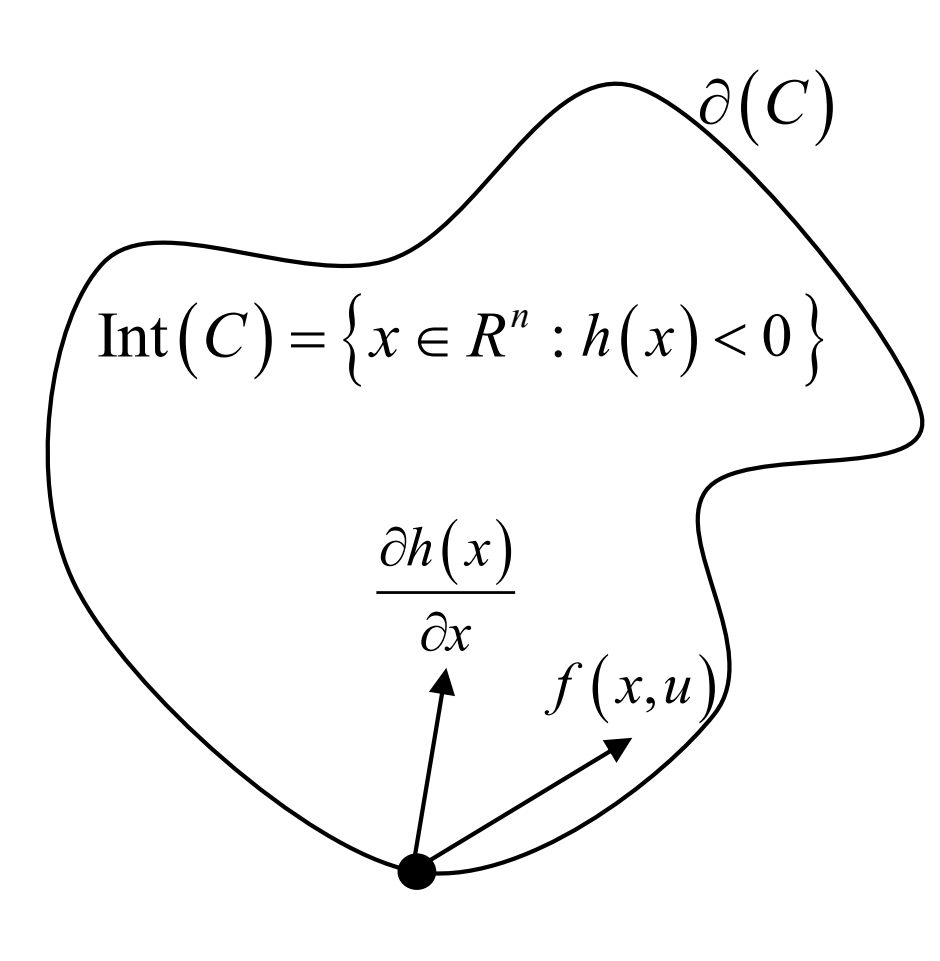

# CBF Augmentation – What and How

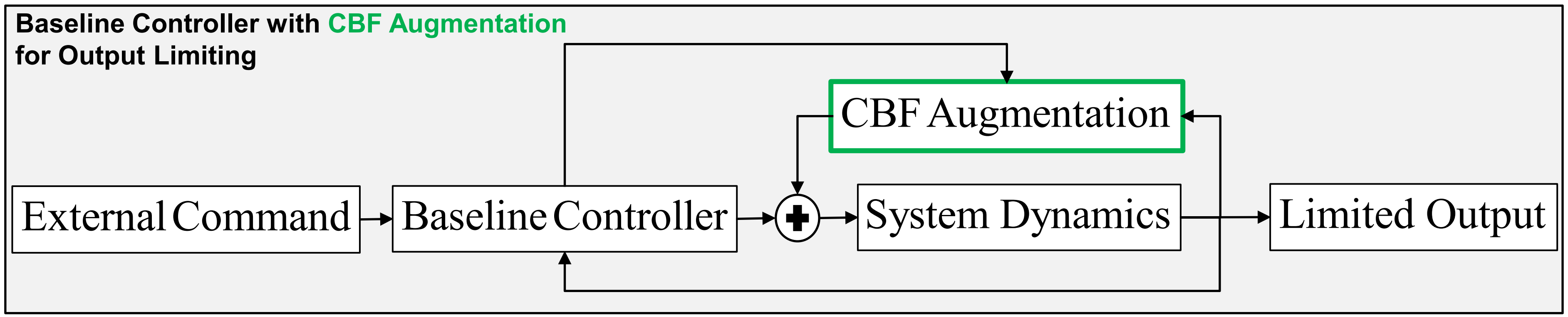

- **CBF Policy = Continuous Static Feedback Augmentation of a Baseline State Feedback Controller**
  - Not a filter
- **Enforces selected output to be uniformly bounded forward in time, via closed-loop feedback, and w/o using "hard" saturation, such as the sat-function.**
  - Not "safe". Classic rules apply: Robustness and margins are the only "guaranteed to be safe" features, as for any control design.
- **For affine-in-control dynamics with linear constraints, CBF solutions can be written and implemented in analytic form. No online optimization is required.**
- **For Linear Time Invariant (LTI) systems with linear constraints, CBF solution is a continuous piece-wise linear state feedback.**
  - Allows analysis and computation of gain and phase margins.
- **Cancels (subtracts out) baseline controller near prescribed output limits.**
  - Similar to Dynamic Inversion near and on boundaries

# Problem Formulation

- **Multi-Input-Multi-Output (MIMO) Controllable Linear Time Invariant (LTI) System Dynamics**

$$\begin{aligned} \text{Open-Loop System Dynamics}: \dot{x} = A\,x + B\,u, \quad x \in R^n \\ \text{Limited Output}: y_{\lim} = C_{\lim} x, \quad y_{\lim} \in R^m \end{aligned}$$

- **Desired Min/Max Operational Output *m*-dim Box Limits, (defined component-wise)**

$$y_{\lim}^{\min} \le y_{\lim} \le y_{\lim}^{\max}$$

- **Control Design Task: State Feedback**
  - Find smallest (min-norm optimal) CBF augmentation control policy to enforce "soft" operational limits
    - via feedback, w/o using "hard" saturation, such as sat-functions

$$u(x) = \underbrace{u_{bl}(x)}_{\text{Baseline Controller}} + \underbrace{\pi(x)}_{\text{CBF Augmentation}}$$

- **Minimize Quadratic Const with (2*m) Linear Constraints ➔ Analytic Solution = Continuous State Feedback**

$$\text{Minimizaion Cost}: J(\pi) = \left(\pi^T R_\pi \, \pi\right) \to \min_{\pi}$$

$$\text{Output Constraints}: \begin{cases} h_1(x) = y_{\lim}^{\min} - C_{\lim} x \le 0 \\ h_2(x) = C_{\lim} x - y_{\lim}^{\max} \le 0 \end{cases}$$

# CBF Design: Preliminaries

- **Assumptions**
  - Controllability
  - Vector Output Relative Degree is well-defined $r = \begin{pmatrix} r_1 & \dots & r_m \end{pmatrix}$

$$\left[\left(C_{\lim}\right)_i A^k B = 0, \forall 0 \le k \le r_i - 2\right] \wedge \left[\left(C_{\lim}\right)_k A^{r_i - 1} B \ne 0\right], \quad i = 1, \dots, m$$

$$\det\left(H_u\right) = \det\begin{pmatrix} \left(C_{\lim}\right)_1 A^{r_1 - 1} B \\ \vdots \\ \left(C_{\lim}\right)_m A^{r_m - 1} B \end{pmatrix}_{m \times m} \ne 0$$

- **Vector Output Relative Degree ➔ Differentiated Outputs ➔ Explicit Control Dependence**

$$\begin{pmatrix} \left(y_{\lim}\right)_1^{(r_1)} \\ \left(y_{\lim}\right)_2^{(r_2)} \\ \vdots \\ \left(y_{\lim}\right)_m^{(r_m)} \end{pmatrix} = \begin{pmatrix} \left(C_{\lim}\right)_1 A^{r_1} \\ \left(C_{\lim}\right)_2 A^{r_2} \\ \vdots \\ \left(C_{\lim}\right)_m A^{r_m} \end{pmatrix} x + \underbrace{\begin{pmatrix} \left(C_{\lim}\right)_1 A^{r_1 - 1} B \\ \left(C_{\lim}\right)_2 A^{r_2 - 1} B \\ \vdots \\ \left(C_{\lim}\right)_m A^{r_m - 1} B \end{pmatrix}}_{H_u \in R^{m \times m}} u$$

# CBF Design: Stable Polynomials for Modified Output Constraints

- **Define Stable Polynomials for each limited output**
  - For each limited $i^{th}$ output, polynomial order = $i^{th}$ output relative degree with negative real eigenvalues

$$\underbrace{\left.\begin{array}{l}\phi_i(s) = \prod_{j=1}^{r_i}(s-\lambda_{ij}) = \sum_{j=0}^{r_i} c_{ij}\, s^j, \quad \lambda_{ij} < 0 \\ \alpha_\pi = \operatorname{diag}\begin{pmatrix} c_{10} & \dots & c_{m0}\end{pmatrix} \in R^{m\times m}\end{array}\right\}}_{\text{Stable Polynomimals}} \Rightarrow \underbrace{\left\{ Y_{\lim} = \underbrace{\begin{pmatrix}\phi_1(s) & \dots & 0 \\ \vdots & \ddots & \vdots \\ 0 & \dots & \phi_m(s)\end{pmatrix}}_{\Phi(s)} y_{\lim} = H_x x + H_u(u_{bl} + \pi)\right.}_{\text{Modified Output}}$$

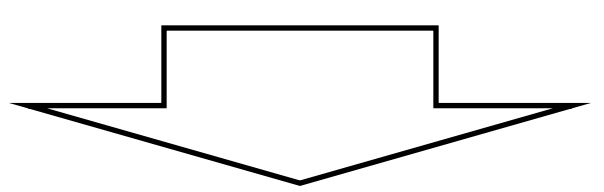

- **Original and Modified Output Constraints**

$$\underbrace{\left.\begin{array}{l} h_1(x) = y_{\lim}^{\min} - C_{\lim} x \le 0 \\ h_2(x) = C_{\lim} x - y_{\lim}^{\max} \le 0 \end{array}\right\}}_{\text{Output Constraints}} \Rightarrow \left\{ H(x, u_{bl}, \pi) = \underbrace{\begin{pmatrix} -Y_{\lim} + \alpha_\pi\, y_{\lim}^{\min} \\ Y_{\lim} - \alpha_\pi\, y_{\lim}^{\max}\end{pmatrix} = \begin{pmatrix} -I_m \\ I_m \end{pmatrix} H_u \pi + \underbrace{\begin{pmatrix} -H_x x - H_u u_{bl} + \alpha_\pi\, y_{\lim}^{\min} \\ H_x x + H_u u_{bl} - \alpha_\pi\, y_{\lim}^{\max}\end{pmatrix}}_{\Delta H(x, u_{bl})} \le 0}_{\text{Modified Output Constraints}} \right.$$

# CBF Design: Modified QP ➔ CBF Analytic Solution

- **Modified QP Formulation**

$$\text{Minimizaion Cost}: J(\pi) = \left( \pi^T \underbrace{R_\pi}_{\left(H_u^T H_u\right)^{-1}} \pi \right) \to \min_\pi$$

$$\text{Output Constraints}: H(x, u_{bl}, \pi) = \begin{pmatrix} -I_m \\ I_m \end{pmatrix} H_u \pi + \Delta H(x, u_{bl}) \le 0$$

- **CBF Analytic Solution derived using Karush-Kuhn-Tucker (KKT) conditions**
  - (Quadratic cost + Linear constraints) ➔ Single optimum
  - Cost function weight is defined to decouple KKT conditions
  - Continuous piece-wise linear state feedback
  - No optimization-in-the-loop required

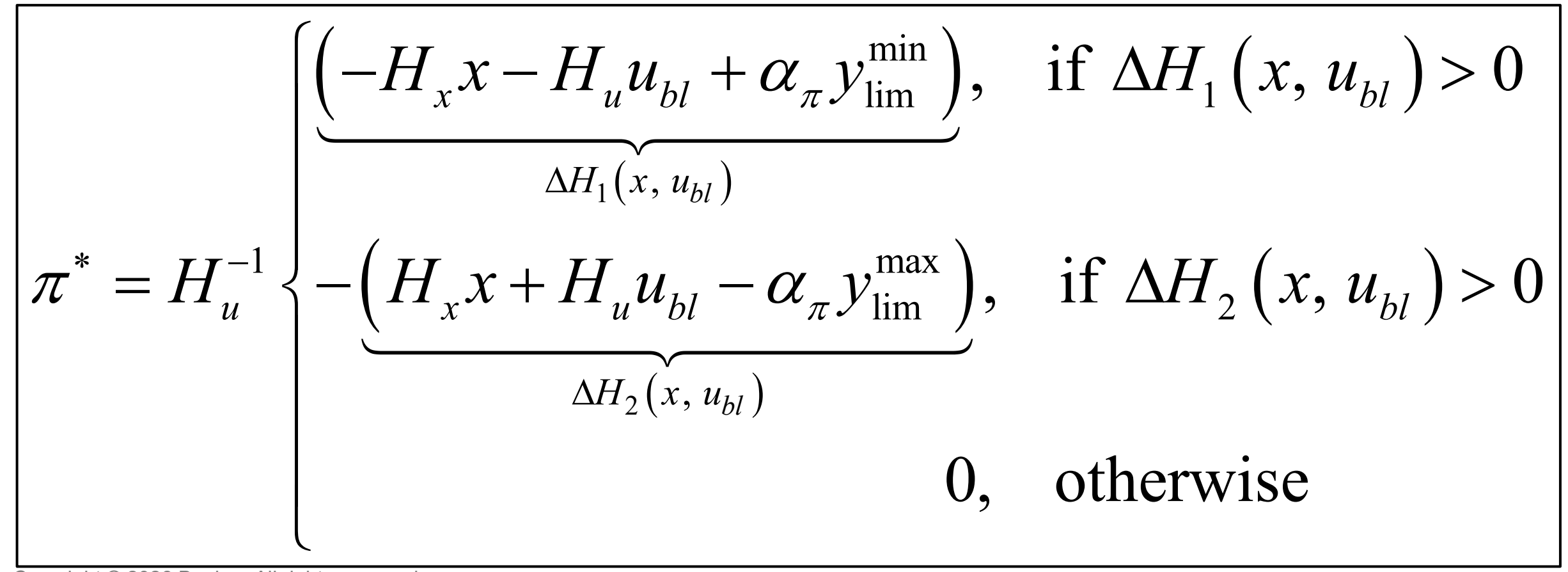

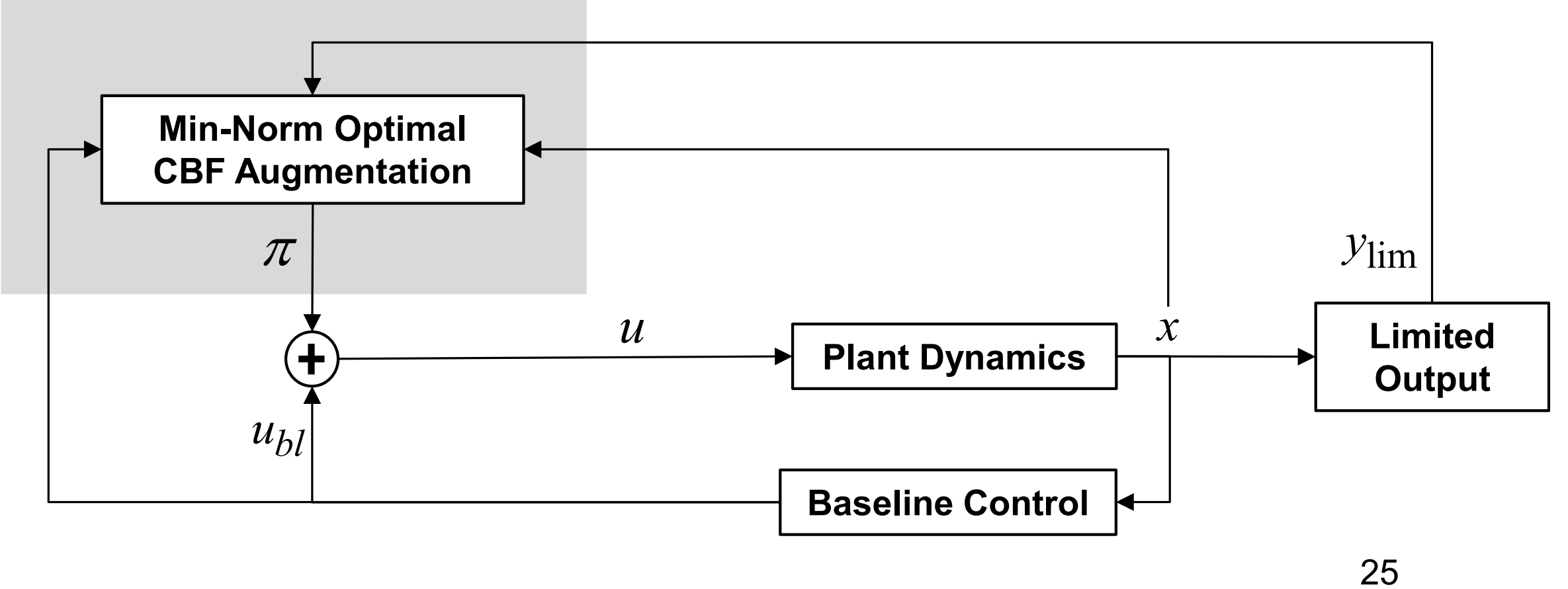

# CBF-Ability Criterion ➔ Closed-Loop System Stability & Boundedness

- **(Baseline + CBF) Controller ➔ Limited Modified Output**

$$Y_{\lim} = H_x x + H_u \underbrace{\left(u_{bl} + \pi^*\right)}_{u} \Rightarrow u = H_u^{-1}\left(Y_{\lim} - H_x x\right) \Rightarrow \underbrace{\left(\alpha_\pi\, y_{\lim}^{\min}\right)}_{Y_{\lim}^{\min}} \le Y_{\lim} \le \underbrace{\left(\alpha_\pi\, y_{\lim}^{\max}\right)}_{Y_{\lim}^{\max}}$$

- **Closed-Loop System Dynamics Driven by Limited Modified Output Signal**

$$\dot{x} = \underbrace{\left(A - B\, H_u^{-1} H_x\right)}_{\tilde{A}_{cl}} x + B\, H_u^{-1} Y_{\lim} = \tilde{A}_{cl} x + B\, H_u^{-1} Y_{\lim}$$

- **<u>CBF-Ability Criterion</u> (iff) for Closed-Loop Stability and Uniform Boundedness under (Baseline + CBF) Controller**
  - Defined by open-loop data and CBF parameters. Independent of baseline controller.
  - Stability margins defined as for gain-scheduled control.

$u_{out}$ → Input Breakpoint → $u_{in}$ → $(s\, I_n - A)^{-1} B$ → $-K_{CBF}$

**Input Breakpoint**

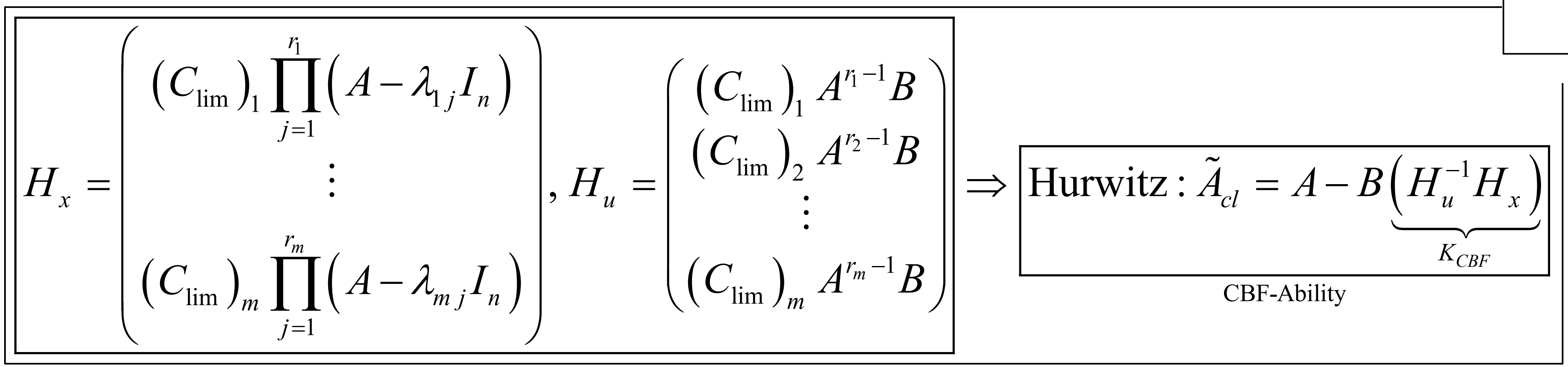

- **CBF = Continuous piece-wise linear state feedback augmentation**
- **Stability Margins defined**
- **Supports standard practices in flight control design and analysis**

# CBF Design Application for Baseline Controllers with Command-Feedforward Connections

- **Total Control = Baseline Servo-Controller + CBF Augmentation**
  - Baseline Proportional Servo-controller + CBF augmentation

$$u = \underbrace{\underbrace{-K_x\, x}_{u_P} + \underbrace{K_{ff}\, y_{cmd}}_{u_{ff}}}_{u_{bl}} = u_{bl} + \pi$$

- **CBF Augmentation Solution**

$$\pi^* = H_u^{-1} \begin{cases} \underbrace{\left(-\left(H_x + H_u K_x\right) x - H_u K_{ff}\, y_{cmd} + \alpha_\pi\, y_{\lim}^{\min}\right)}_{\Delta H_1}, & \text{if } \Delta H_1 > 0 \\ -\underbrace{\left(\left(H_x + H_u K_x\right) x + H_u K_{ff}\, y_{cmd} - \alpha_\pi\, y_{\lim}^{\max}\right)}_{\Delta H_2}, & \text{if } \Delta H_2 > 0 \\ 0, & \text{otherwise} \end{cases}$$

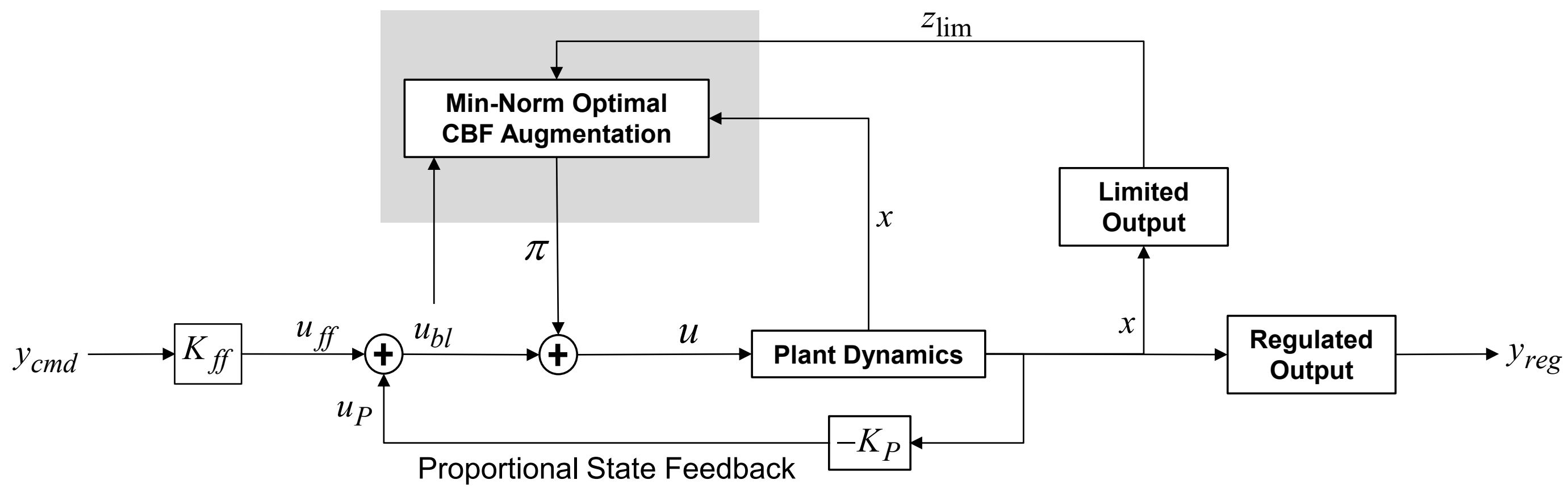

# Example: Simulation Data: CBF vs Projection Operator: Soft Min/Max Limits on System State

- **Open-Loop Scalar Unstable System**
- **Baseline Servo-Control Gains to Track Commands**
- **CBF Augmentation Analytic Solution**
  - CBF eigenvalue: -1
- **Simulation Test Data Comparison**
  - CBF Augmentation vs Rectangular Projection Operator ➔ Soft limits (via feedback) on system state

$$\text{Open-Loop Plant}: \dot{x} = x + u$$

$$\text{State Constraints}: \underbrace{x_{\min} = -0.5}_{y_{\lim}^{\min}} \le \underbrace{x}_{y_{\lim}} \le \underbrace{x_{\max} = 0.5}_{y_{\lim}^{\max}}$$

$$\text{Baseline Control}: u_{bl} = \underbrace{-4\,x}_{\text{Stabilizing Feedback}} + \underbrace{3x_{cmd}}_{\text{Command-Feedforward}}$$

$$\text{Total Control}: u = u_{bl} + \pi^*$$

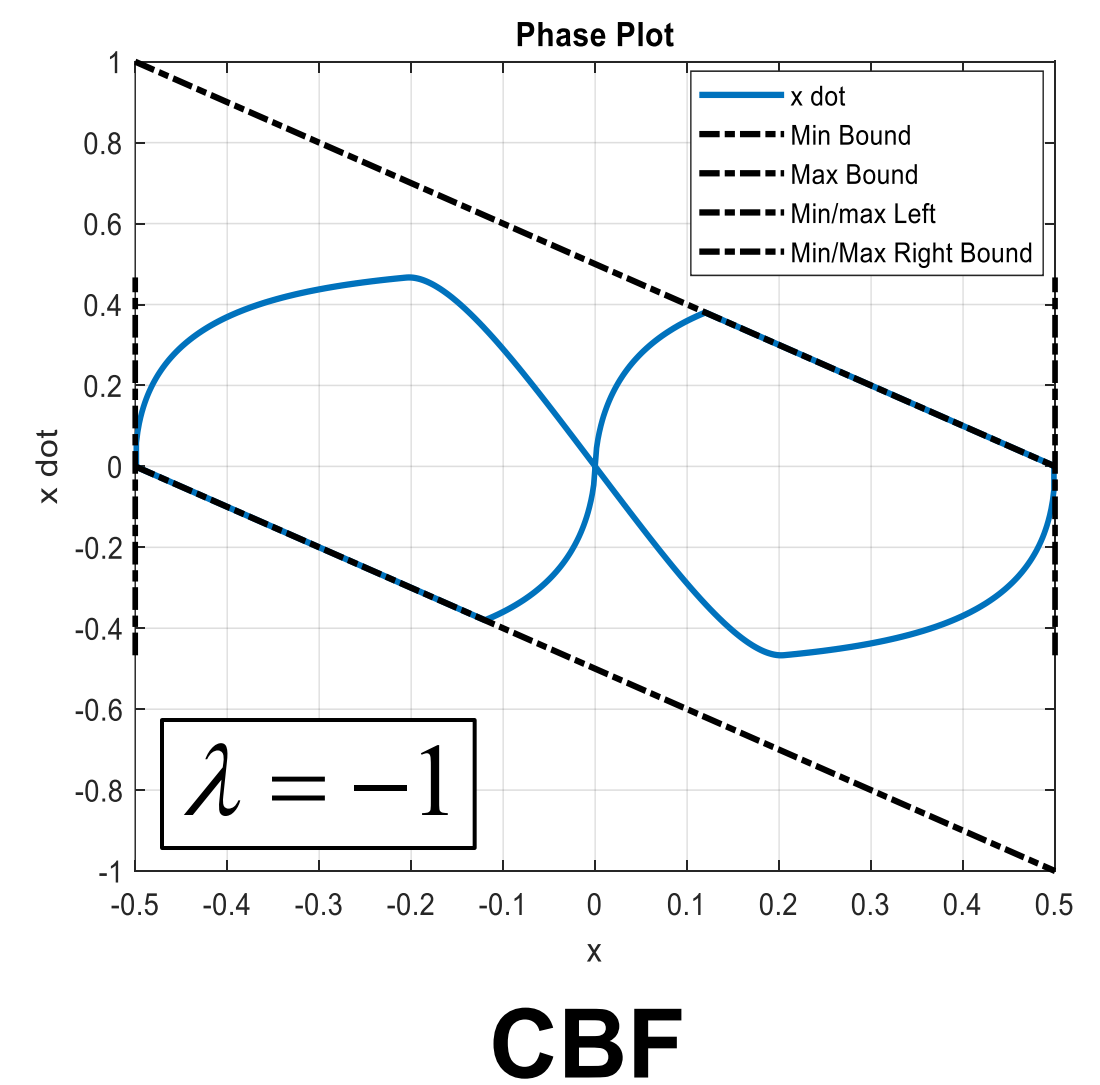

**CBF**

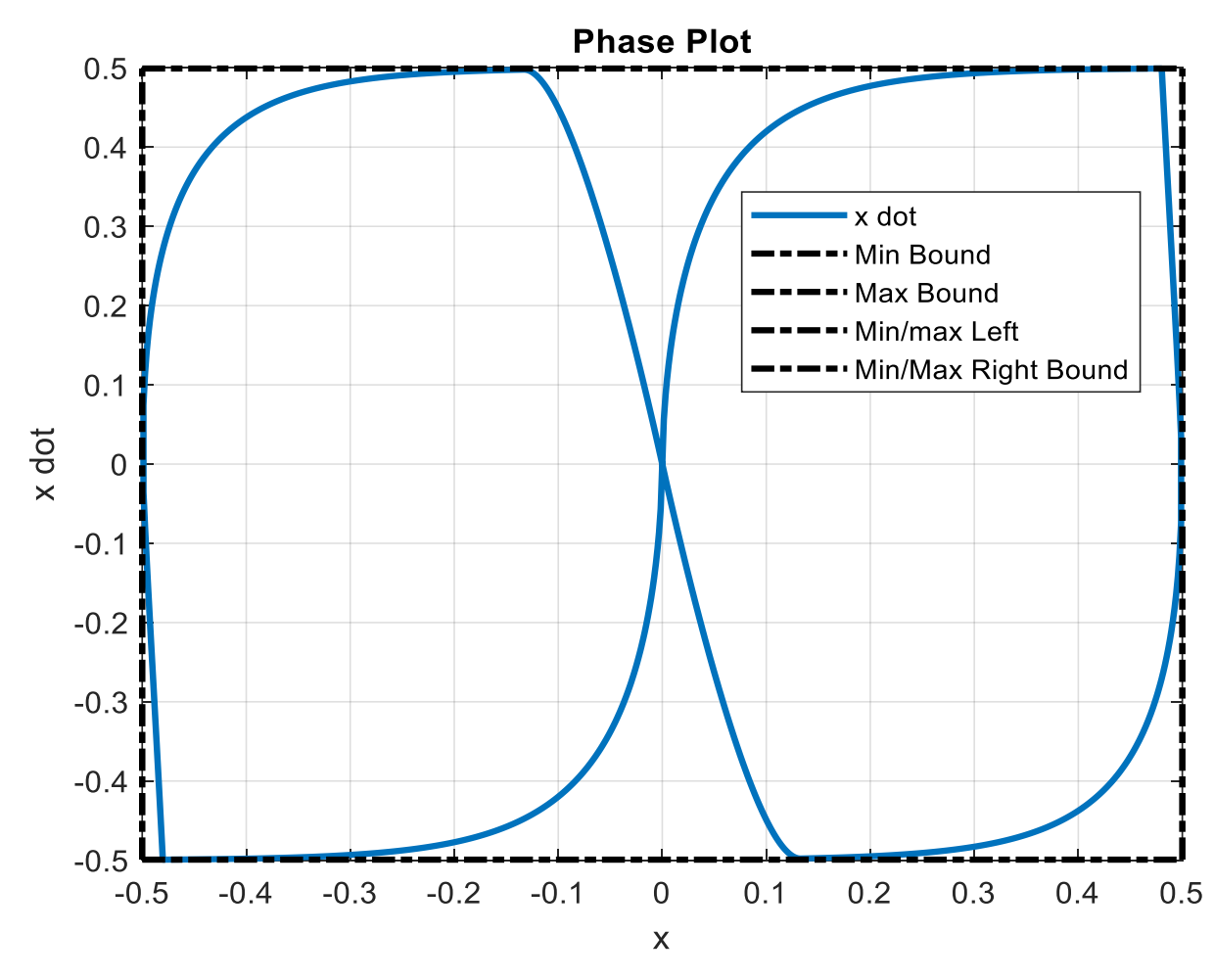

**Projection Operator**

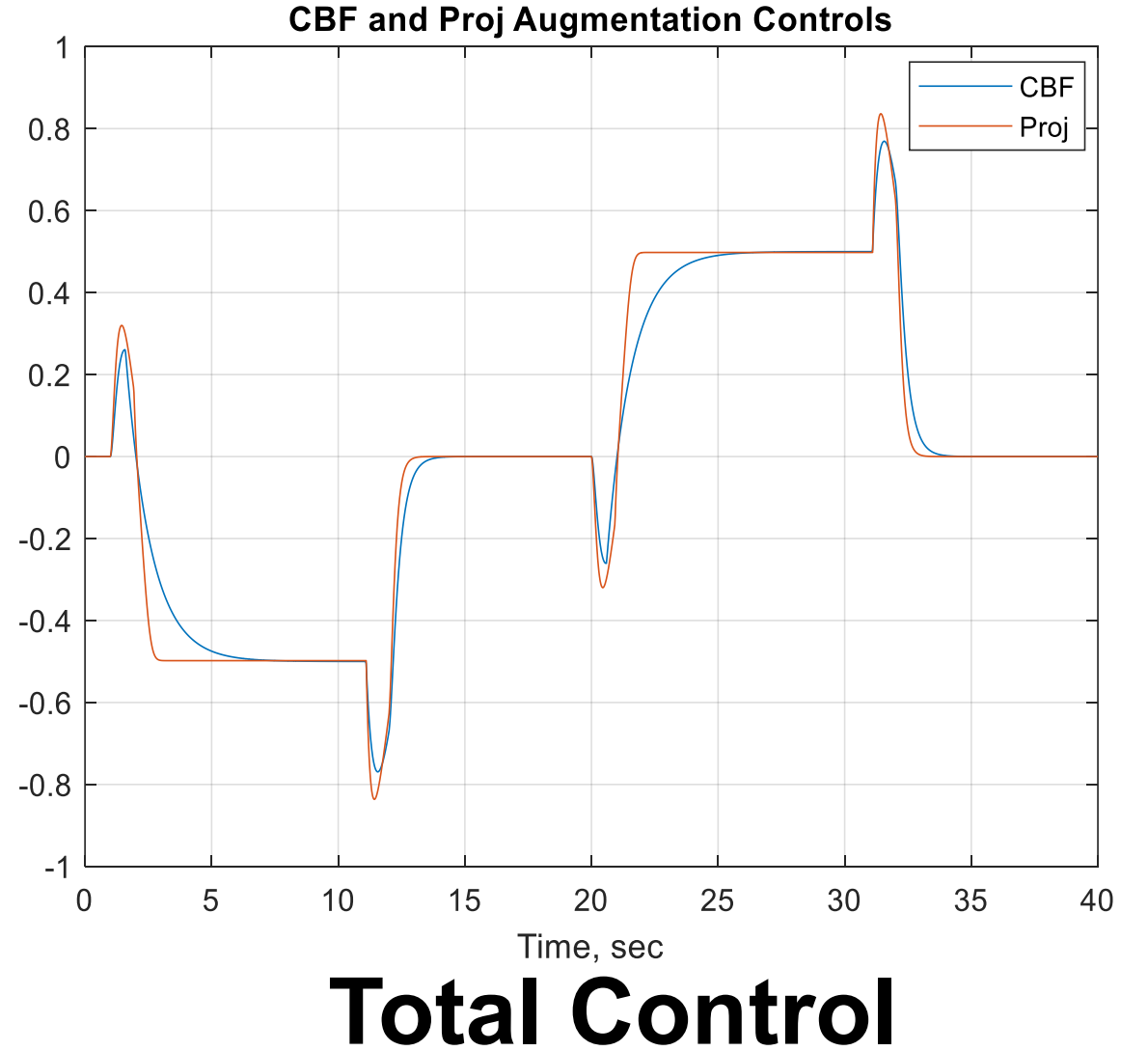

**Total Control**

# CBF Design Application for Servo-Controllers with (Proportional + Integral) Connections

- **Multi-Input-Multi-Output (MIMO) Linear Time Invariant (LTI) Dynamics**

$$\begin{aligned} \text{Open-Loop System Dynamics}: \dot{x}_p &= A_p x_p + B_p u, \quad x_p \in R^{n_p} \\ \text{Regulated Output}: y_{reg} &= C_{p\,reg} x_p + D_{p\,reg} u, \quad y_{reg} \in R^m \\ \text{Limited Output}: z_{\lim} &= C_{p\lim} x_p, \quad z_{\lim} \in R^m \end{aligned}$$

- **Input and Output Min/Max (Box) Operational Limits, (defined component-wise)**

$$u^{\min} \le u \le u^{\max}, \quad z_{\lim}^{\min} \le z_{\lim} \le z_{\lim}^{\max}$$

**2*m Double-Inequalities**

- **Control Design Task**
  - Find state feedback proportional-integral (PI) servo-controller to track external bounded commands in the presence of “soft” operational limits
    - via feedback, w/o using “hard” saturation logic

- **Assumptions**
  - System is controllable.
  - No transmission zeros at the origin, (for the input-to-regulated output dynamics).
  - Input / output constraints are feasible.

# Extended Open-Loop Dynamics with CBF AW Mod and Robust PI Servo-Controller with CBF Augmentation

- **Tracking Error Dynamics Augmented with <u>CBF AW input</u>**

$$\dot{e}_{yI} = \underbrace{y_{reg} - y_{cmd}}_{\text{Tracking Error}: e_y} + \underbrace{\boxed{v}}_{\text{CBF AW Augmentation}}$$

- **Extended Open-Loop System Dynamics**

$$\underbrace{\begin{pmatrix} \dot{e}_{yI} \\ \dot{x}_p \end{pmatrix}}_{\dot{x}} = \underbrace{\begin{pmatrix} 0_{m\times m} & C_{p\,reg} \\ 0_{n_p\times m} & A_p \end{pmatrix}}_{A} \underbrace{\begin{pmatrix} e_{yI} \\ x_p \end{pmatrix}}_{x} + \underbrace{\begin{pmatrix} D_{p\,reg} \\ B_p \end{pmatrix}}_{B} u + \underbrace{\begin{pmatrix} -I_m \\ 0 \end{pmatrix}}_{B_{cmd}} \left( y_{cmd} - v \right)$$

- **Robust PI Servo-Controller Solution w/o Operational Limits**

$$u_{bl} = -\underbrace{K_x}_{(K_I \quad K_P)} \underbrace{x}_{\begin{pmatrix} e_{yI} \\ x_p \end{pmatrix}} = \underbrace{-K_I\, e_{yI}}_{\text{Integral Output Feedback}: u_I} \underbrace{-K_P\, x_p}_{\text{Proportional State Feedback}: u_P} = u_I + u_P$$

- **<u>CBF Control Augmentation</u>**

$$u^{cmd} = \underbrace{u_{bl}}_{\text{Baseline Control Command}} + \underbrace{\boxed{w}}_{\text{CBF Control Augmentation}}$$

CBF AW and Control Augmentation Signals

# CBF Augmentation Solution for PI Servo-Controllers

- **State-Feedback Baseline Control Augmentation to Enforce Operational Constraints**
  - Continuous piece-wise linear state feedback analytic solution
    - No optimization-in-the-loop required
    - Applicable to MIMO LTI State Feedback Servo-Controllers with Proportional and Integral External Command connections
  - Component-wise min/max (“box”) limits on control and selected output
  - Limits enforced by feedback (closed-loop) vs sat-function (open-loop)
  - Provides anti-windup management for MIMO-coupled integrators
  - Supports linear analysis: Stability, Robustness and Performance

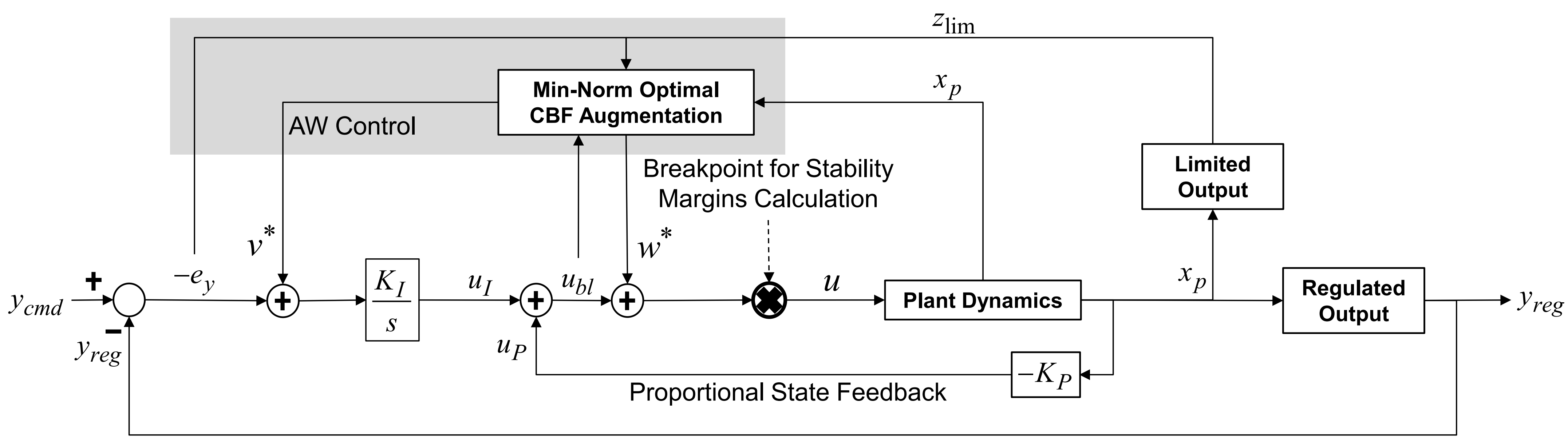

# Reduced Order Modeling of Unsteady Aerodynamics for Flight Control Applications

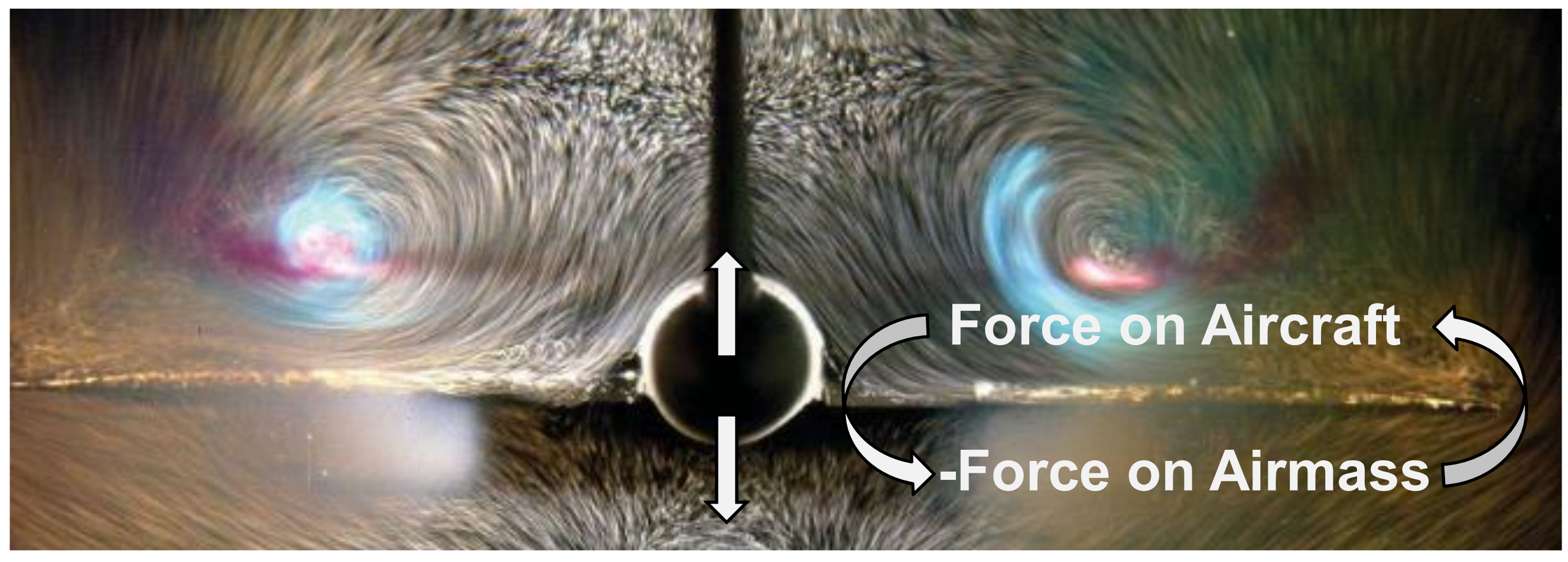

**<u>3rd Law of Newton</u>: For every action (force), there is an equal and opposite reaction**

- Airmass vorticity creates lift force
- Negative of Lift force pushes down on airmass ➔ Change in Vorticity ➔ Change in Lift Force on Aircraft
- Tightly coupled dynamical system: **<u>Aircraft Dynamics ← ➔ Vorticity Gust-like Dynamics</u>**

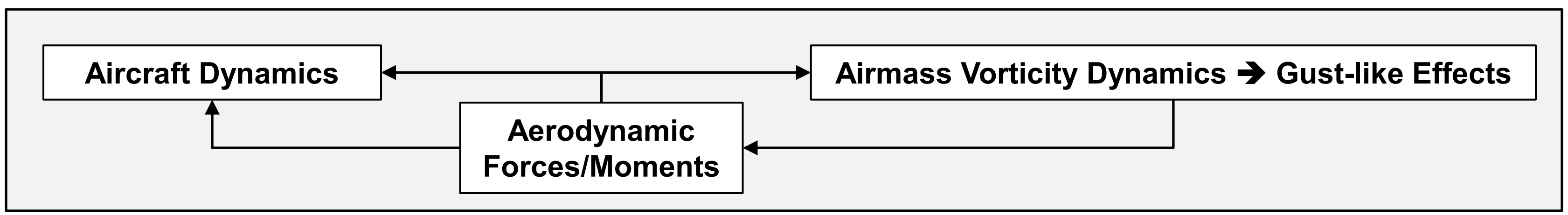

**Edward Norton Lorenz**
**1917-2008**

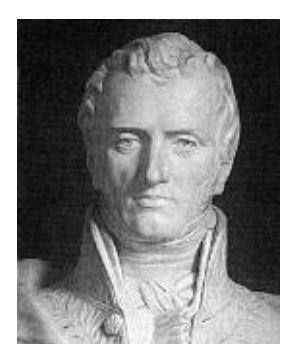
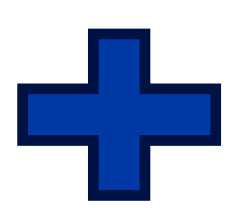
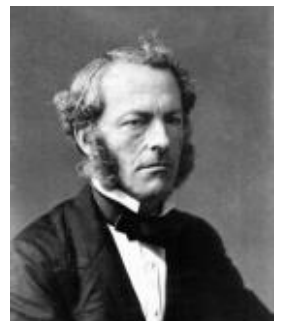
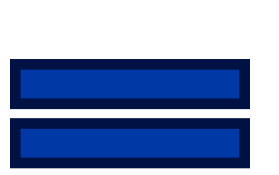

**The Navier-Stokes Equations**
1822 (Navier) to 1842–1850 (Stokes)

**Claude-Louis Navier**
**1785-1836**

**Sir George Gabriel Stokes, 1st Baronet**
**1819-1923**

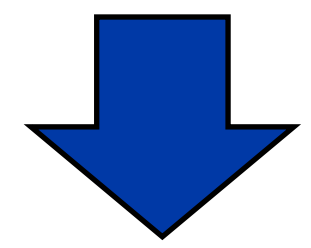

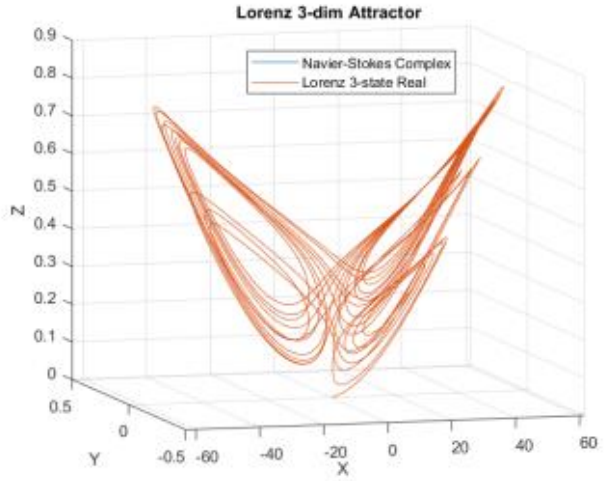

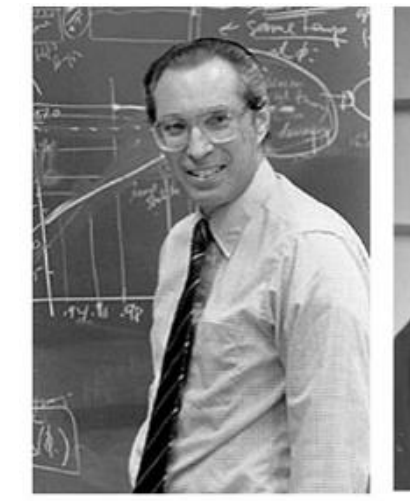

Barry Saltzman

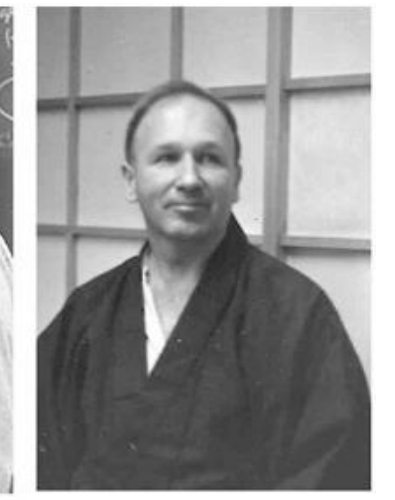

Edward Lorenz

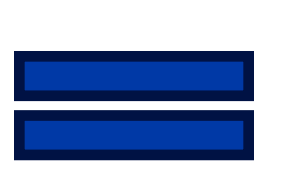

**The Deterministic Chaos Theory**
Reduced Order Model of Navier-Stokes Equations

# Development Roadmap

Lift Force
on Aircraft
Downwash

- **Navier-Stokes (NS) Fluid Dynamics ➔ Oberbeck-Boussinesq (OB) Model**
  - Unsteady fluid convection between two horizontal boundaries driven by external force
    - **3rd Law of Newton: Forces acting on Airmass = -Forces acting on Aircraft**
      - Two-Body Free Diagram

- **Fourier-Galerkin (FG) Approximations for <u>OB Dynamics</u>**
  - High order stable modes truncated

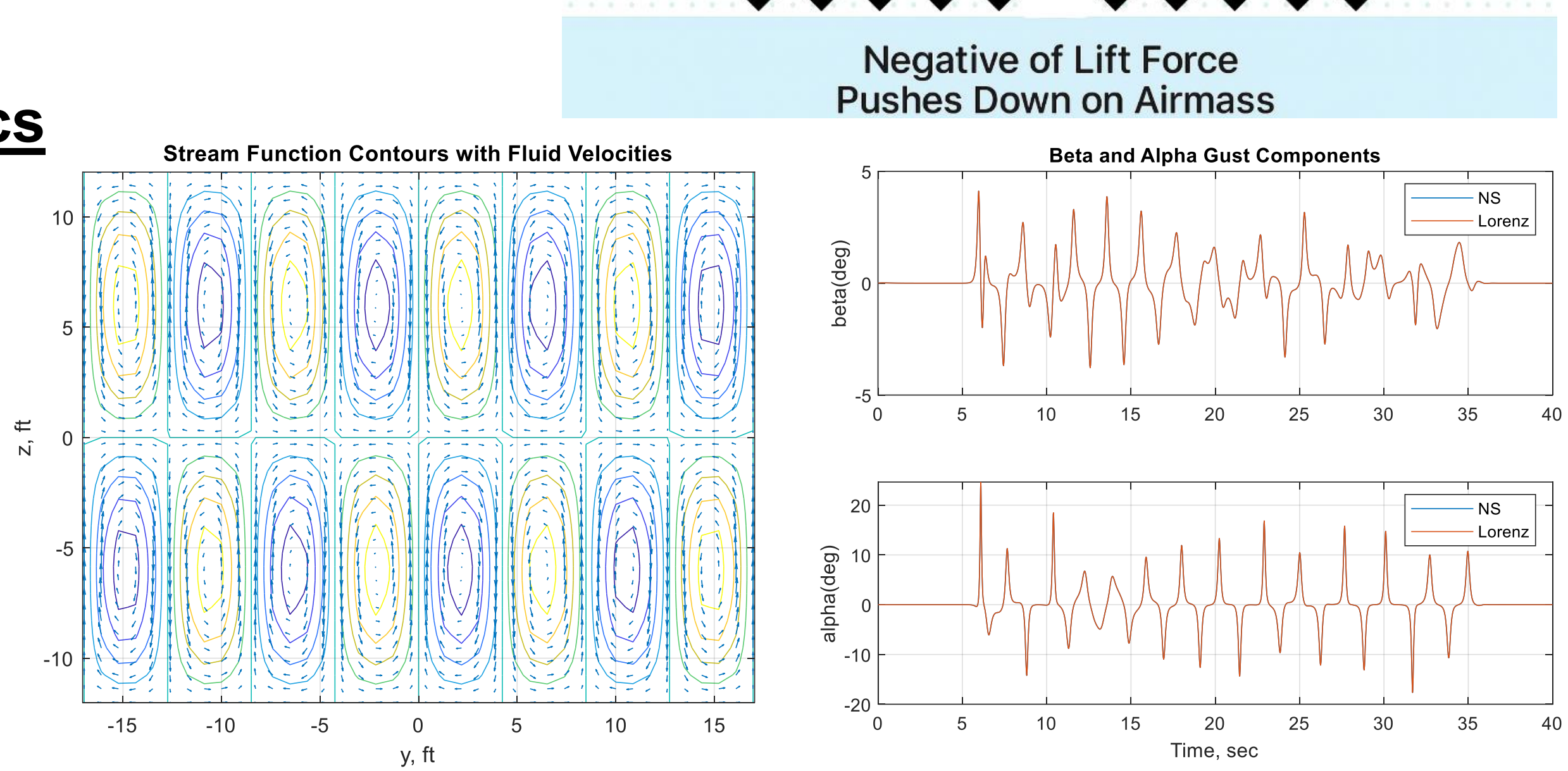

- **Lorenz Model and Strange Attractors**
  - Minimal (3-state) reduced order model (ROM)
    - Three coupled ordinary differential equations
    - Unsteady fluid convection dynamics
  - Driven by external force (such as negative of lift)
  - Simulates gust-like disturbances in aircraft body-fixed axes
    - Deterministic chaos

- **Aircraft Flight Dynamics with Unsteady Aerodynamic Effects**
  - Coupled system: Aircraft flight simulation with Lorenz-Based Gust Profile
  - Supports applications with faster than real time evaluation and testing

**gust-like airmass vorticity**
6-DoF
**lagged state-controls-dot-effects**
**states and controls**
Internal Dynamics
**aerodynamic forces**
Lorenz

# Unsteady Aerodynamics Modeling in Aircraft Body-Fixed Frame

- **Navier-Stokes (NS) Airmass Velocity Dynamics in Aircraft Body-Fixed Axes**
  - Driven by External Force + (negative of Aircraft Aerodynamic Force)

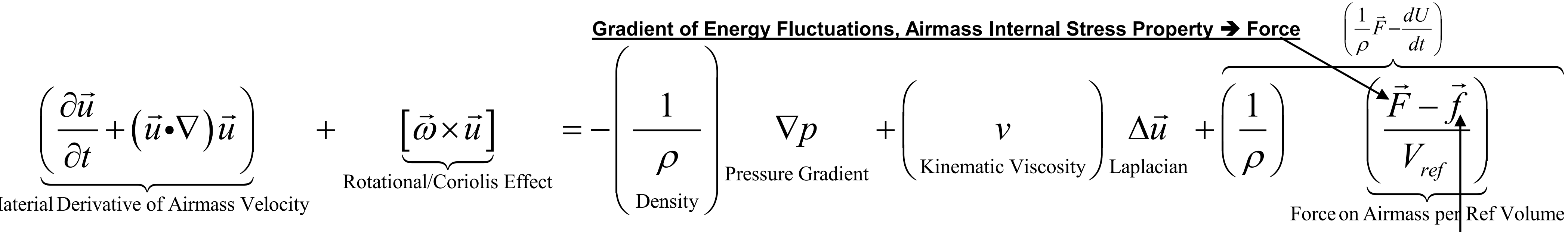

$$\left(\frac{\partial \vec{u}}{\partial t} + (\vec{u}\bullet\nabla)\vec{u}\right) + [\vec{\omega}\times\vec{u}] = -\left(\frac{1}{\rho}\right)\nabla p + (\nu)\Delta\vec{u} + \left(\frac{1}{\rho}\right)\left(\frac{\vec{F}-\vec{f}}{V_{ref}}\right), \quad \left(\frac{1}{\rho}\vec{F} - \frac{dU}{dt}\right)$$

- **Aircraft 6-DoF Dynamics**

$$\text{Translational}: \dot{\vec{U}} = -[\vec{\omega}\times\vec{U}] + \vec{g} + \frac{1}{m}\vec{f}$$

$$\text{Rotational}: \dot{\vec{\omega}} = J^{-1}\left([\vec{\omega}\times J\vec{\omega}] + \vec{M}\right)$$

**Newton's Third Law of Motion: Total force acting on airmass includes force component equal to the opposite of that acting on the aircraft**

- **Oberbeck-Boussinesq (OB) Dynamics: Stream Function and External Force Deviation Function**

$$\text{Velocity Curl}: \omega = [\nabla\times\vec{u}] \Rightarrow \text{Stream Function}: \frac{D}{Dt}(\nabla^2\psi) = \nu\nabla^4\psi - \frac{1}{\rho}\frac{\partial\varphi}{\partial y}$$

$$\text{Force Diffusivity}: \frac{D}{Dt}(\varphi) = \kappa\nabla^2\varphi - \frac{F_0}{H}\frac{\partial\psi}{\partial y}$$

Force Diffusivity from energy conservation. $F_0$: Force per z – direction

- **Fourier-Galerkin (FG) Approximation for OB Solutions → Reduced Order Model (ROM) → Lorenz (LZ) Dynamics for FG Expansion Coefficients of (*n*, *m*) – wave modes**

$$f(y,z,t) = \sum_n\sum_m c_{n,m}(t)\, e^{i\pi\left(\frac{y}{L}n + \frac{z}{H}m\right)}, \quad \text{where}: f = \{\psi, \varphi\}$$

Nondimensional LZ Dynamics

$$\dot{X}^* = -\sigma X^* + \sigma Y^*$$
$$\dot{Y}^* = r X^* - Y^* - X^* Z^*$$
$$\dot{Z}^* = -b Z^* + X^* Y^*$$

Large values (of $r$) start Chaos → Unsteady Convection

# Flight Control Technical Challenges and Open Problems

# Gain Scheduling and Closed-loop System Identification

- **Flight Control Gain Scheduling**
  - No theoretical justification
    - Works well when designed by subject matter experts
  - Baseline architecture in aerospace and other industries

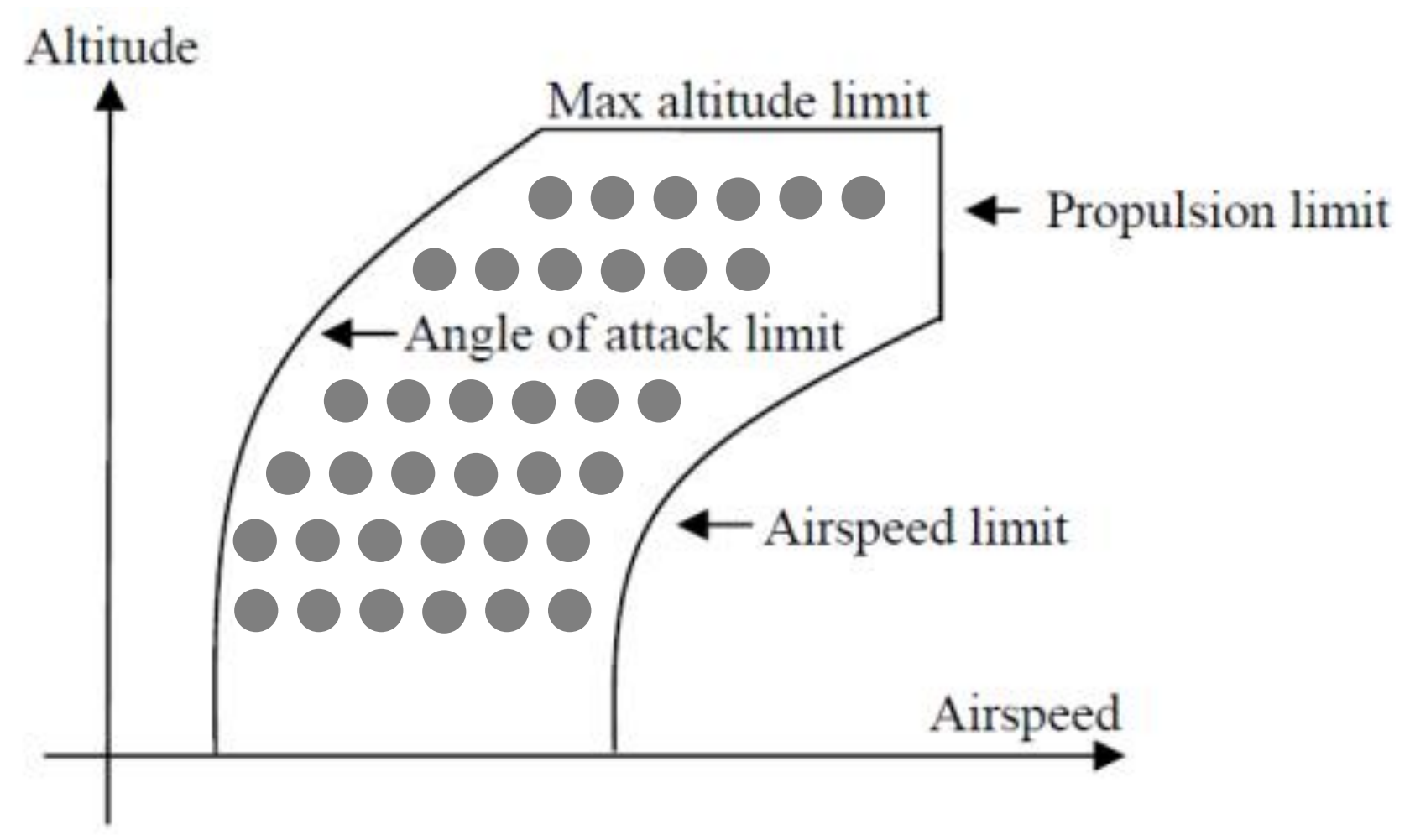

- **System Identification: Offline and Real-time**
  - Persistency of Excitation (PE) monitoring
  - Aerodynamics
    - In-Ground Effects (IGE)
    - Interlacing propeller wakes for over-actuated aircraft like VTOL
    - High Angle of Attack (AOA) flight conditions
  - Aircraft on-ground dynamics and in-air transitions: Taxi, Takeoff, and Landing
  - Structural dynamics and coupling with aerodynamics
    - Very and semi flexible aircraft

- **Sys ID Algorithms**
  - Closed-loop to Open-Loop
  - Must be robust to “unknown unknowns”
    - Actuator and Sensor Uncertainties
    - Environmental disturbances
    - Sensor noise, Turbulence, Gust, Time Delays, etc

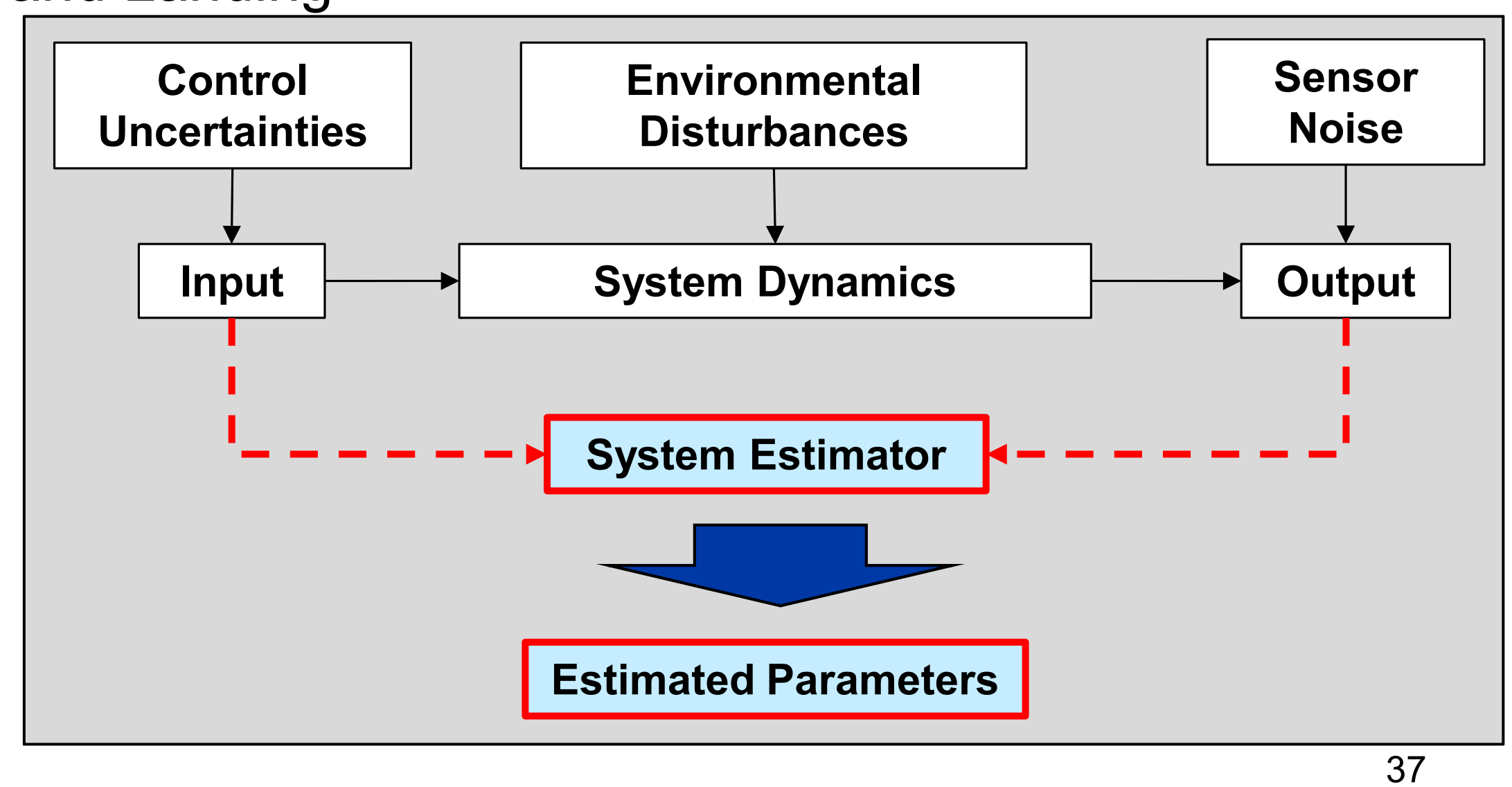

# Gain and Time-Delay Margins for Nonlinear Systems

- **Gain Margin**
  - Find min and max *k* such that closed-loop stability and tracking are preserved

- **Time-Delay Margin**
  - Find max $\tau$ such that closed-loop stability and tracking are preserved

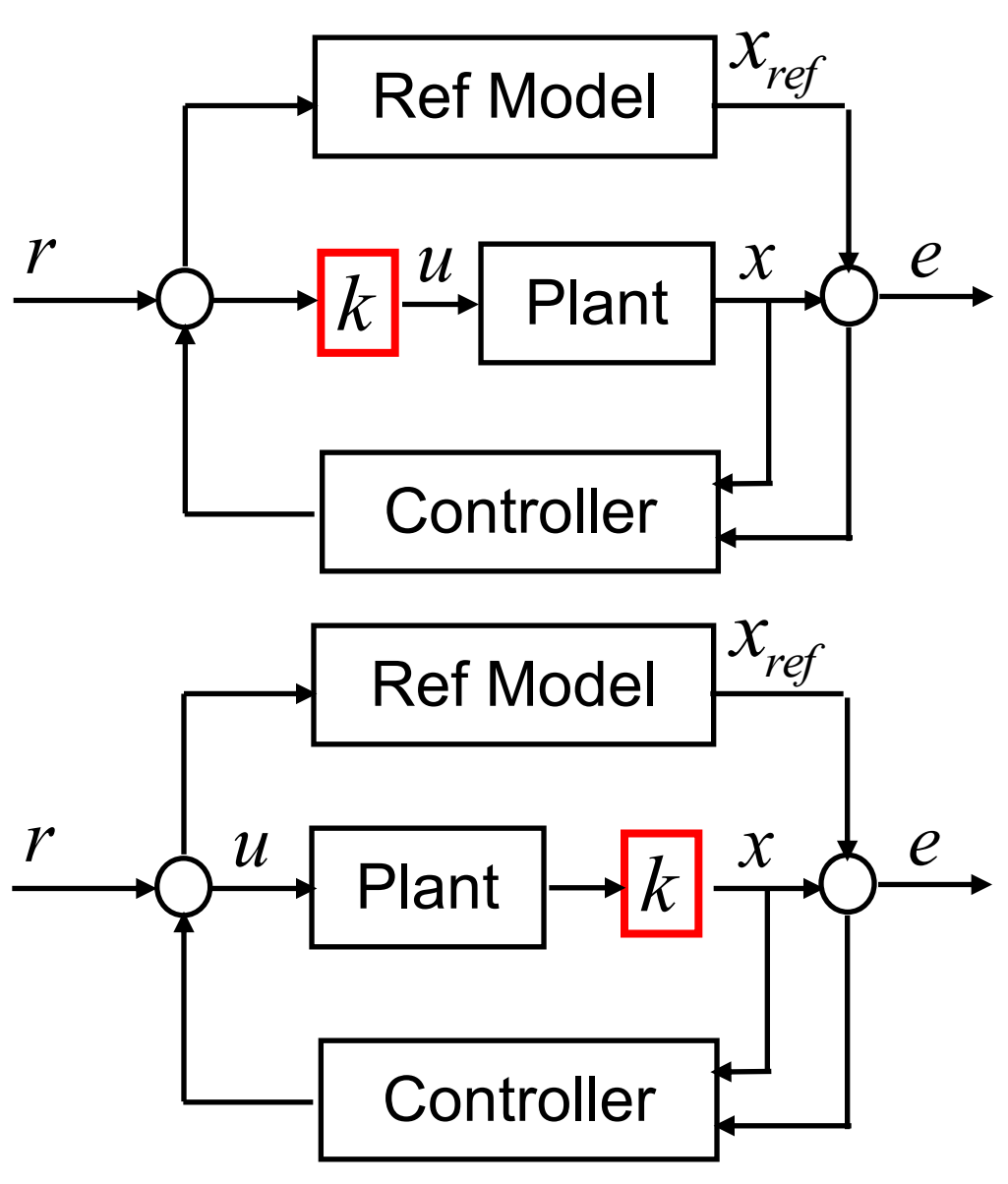

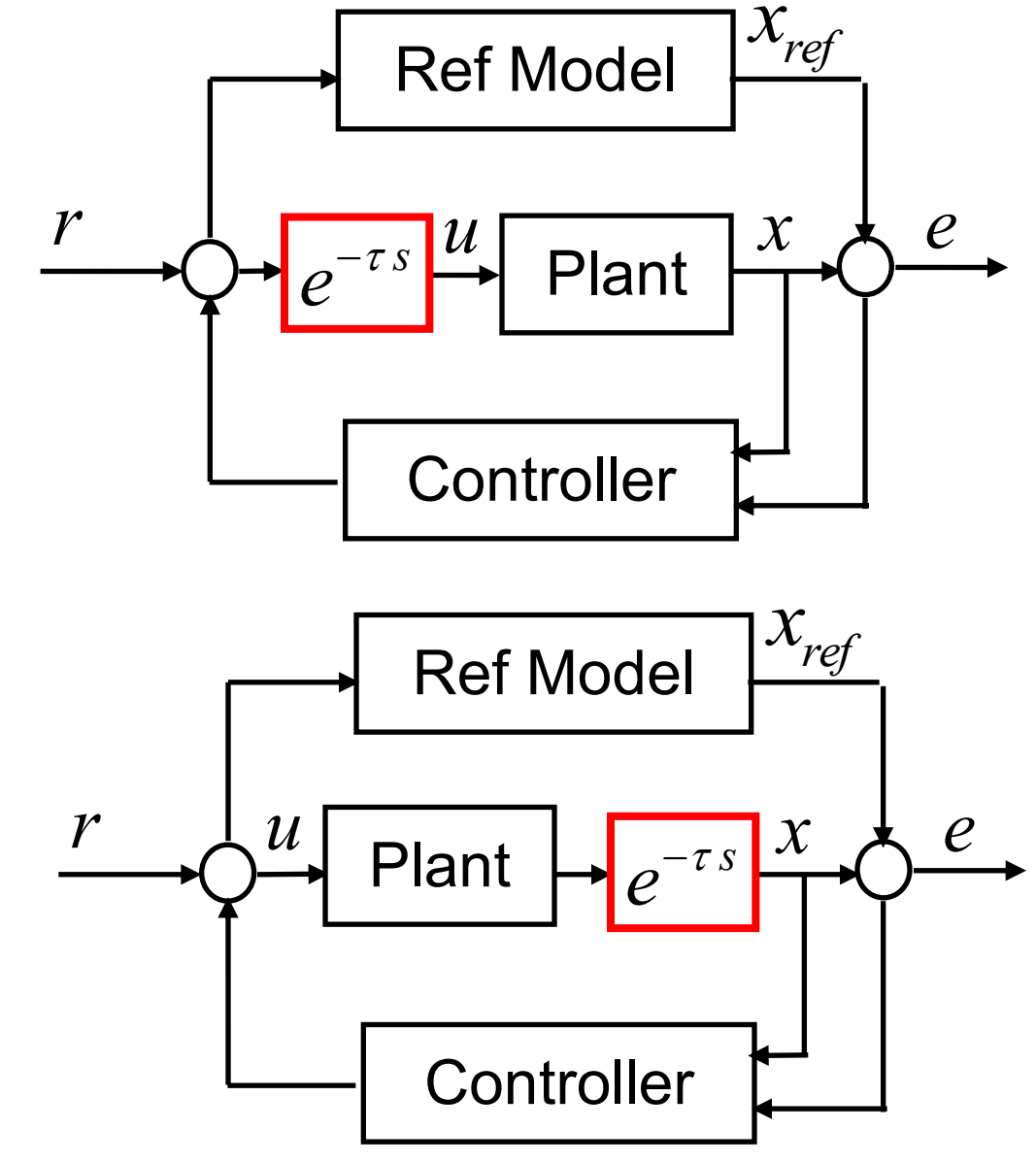

**Gain and Time-Delay Margins:**

Find min / max boundaries for $k_1$, $k_2$, $\tau_1$ and $\tau_2$ such that closed-loop stability and tracking are preserved

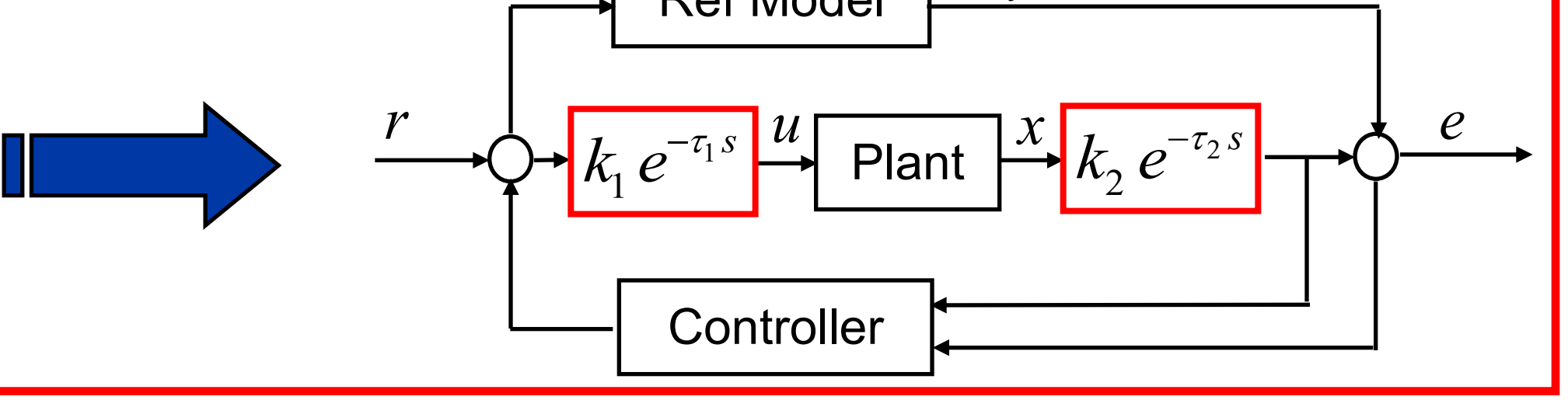

# The End … and The Beginning

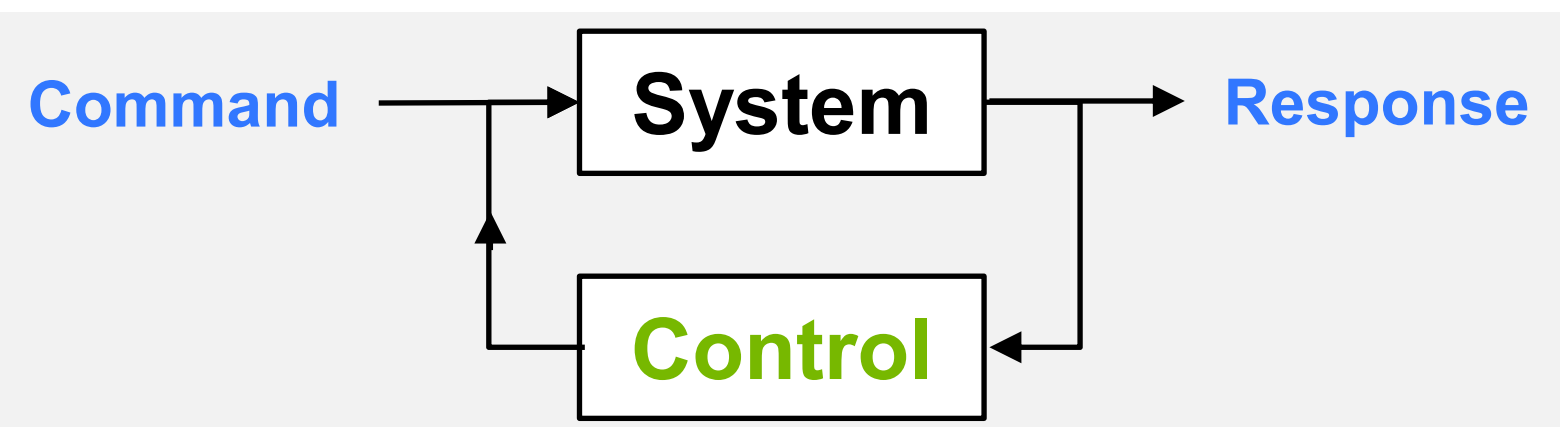

# Thank you!